\PassOptionsToPackage{dvipsnames}{xcolor}

\documentclass[
  sigconf,screen,
nonacm,dvipsnames
]{acmart}

\usepackage[utf8]{inputenc}
\usepackage[T1]{fontenc}
\usepackage{wrapfig}
\usepackage{mathpartir}
\usepackage{amsmath}
\usepackage{amsfonts}
\usepackage{amsthm}
\usepackage{mathtools}
\usepackage{stmaryrd}
\usepackage{dsfont}
\usepackage{bbold}
\usepackage{array}
\usepackage{siunitx}
\usepackage{paralist}
\usepackage[inline]{enumitem}
\usepackage{subcaption}

\RequirePackage{paralist}

\usepackage[noabbrev,capitalize]{cleveref}
\makeatletter
\DeclareRobustCommand{\crefnosort}[1]{\begingroup\@cref@sortfalse\cref{#1}\endgroup }
\makeatother

\usepackage{xcolor}
\usepackage{booktabs}
\usepackage{microtype}

\RequirePackage{graphicx}
\graphicspath{{./img/}}

\usepackage{tikz}

\usetikzlibrary{shapes.geometric,positioning,arrows.meta,fit,backgrounds,calc}
\usepackage{fancybox}

\usepackage{bm}

\DeclareUnicodeCharacter{03C4}{\ensuremath{\tau}}

\NewCommandCopy{\mathphi}{\phi}
\renewcommand{\phi}{\ensuremath{{\mathphi}}}

\makeatletter
\newcommand\ifanon[2]{\if@ACM@anonymous#1\else#2\fi}
\makeatother

\newcommand\st[1][]{\ensuremath{{\sigma#1}}}
\newcommand\pre[1][]{\ensuremath{{\Phi#1}}}
\newcommand\post[1][]{\ensuremath{{\Psi#1}}}
\newcommand\relpre[3][]{\ensuremath{{{#2} \mathrel{\pre[#1]} {#3}}}}
\newcommand\relpost[3][]{\ensuremath{{{#2} \mathrel{\post[#1]} {#3}}}}

\newcommand\Ip[1][]{\ensuremath{{I_{P}#1}}}
\newcommand\Iq[1][]{\ensuremath{{I_{Q}#1}}}
\newcommand\domain[1][]{\ensuremath{{\alpha#1}}}
\newcommand\ans[1][]{\ensuremath{{\mathsf{A}#1}}}
\newcommand\events[1][]{\ensuremath{{\mathsf{E}#1}}}
\newcommand\results[1][]{\ensuremath{{\mathsf{R}#1}}}

\newcommand\lv[1][]{\ensuremath{{\ell#1}}}

\newcommand\csimname[1][]{\ensuremath{{\preccurlyeq#1}}}
\newcommand\csim[3][]{\ensuremath{{{#2} \mathrel{\csimname[#1]} {#3}}}}
\newcommand\prname{\operatorname{Pr}}
\newcommand\pr[2][]{\ensuremath{{\prname_{#1}\!\left[{#2}\right]}}}

\newcommand\euttname[1][]{\ensuremath{{\approx#1}}}
\newcommand\ruttname[1][]{\ensuremath{{\raisebox{-1pt}{\ensuremath{{\overset{\ensuremath{\text{\tiny{\ensuremath{e}}}}}{\approx}}}}#1}}}

\makeatletter
\newcommand\rhltuple[5][]{\ensuremath{{#1{#2} \mathrel{\sim} #1{#3} \mathrel{:} {#4} \mathrel{\Rightarrow}
  {#5}}}}

\newcommand\rhl@gen[7][]{\ensuremath{{{#6}^{#7}~\rhltuple[#1]{#2}{#3}{#4}{#5}}}}
\newcommand\rhl@valid[5][]{\rhl@gen[#1]{#2}{#3}{#4}{#5}{\models}{}}
\newcommand\rhl@provable[5][]{\rhl@gen[#1]{#2}{#3}{#4}{#5}{\vdash}{}}

\newcommand\rhl{\@ifstar{\rhl@valid}{\rhl@provable}}

\newcommand\rhlPQ@valid[6][]{\ensuremath{{{#2}~\rhl@valid[#1]{#3}{#4}{#5}{#6}}}}
\newcommand\rhlPQ@provable[6][]{\ensuremath{{{#2}~\rhl@provable[#1]{#3}{#4}{#5}{#6}}}}
\newcommand\rhlPQ{\@ifstar{\rhlPQ@valid}{\rhlPQ@provable}}

\newcommand\compose[2]{\ensuremath{{{#1} \circ {#2}}}}
\newcommand\composename{\ensuremath{{\circ}}}

\newcommand\rhle{\rhl@valid[]}

\newcommand\rhllv[5]{\rhl@valid[]{#1}{#2}{{#3}/{#4}}{#5}}

\newcommand\rhlfun[6]{\rhl@provable{#1}{#2}{{#3} \mid {#4}}{{#5} \mid {#6}}}
\makeatother

\newcommand\fbody[2][]{\ensuremath{{{#2}_{\text{\tiny{\textsf{body}}}}#1}}}
\newcommand\farg[2][]{\ensuremath{{{#2}_{\text{\tiny{\textsf{arg}}}}#1}}}
\newcommand\fres[2][]{\ensuremath{{{#2}_{\text{\tiny{\textsf{res}}}}#1}}}

\newcommand\adversary[1][]{\ensuremath{{\mathscr{A}#1}}}
\newcommand\challenger[1][]{\ensuremath{{\mathscr{C}#1}}}

\newcommand\Decapname{\ensuremath{{\operatorname{Decap}}}}
\newcommand\Decap[3][]{\ensuremath{{\Decapname^{#1}\mathchoice{\!}{\!}{}{}\left({#2}, {#3}\right)}}}
\newcommand\Decapsk{\ensuremath{{\Decapname\mathchoice{\!}{\!}{}{}\left(\ensuremath{{\mathit{sk}}}\right)}}}

\newcommand\msg[1][]{\ensuremath{{m#1}}}

\newcommand\maxtime[1][]{\ensuremath{{T#1}}}

\newcommand\rocqlogo{\raisebox{-0.2ex}{\includegraphics[height=0.7em]{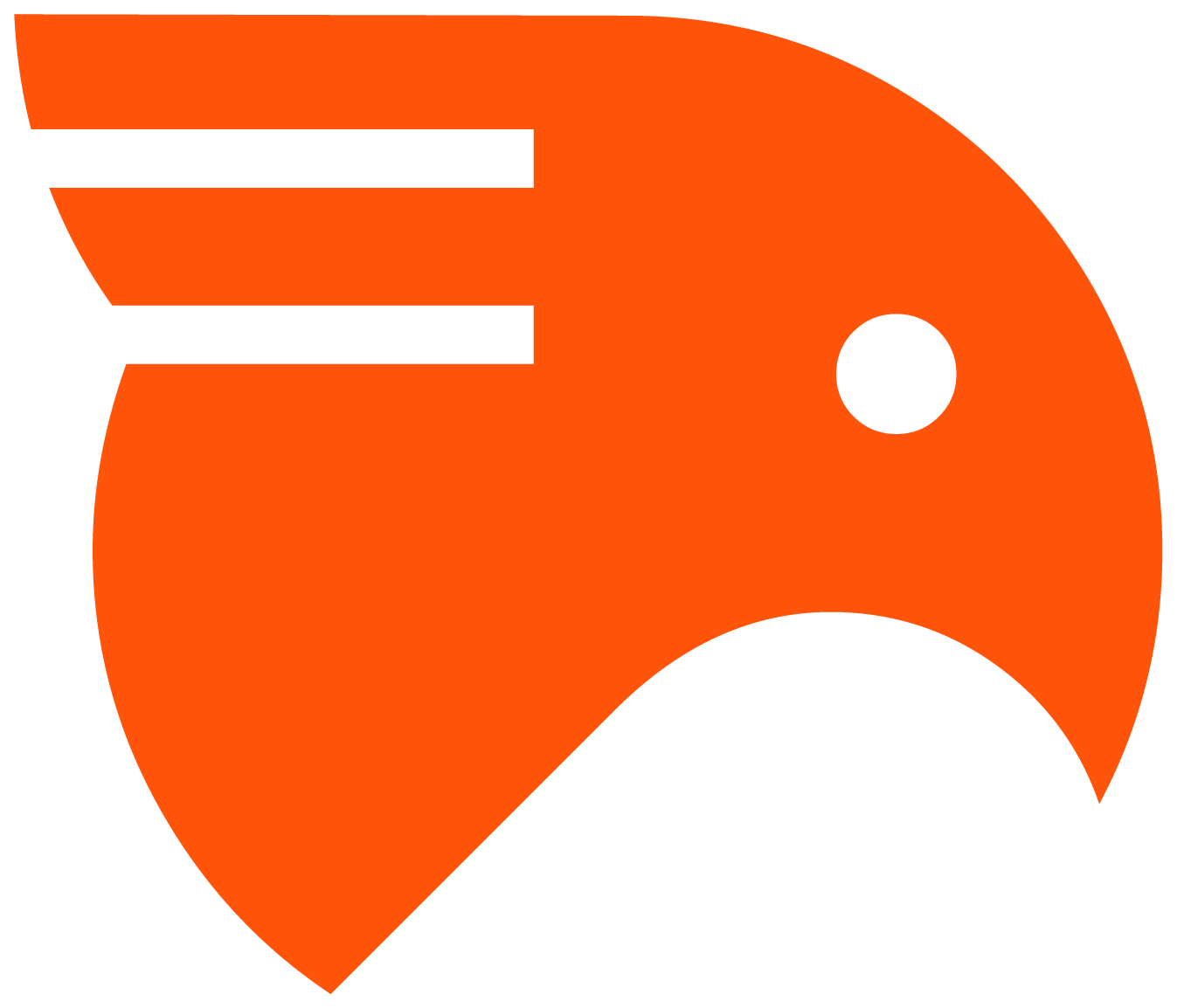}}}
\newcommand\inartifact[1][cite.\ifanon{anonymous-artifact}{public-artifact}]{\hyperlink{#1}{\rocqlogo}}

\DeclareMathOperator{\dom}{dom}

\begin{document}

\title{KEM-IND-CCA{}-Preserving Compilation of \textsc{Jasmin}{}'s ML-KEM}

\author{Santiago Arranz-Olmos}
\orcid{0009-0007-7425-570X}
\affiliation{\institution{MPI-SP}
  \city{Bochum}
  \country{Germany}
}
\email{santiago.arranz-olmos@mpi-sp.org}

\author{Gilles Barthe}
\orcid{0000-0002-3853-1777}
\affiliation{\institution{MPI-SP}
  \city{Bochum}
  \country{Germany}
}
\affiliation{\institution{IMDEA Software Institute}
  \city{Madrid}
  \country{Spain}
}
\email{gilles.barthe@mpi-sp.org}

\author{Lionel Blatter}
\orcid{0000-0001-9058-2005}
\affiliation{\institution{MPI-SP}
  \city{Bochum}
  \country{Germany}
}
\email{lionel.blatter@mpi-sp.org}

\author{Benjamin Grégoire}
\orcid{0000-0001-6650-9924}
\affiliation{\institution{Inria}
  \city{Sophia Antipolis}
  \country{France}
}
\email{benjamin.gregoire@inria.fr}

\author{Vincent Laporte}
\orcid{0000-0002-3468-352X}
\affiliation{\institution{Inria}
  \city{Nancy}
  \country{France}
}
\affiliation{\institution{Université de Lorraine}
  \city{Nancy}
  \country{France}
}
\email{Vincent.Laporte@inria.fr}

\author{Paolo Torrini}
\orcid{}
\affiliation{\institution{Inria}
  \city{Sophia Antipolis}
  \country{France}
}
\email{paolo.torrini@inria.fr}

\begin{abstract}
High-assurance cryptography provides strong guarantees that source
implementations are functionally correct and provably secure. In
this paper, we demonstrate that the \textsc{Jasmin}{} compiler preserves
functional correctness and KEM-IND-CCA{} security (which were established
in prior work) of a highly optimized \textsc{Jasmin}{} implementation of ML-KEM
used in the popular messenger Signal.  Our proof of preservation is
fully mechanized in the \textsc{Rocq}{} prover and is based on three general
contributions:
\begin{enumerate*}
\item A general framework for modeling game-based security and for
  reasoning about preservation of game-based security under
  compilation.
\item A new, interaction-trees-based semantics of \textsc{Jasmin}{} and
  assembly programs. Our new semantics supports features required by
  ML-KEM, such as probabilistic computations and rejection sampling
  routines.
\item A new relational Hoare logic for interaction trees, which we use
  to prove correctness of the \textsc{Jasmin}{} compiler under our new
  semantics.
\end{enumerate*}
 \end{abstract}



\maketitle

\section{Introduction}\label{sec:introduction}
Formally verified cryptography~\cite{DBLP:conf/sp/BarbosaBBBCLP21}
provides strong guarantees that cryptographic schemes are provably
secure and that implementations are functionally correct with respect
to these schemes. Moreover, these guarantees can be composed to yield
stronger assurance that implementations are both correct and
secure. However, a final and significant step remains: proving that
correctness and security guarantees carry to assembly code. This final
step eliminates relying on compilers, which may contain bugs or
introduce security
concerns~\cite{DBLP:conf/sp/DSilvaPS15,DBLP:conf/uss/XuLDDLWPM23}.

This paper reports on an experiment to carry out this step for a
concrete case: a highly optimized implementation of
ML-KEM~\cite{formosa-mlkem}.  ML-KEM
is the newly adopted FIPS 203 post-quantum key encapsulation
mechanism~\cite{NIST-FIPS-203-2024}. ML-KEM is being actively deployed
as part of the post-quantum transition, which is expected to be
completed around 2035~\cite{NISTpqctransition}.  For example, Google
recently announced they aim to complete their transition by
2029~\cite{google-pqc}. The implementation we target is written in
\textsc{Jasmin}{}, a programming language for high-assurance
cryptography~\cite{DBLP:conf/ccs/AlmeidaBBBGLOPS17,DBLP:conf/sp/AlmeidaBBGKL0S20},
and has been formally verified for correctness and
indistinguishability against chosen ciphertext attacks
(KEM-IND-CCA{})~\cite{DBLP:journals/tches/AlmeidaBBGLL00Q23,DBLP:conf/crypto/AlmeidaOBBDGLLLOPQSS24}---both
proofs in the \textsc{EasyCrypt}{} proof
assistant~\cite{DBLP:conf/crypto/BartheGHB11,DBLP:conf/fosad/BartheDGKSS13}.
As of 2026, the \textsc{Jasmin}{} implementation is used in the popular secure
messaging application Signal~\cite{almeida2026formosa}.
\looseness=-1

Unfortunately, the current infrastructure of \textsc{Jasmin}{} does not
suffice for carrying functional correctness and KEM-IND-CCA{} security
from source \textsc{Jasmin}{} programs, such as ML-KEM, to generated assembly.
The foremost reason is
that ML-KEM performs probabilistic computations and in particular uses
rejection sampling (a.k.a.\ repeat until success). Hence, executions
of ML-KEM are potentially unbounded---in probabilistic programming
jargon, ML-KEM satisfies a probabilistic notion of termination known as
\emph{almost sure termination}. However, the original semantics of
\textsc{Jasmin}{} programs is limited to deterministic and terminating
programs. To overcome these issues, we equip both \textsc{Jasmin}{} and assembly
programs with a new denotational semantics that models probabilistic
behavior, and we prove compiler correctness for these new semantics.

The second reason is that game-based security notions such as
KEM-IND-CCA{} are modeled as adversarial, probabilistic (hyper)properties
that fall outside the scope typically considered in compiler correctness
and secure compilation~\cite{handbook:pa}, and in particular in the
context of the \textsc{Jasmin}{} compiler.  We address this limitation by
defining a general framework to define game-based security and by
lifting the compiler correctness result to show its preservation.

We address both challenges using interaction trees
(ITrees)~\cite{DBLP:journals/pacmpl/XiaZHHMPZ20,DBLP:journals/pacmpl/ZakowskiBYZZZ21}, a semantic framework well-suited
to capture game-based security.
Intuitively, the reason is that \emph{interactions} with an
environment---which are a central notion in ITrees, realized as
``events'' and discussed later---support all relevant features of
game-based security (queries, randomness, nontermination, statefulness,
etc.).

\begin{figure}
  \centering
  \scalebox{0.95}{\begin{tikzpicture}[
  font=\footnotesize,
  repr/.style={
    draw,rounded corners=4pt,align=center,inner sep=5pt,
  },
  prop/.style={
    draw,rounded corners=4pt,align=center,inner sep=4pt,
    thick, dotted, minimum width=8em,
  },
]

  \node[prop] (fcL) {\textbf{Functional}\\\textbf{Correctness}\\
    (\textsc{EasyCrypt}{} \cite{DBLP:journals/tches/AlmeidaBBGLL00Q23})};

  \node[prop, below=16pt of fcL] (secL) {\textbf{KEM-IND-CCA{}}\\
    (\textsc{EasyCrypt}{} \cite{DBLP:conf/crypto/AlmeidaOBBDGLLLOPQSS24})};

  \node[above=6pt of fcL, inner sep=2pt] (titleL) {\vphantom{\textbf{Jg}}};

  \node[prop, right=9em of fcL] (fcR) {\textbf{Assembly-Level}\\\textbf{Functional}\\\textbf{Correctness}\\
    (\textsc{Rocq}{})};

  \node[prop] (secR) at (secL -| fcR) {\textbf{Assembly-Level}\\\textbf{KEM-IND-CCA{}}\\
    (\textsc{Rocq}{})};

  \node[inner sep=2pt] (titleR) at (titleL -| fcR) {\vphantom{\textbf{Jg}}};

  \draw[semithick, line cap=butt, shorten <=2pt, shorten >=5pt]
    ([yshift= 1.3pt]fcL.east) -- ([yshift= 1.3pt]fcR.west);
  \draw[semithick, line cap=butt, shorten <=2pt, shorten >=5pt]
    ([yshift=-1.3pt]fcL.east) -- ([yshift=-1.3pt]fcR.west);
  \draw[semithick, double distance=2pt, -{Implies}, line cap=butt,
        shorten >=2pt]
    ($(fcR.west)+(-5pt,0)$) -- (fcR.west);

  \path (fcL) -- (fcR)
    node[midway,above=2pt] {
      \begin{Bcenter}Compiler\\Correctness\end{Bcenter}
    }
    node[midway,below=2pt] {
      \Cref{thm:compiler-correctness} (\textsc{Rocq}{})
    };

  \draw[semithick, line cap=butt, shorten <=2pt, shorten >=5pt]
    ([yshift= 1.3pt]secL.east) -- ([yshift= 1.3pt]secR.west);
  \draw[semithick, line cap=butt, shorten <=2pt, shorten >=5pt]
    ([yshift=-1.3pt]secL.east) -- ([yshift=-1.3pt]secR.west);
  \draw[semithick, double distance=2pt, -{Implies}, line cap=butt,
        shorten >=2pt]
    ($(secR.west)+(-5pt,0)$) -- (secR.west);

  \path (secL) -- (secR)
    node[midway,above=2pt] {
      \begin{Bcenter}Security\\Preservation\end{Bcenter}
    }
    node[midway,below=2pt] {
      \Cref{thm:ind-cca-preservation} (\textsc{Rocq}{})
    };

    \node[repr, inner sep=6pt,
      fit={(titleL)(fcL)(secL)(fcL |- fcR.north)(fcL |- secR.south)($(fcL.east)+(2pt,0)$)}]
      (jasminbox) {};
    \node[repr, inner sep=6pt,
      fit={(titleR)(fcR)(secR)(fcR |- fcL.north)(fcR |- secL.south)($(fcR.west)+(-2pt,0)$)}]
      (asmbox) {};

  \node[anchor=north, inner sep=3pt] at (jasminbox.north)
    {\textbf{\textsc{Jasmin}{} ML-KEM}};
  \node[anchor=north, inner sep=3pt] at (asmbox.north)
    {\textbf{Assembly ML-KEM}};

  \begin{scope}[on background layer]
    \node[
      fill=Green!15,
      fill opacity=0.5,
      rounded corners=2pt,
      inner sep=4pt,
      fit={($(jasminbox.east)+(-4pt,0)$)(asmbox)
           ($(asmbox.north)+(0,7pt)$)},
    ] (thisbox) {};
    \node[
      fill=Orange!15,
      fill opacity=0.5,
      rounded corners=2pt,
      inner sep=4pt,
      fit={(jasminbox.north west)(jasminbox.south west)
           ($(thisbox.west)+(-5pt,0)$)
           ($(jasminbox.north)+(0,7pt)$)},
    ] (priorbox) {};
  \end{scope}

  \node[
    anchor=north west,
    inner sep=3pt,
  ] at (thisbox.north west)
    {\scriptsize\begin{Bflushleft}This Paper\end{Bflushleft}};

  \node[
    anchor=north east,
    inner sep=3pt,
  ] at (priorbox.north east)
    {\scriptsize\begin{Bflushright}Prior Work\end{Bflushright}};
\end{tikzpicture}}{}
  \caption{Prior ML-KEM verification efforts and this work.}\label{fig:overview}
\end{figure}

\paragraph*{Contributions}
This paper makes the following contributions.
\begin{asparaitem}
\item[\textbf{(C1)}]
  We define a general notion of security game, based on oracle
  systems~\cite{DBLP:conf/ccs/BartheDKL10,DBLP:conf/eurocrypt/BellareR06,verifiedzkevm2024vcvio}, that captures
  common game-based security notions (e.g., indistinguishability and
  unforgeability) and provides a unified foundation for reasoning about
  source and target programs.

\item[\textbf{(C2)}]
  We equip \textsc{Jasmin}{} and assembly programs with a denotational,
  probabilistic semantics, and we prove that the \textsc{Jasmin}{} compiler
  is correct with respect to this semantics.

\item[\textbf{(C3)}]
  We instantiate game-based security notions to \textsc{Jasmin}{} and
  assembly programs and show that compiler correctness implies
  preservation of all game-based security properties
  (\cref{thm:generic-preservation}). As a consequence,
  we get preservation of KEM-IND-CCA{} for the \textsc{Jasmin}{} vectorized
  implementation of ML-KEM, as shown in \cref{fig:overview} (prior work
  established KEM-IND-CCA{} security and functional correctness at the
  source level, our work defines these properties at assembly level and
  bridges the gap).

\item[\textbf{(C4)}]
  We develop a variant of relational Hoare logic for interaction trees.
  The logic, which is of independent interest and tailored for compiler
  correctness proofs, yields compact proofs.
\end{asparaitem}

\paragraph*{Artifact}
The \textsc{Rocq}{} mechanization of our framework and the version of \textsc{Jasmin}{}
with its new proofs are open-source \cite{\ifanon{anonymous-artifact}{public-artifact}}{}.

\paragraph*{Trusted Computing Base (TCB)}
The TCB for our work comprises \textsc{Jasmin}{}'s preprocessor and printer;
its formalization of x86-64; and, naturally, the \textsc{Rocq}{} kernel and
OCaml compiler.
\looseness=-1

\paragraph*{Future Work}
This paper specifically focuses on compiler preservation of game-based
security. We consider the classic setting of game-based security, in
which adversaries interact with cryptographic schemes through their
specified interfaces, but cannot observe their side-channel leakage
and cannot interfere with their execution, e.g., by controlling the
cache or the branch predictors.  While such stronger adversaries are
out of scope of our work, the contributions of this paper provide a
foundation upon which side-channel security guarantees can be built.

In follow-up work, we plan to extend the compiler security guarantees to
stronger, system-level adversaries (e.g., that observe or carry out
Spectre attacks).
The main steps toward this goal are:
\begin{enumerate*}
\item formalizing a (speculative) constant-time type
  system~\cite{DBLP:conf/pldi/CauligiDGTSRB20} \emph{but at assembly
  level};
\item proving that said type system enforces noninterference \emph{in a
  system-level model of execution}~\cite{DBLP:conf/ccs/BartheBCLP14};
\item connecting the guarantees of the type system with the compiler's.
\end{enumerate*}
Further details are discussed in the conclusion.

\paragraph*{Outline}
\Cref{sec:crypto} introduces our framework for game-based security
and reductions using ITrees.  Throughout the paper, we use
KEM-IND-CCA{} as an illustrative example, discussing the generalization
to game-based security notions afterward.  \Cref{sec:semantics}
defines the ITree semantics of the \textsc{Jasmin}{} language and the notion
of security for \textsc{Jasmin}{} programs.  \Cref{sec:preservation} then
states compiler correctness and leverages it to prove that \textsc{Jasmin}{}
preserves game-based security.  \Cref{sec:rhl} presents the key
proof technique used to prove compiler correctness: relational Hoare
logic for ITrees.  Lastly, \cref{sec:verification} discusses the
verification of the \textsc{Jasmin}{} compiler, illustrating each key idea by
discussing a concrete compiler pass.

 \section{Game-Based Security with Interaction Trees}\label{sec:crypto}
This section introduces our formalization of game-based security and
reductions.
For concreteness, we first focus on key encapsulation mechanisms
(KEMs) and extend our methodology to arbitrary oracle systems
afterward.  Our approach is based on interaction
trees~\cite{DBLP:journals/pacmpl/XiaZHHMPZ20,DBLP:journals/pacmpl/ZakowskiBYZZZ21},
a powerful semantic framework for modeling and reasoning about
effectful computations.  In order to keep our exposition accessible to
a general audience, we introduce instances and features of interaction
trees as needed. A more technical introduction is provided in the
following sections and in \cref{appendix:itrees}.

\subsection{Key Encapsulation Mechanisms}\label{sec:crypto:kem}
Cryptographic schemes are modeled as sets of probabilistic algorithms.
For instance, a KEM is given by three algorithms for key generation,
encapsulation and
decapsulation. Informally, key generation generates a key pair
consisting of a public key and a secret key. Encapsulation takes as
input a public key and generates a message and its ciphertext. Finally,
decapsulation takes as input a secret key and a ciphertext and returns
a message. We let $\ensuremath{{\mathcal{PK}}}$, $\ensuremath{{\mathcal{SK}}}$, $\ensuremath{{\mathcal{M}}}$ and
$\ensuremath{{\mathcal{C}}}$ denote the types of public keys, secret keys, messages,
and ciphertexts, respectively. Our modeling of KEMs uses ITrees of type
\ensuremath{{\mathsf{itree}~{\ensuremath{{\mathsf{Rnd}}}}~{\results}}}, which represent probabilistic computations
(\ensuremath{{\mathsf{Rnd}}}{} corresponds to random sampling) that output results
in~\results{}. Thus, KEMs are triples of the form:
\begin{align*}
  \ensuremath{{\operatorname{GenKey}}} &:
    () \to \ensuremath{{\mathsf{itree}~{\ensuremath{{\mathsf{Rnd}}}}~{\left(\ensuremath{{\mathcal{PK}}} \times \ensuremath{{\mathcal{SK}}}\right)}}}\text,
  \\
  \ensuremath{{\operatorname{Encap}}} &:
    \ensuremath{{\mathcal{PK}}} \to
    \ensuremath{{\mathsf{itree}~{\ensuremath{{\mathsf{Rnd}}}}~{\left(\ensuremath{{\mathcal{M}}} \times \ensuremath{{\mathcal{C}}}\right)}}}\text,
  \\
  \Decapname &:
    \ensuremath{{\mathcal{SK}}} \times \ensuremath{{\mathcal{C}}} \to \ensuremath{{\mathsf{itree}~{\ensuremath{{\mathsf{Rnd}}}}~{\ensuremath{{\mathcal{M}}}}}}\text.
\end{align*}
That is, \ensuremath{{\operatorname{GenKey}}}{} is a computation that samples randomness
and returns a key pair. Similarly, decapsulation takes a secret and a
ciphertext and returns a message; note that our model corresponds to
KEMs with implicit rejection, in which \Decapname{} never fails.
Modeling explicit rejection entails changing the return type
of \Decapname{} to \ensuremath{{\mathsf{itree}~{\ensuremath{{\mathsf{Rnd}}}}~{\left(\ensuremath{{\mathcal{M}}} \cup
  \ensuremath{{\{{\bot}\}}}\right)}}}, to indicate that computations may fail.

Critically, ITrees do not constrain the termination behavior of
probabilistic computations; in particular, it is possible to model
rejection sampling (or repeat until success), which is at the core of
ML-KEM\@. The ability to model rejection sampling is a main challenge
for our development.

\subsection{Chosen Ciphertext Attack Security}\label{sec:crypto:ind-cca}
The standard notion of security for a KEM is that an attacker cannot
distinguish between a real message and a random one, given
ciphertexts: this is called indistinguishability against chosen
ciphertext attacks.  The notion is defined using a probabilistic
game between a challenger~\challenger{} with access to a KEM and
an adversary~\adversary{}. We model adversaries using interaction
trees of the form \ensuremath{{\mathsf{itree}~{\ensuremath{{\left(\ensuremath{{\mathsf{Dec}}} \mathop{+'} \ensuremath{{\mathsf{Rnd}}}\right)}}}~{\results}}}, which informally
represent probabilistic computations that can perform decapsulation
queries~\ensuremath{{\mathsf{Dec}}}{} that are answered with the KEM's decapsulation oracle
(\(\mathop{+'}\) denotes the disjoint union of event types). This
adversary is modeled as a pair of probabilistic algorithms:
\begin{align*}
  \adversary[_1] &: \ensuremath{{\mathcal{PK}}} \to \ensuremath{{\mathsf{itree}~{\ensuremath{{\left(\ensuremath{{\mathsf{Dec}}} \mathop{+'} \ensuremath{{\mathsf{Rnd}}}\right)}}}~{\ensuremath{{\mathcal{S}}}}}}
  \\
  \adversary[_2] &: \ensuremath{{\mathcal{S}}} \times \ensuremath{{\mathcal{C}}} \times \ensuremath{{\mathcal{M}}} \to
    \ensuremath{{\mathsf{itree}~{\ensuremath{{\left(\ensuremath{{\mathsf{Dec}}} \mathop{+'} \ensuremath{{\mathsf{Rnd}}}\right)}}}~{\ensuremath{{\{{0, 1}\}}}}}}\text,
\end{align*}
where \ensuremath{{\mathcal{S}}}{} represents the adversary's state.

\begin{figure}
  
  \begin{gather*}
    \ensuremath{{\ensuremath{{\operatorname{KEM-IND-CCA}}}_{{\challenger}, {\adversary}}}} \triangleq
    \begin{array}[t]{@{}l@{}}
      \ensuremath{{\mathit{pk}}}, \ensuremath{{\mathit{sk}}} \mathop{\ensuremath{{\leftarrow}}} \ensuremath{{\ensuremath{{\operatorname{GenKey}}}\mathchoice{\!}{\!}{}{}\left(\right)}}{}\mathop{;}\\
      \ensuremath{{s}} \mathop{\ensuremath{{\leftarrow}}} \ensuremath{{\langle {\adversary[_1](\ensuremath{{\mathit{pk}}})} \mathrel{\ensuremath{{\bowtie}}} {\Decapsk} \rangle }}\mathop{;}\\
      \msg[_0], \ensuremath{{\mathit{c}}} \mathop{\ensuremath{{\leftarrow}}} \ensuremath{{\ensuremath{{\operatorname{Encap}}}\mathchoice{\!}{\!}{}{}\left({\ensuremath{{\mathit{pk}}}}\right)}}\mathop{;}\\
      \msg[_1] \mathop{\ensuremath{{\leftarrow}}} \ensuremath{{\mathcal{U}\mathchoice{\!}{\!}{}{}\left({\ensuremath{{\mathcal{M}}}}\right)}}\mathop{;}\\
      b \mathop{\ensuremath{{\leftarrow}}} \ensuremath{{\mathcal{U}\mathchoice{\!}{\!}{}{}\left({\ensuremath{{\{{0,1}\}}}}\right)}}\mathop{;}\\
      b' \mathop{\ensuremath{{\leftarrow}}} \ensuremath{{\langle {\adversary[_2](\ensuremath{{s}}, \ensuremath{{\mathit{c}}}, \msg[_b])} \mathrel{\ensuremath{{\bowtie}}} {\Decapsk} \rangle }}\mathop{;}\\
      \ensuremath{{\ensuremath{{\mathsf{Ret}}}\mathchoice{\!}{\!}{}{}\left({b = b'}\right)}}
    \end{array}
  \end{gather*}
{}
  \caption{KEM-IND-CCA{} using our ITree language.}\label{fig:ind-cca-cca}\end{figure}

\Cref{fig:ind-cca-cca}{} presents the standard game for KEM-IND-CCA{}. In the
first stage, the challenger generates a key pair \ensuremath{{\mathit{pk}}}{}~and~\ensuremath{{\mathit{sk}}}{} and
reveals the public key to the adversary~\adversary[_1]{}.  The adversary
queries the decapsulation oracle (potentially many times, the
interaction is denoted \ensuremath{{\bowtie}}{}), and writes to their
state~\ensuremath{{s}}{}.  Note that the adversary can produce encapsulations
at will, since the public key is known.  Then, the challenger calls
the encapsulation oracle to generate a message-ciphertext pair
\msg[_0]~and~\ensuremath{{\mathit{c}}}{}.  The challenger then samples a random
message~\msg[_1] and a bit~\(b\) uniformly, giving
\msg[_b]~and~\ensuremath{{\mathit{c}}}{} to~\adversary[_2]{}, which returns a bit~\(b'\).  The
adversary wins if the guess~\(b'\) is~\(b\) and \adversary[_2]{}~did not
query the decapsulation oracle on the challenge ciphertext~\ensuremath{{\mathit{c}}}{}: the
adversary's \emph{advantage} is defined as
\begin{equation*}
  \ensuremath{{\ensuremath{{\operatorname{Adv}}}_{\ensuremath{\text{\tiny{\ensuremath{\ensuremath{{\operatorname{KEM-IND-CCA}}}}}}}}\mathchoice{\!}{\!}{}{}\left({\challenger}, {\adversary}\right)}} \triangleq
    \ensuremath{{\left| {\pr{
      \begin{array}{@{}c@{}}
        \ensuremath{{\ensuremath{{\operatorname{KEM-IND-CCA}}}_{{\challenger}, {\adversary}}}} = 1 \text{ and}
        \\
        \ensuremath{{\mathit{c}}} \notin \mathcal{L}(\adversary[_2])
      \end{array}
    } - \frac{1}{2}} \right|}}\text,
\end{equation*}
where \ensuremath{{\mathcal{L}(\adversary[_2])}} is the set of ciphertexts that
\adversary[_2]{} queries \Decapname{} on.

\subsection{Probabilistic Semantics for Games}\label{sec:crypto:semantics}
Formalizing the adversary's advantage requires a semantics for our
ITree language, defined below. This section defines a semantics for
games that will allow us to define the adversary's advantage formally.
Concretely, we first introduce a concrete representation of ITrees,
define a generic probabilistic semantics for them, and finally use it on
games.

\paragraph*{Interaction Trees}
Given a family of event types \ensuremath{{\events : \ensuremath{{\mathsf{Type}}} \to \ensuremath{{\mathsf{Type}}}}} and a
type of results~\results{}, the type \ensuremath{{\mathsf{itree}~{\events}~{\results}}}
represents a computation as a (potentially) infinite tree of all its
behaviors, where leaf nodes are results~\results{} and non-leaf nodes
are either silent steps or visible events~\events{}.
Concretely, ITrees are defined coinductively as
\begin{equation*}
  \ensuremath{{\mathsf{itree}~{\events}~{\results}}} \ni t \Coloneqq
    \ensuremath{{\ensuremath{{\mathsf{Ret}}}\mathchoice{\!}{\!}{}{}\left({r}\right)}} \mid \ensuremath{{\ensuremath{{\mathsf{Tau}}}\mathchoice{\!}{\!}{}{}\left({t}\right)}} \mid \ensuremath{{\ensuremath{{\mathsf{Vis}}}\mathchoice{\!}{\!}{}{}\left({\ans, e}, {k}\right)}}\text,
\end{equation*}
where \ensuremath{{r : \results}}, \ans{} is a type, \ensuremath{{e : \events~\ans}},
and \ensuremath{{k : \ans \to \ensuremath{{\mathsf{itree}~{\events}~{\results}}}}}.
The \emph{return} constructor~\ensuremath{{\ensuremath{{\mathsf{Ret}}}\mathchoice{\!}{\!}{}{}\left({r}\right)}} labels the leaves of the tree:
each leaf denotes a completed computation that yields the
result~\(r\). Silent nodes~\ensuremath{{\ensuremath{{\mathsf{Tau}}}\mathchoice{\!}{\!}{}{}\left({t}\right)}} perform an unobservable
transition before continuing as~\(t\) (i.e., there is only one branch).
Visible event nodes~\ensuremath{{\ensuremath{{\mathsf{Vis}}}\mathchoice{\!}{\!}{}{}\left({\ans, e}, {k}\right)}} trigger the event~\(e\), expect
an answer~\ensuremath{{a : \ans}}, and continue as~\(k(a)\) (i.e., there is one
branch per answer).
Note that each~\ensuremath{{\mathsf{Vis}}}{} node has its own type~\ans{}.
We leave~\ans{} implicit when it is clear from the context.

\begin{figure}
  \small
  
  \begin{align*}
    \ensuremath{{\left\llbracket t\right\rrbracket^{{0}}}} &\triangleq \ensuremath{{\mathbb{0}}}{}
    \\
    \ensuremath{{\left\llbracket\ensuremath{{\ensuremath{{\mathsf{Ret}}}\mathchoice{\!}{\!}{}{}\left({r}\right)}}\right\rrbracket^{{n + 1}}}} &\triangleq \ensuremath{{\mathds{1}_{r}}}
    \\
    \ensuremath{{\left\llbracket\ensuremath{{\ensuremath{{\mathsf{Tau}}}\mathchoice{\!}{\!}{}{}\left({t}\right)}}\right\rrbracket^{{n + 1}}}} &\triangleq \ensuremath{{\left\llbracket t\right\rrbracket^{{n}}}}
    \\
    \ensuremath{{\left\llbracket\ensuremath{{\ensuremath{{\mathsf{Vis}}}\mathchoice{\!}{\!}{}{}\left({\ensuremath{{\mathsf{Rnd}}}(X)}, {k}\right)}}\right\rrbracket^{{n + 1}}}} &\triangleq
      \ensuremath{{\lambda{r}.\, {\frac{1}{\ensuremath{{\left| {X} \right|}}} \sum_{x \in X} \ensuremath{{\left\llbracket k(x)\right\rrbracket^{{n}}}}(r)}}}
    \\
    \ensuremath{{\left\llbracket t\right\rrbracket}} &\triangleq \lim_{n \to \infty} \ensuremath{{\left\llbracket t\right\rrbracket^{{n}}}}
  \end{align*}
{}
  \caption{Probabilistic semantics for ITrees.}\label{fig:semp}
\end{figure}

\paragraph*{Probabilistic Semantics}
\Cref{fig:semp}{} presents the probabilistic interpretation of ITrees.  Given
a tree~\ensuremath{{t : \ensuremath{{\mathsf{itree}~{\ensuremath{{\mathsf{Rnd}}}}~{\results}}}}}, we first define its
\(n\)-step approximation semantics \ensuremath{{\ensuremath{{\left\llbracket t\right\rrbracket^{{n}}}} : \results{}
  \to \ensuremath{{\mathopen[0, 1\mathclose]}}}} (a distribution), and then take its limit.  All
zero-step approximations are the null
distribution~\ensuremath{{\mathbb{0}}}{}, which assigns zero probability to all results.
All positive approximations of~\ensuremath{{\ensuremath{{\mathsf{Ret}}}\mathchoice{\!}{\!}{}{}\left({r}\right)}} are the unit distribution on
the result~\(r\), denoted~\ensuremath{{\mathds{1}_{r}}}, which assigns probability one
to~\(r\) and zero to all other values. The \((n+1)\)-step approximation
of~\ensuremath{{\ensuremath{{\mathsf{Tau}}}\mathchoice{\!}{\!}{}{}\left({t}\right)}} is the \(n\)-step approximation of~\(t\)---note that this
means that, as expected, diverging executions have probability
zero. Then, the \((n+1)\)-step approximation of~\ensuremath{{\ensuremath{{\mathsf{Vis}}}\mathchoice{\!}{\!}{}{}\left({\ensuremath{{\mathsf{Rnd}}}(X)}, {k}\right)}}
accumulates the probabilities over all possible random
samples~\ensuremath{{x \in X}} and divides by the number of such possible
samples; technically, this clause implements a monadic bind where the
first argument is the uniform distribution over the random space~\(X\).
Finally, the probabilistic interpretation of~\(t\), denoted~\ensuremath{{\left\llbracket t\right\rrbracket}},
is the limit over~\(n\) of \ensuremath{{\left\llbracket t\right\rrbracket^{{n}}}} (which exists because
\ensuremath{{\left\llbracket t\right\rrbracket^{{n}}}} is increasing on~\(n\)).  This semantics is
denotational---i.e., it is a distribution---as is the natural definition
in pen-and-paper presentations.
As expected, this semantics supports almost-sure termination (programs
with termination probability one, e.g., rejection sampling, give rise to
proper distributions) and assigns null probability to diverging
computations (in such cases, the semantics is a \emph{subdistribution}
rather than a proper one; our formalization is generalized to this case
as well).
\looseness=-1

\paragraph*{Semantics for Security Games}
The semantics of \ensuremath{{\operatorname{KEM-IND-CCA}}}{} interprets assignments~\ensuremath{{{x} \mathop{\ensuremath{{\leftarrow}}} {t} \mathop{;} {k}}} with
the ITree sequencing operator (i.e., the monadic bind), returns~\ensuremath{{\ensuremath{{\mathsf{Ret}}}\mathchoice{\!}{\!}{}{}\left({r}\right)}}
with the return constructor, \ensuremath{{\ensuremath{{\mathsf{trigger}}}\mathchoice{\!}{\!}{}{}\left({e}\right)}} as visible event nodes, and
interactions~\ensuremath{{\langle {t} \mathrel{\ensuremath{{\bowtie}}} {o} \rangle }} using the~\ensuremath{{\mathsf{interp}}}{} ITree operator
(which replaces \(t\)'s events~\ensuremath{{\ensuremath{{\mathsf{Dec}}}\mathchoice{\!}{\!}{}{}\left({\ensuremath{{\mathit{c}}}}\right)}} with the result of calling
the oracle~\(o\) on~\ensuremath{{\mathit{c}}}{}).
Crucially, the \ensuremath{{\bowtie}}{}~operator keeps track of all queries
that \(t\)~performs to the oracle~\(o\).
This allows us to define \ensuremath{{\ensuremath{{\operatorname{KEM-IND-CCA}}}_{{\challenger}, {\adversary}}}} as an
\ensuremath{{\mathsf{itree}~{\ensuremath{{\mathsf{Rnd}}}}~{\left(
  \ensuremath{{\{{0,1}\}}} \times \ensuremath{{\mathcal{C}}} \times \ensuremath{{\mathcal{P}\mathchoice{\!}{\!}{}{}\left({\ensuremath{{\mathcal{C}}}}\right)}}
\right)}}}, where the first component is the correctness of the adversary's
guess, the second one is the challenge ciphertext~\ensuremath{{\mathit{c}}}{}, and the third
one is the set of ciphertexts that \adversary[_2]{}~queried to the
decapsulation oracle (i.e., corresponds to
\ensuremath{{\mathcal{L}(\adversary[_2])}}).
Finally, the adversary's advantage is
\begin{equation*}
  \ensuremath{{\ensuremath{{\operatorname{Adv}}}_{\ensuremath{\text{\tiny{\ensuremath{\ensuremath{{\operatorname{KEM-IND-CCA}}}}}}}}\mathchoice{\!}{\!}{}{}\left({\challenger}, {\adversary}\right)}} \triangleq
    \ensuremath{{\left| {\pr{
      \begin{array}{@{}c@{}}
        \ensuremath{{\left\llbracket\ensuremath{{\ensuremath{{\operatorname{KEM-IND-CCA}}}_{{\challenger}, {\adversary}}}}\right\rrbracket}} = 1, \ensuremath{{\mathit{c}}}, \ensuremath{{\vec{q}}}
        \\
        \text{and } \ensuremath{{\mathit{c}}} \notin \ensuremath{{\vec{q}}}
      \end{array}
    } - \frac{1}{2}} \right|}}\text.
\end{equation*}

\subsection{Relating Security Games}\label{sec:crypto:reductions}

Most security proofs are based on \emph{reductionist arguments}, which
reduce the security of one cryptographic system to the security of
another system or a hardness assumption.
In the case of compilers, the aim is to reduce the security of the
compiled program to the source's, such that users need only reason about
source programs in their security proofs.
Essentially, reductionist arguments state that attacks for the compiled
program can be turned into attacks for the source program with
acceptable advantage.

Our framework allows stating such reductions and proving them.
These proofs fundamentally rely on the following result about the
probabilistic semantics of equivalent ITrees.

\begin{theorem}[\inartifact{}]\label{thm:eutt-semp}
  Equivalent ITrees \ensuremath{{{t} \mathrel{\euttname[_{}]} {t'}}} give the same distribution,
  i.e., \ensuremath{{\pr{\ensuremath{{\left\llbracket t\right\rrbracket}} = \st} = \pr{\ensuremath{{\left\llbracket t'\right\rrbracket}} = \st}}} for
  each~\st{} (that is, \ensuremath{{\ensuremath{{\left\llbracket t\right\rrbracket}} = \ensuremath{{\left\llbracket t'\right\rrbracket}}}}).
\end{theorem}

The equivalence relation is the standard one in the ITree library (weak
bisimilarity, discussed in more detail in \cref{sec:rhl:xrutt}) and is
what the \textsc{Jasmin}{} compiler achieves after our changes.

 \section{Interaction Trees Semantics of \textsc{Jasmin}{}}\label{sec:semantics}
This section presents an interaction tree semantics for the \textsc{Jasmin}{}
language and instantiates the notions of game-based security.

\subsection{Background on the \textsc{Jasmin}{} Language}\label{sec:semantics:jasmin}
\textsc{Jasmin}{}~\cite{DBLP:conf/ccs/AlmeidaBBBGLOPS17,DBLP:conf/sp/AlmeidaBBGKL0S20} is a framework for high-assurance
high-speed cryptography. The main components of the framework are the
\textsc{Jasmin}{} programming language and its compiler.

The language supports ``assembly in the head,'' empowering
programmers to write highly optimized code that remains amenable to
formal analysis and verification.  Concretely, the language features a
combination of low-level constructs (such as vectorized instructions)
and high-level constructs (such as structured control flow, e.g.,
conditionals, loops, and functions), and zero-cost abstractions (such
as explicit variable names and functional arrays).  The latter are
called zero-cost because they introduce no run-time overhead; for
example, functional array manipulations are always compiled to
in-place memory operations that involve no copies.
The core syntax of \textsc{Jasmin}{} from this perspective is as follows:
\begin{align*}
  \ensuremath{{\mathsf{Expr}}} \ni e &\Coloneqq
    n \mid b \mid x
    \mid \left(e, \cdots, e\right)
    \mid {\oplus}(e)
    \mid \ensuremath{{{x}\!\left[{e}\right]}}
    \mid \ensuremath{{\left[{e}\right]}}
  \\
  \ensuremath{{\mathsf{LVal}}} \ni \lv &\Coloneq
    x
    \mid \ensuremath{{{x}\!\left[{e}\right]}}
    \mid \ensuremath{{\left[{e}\right]}}
    \mid \left(\lv, \cdots, \lv\right)
  \\
  \ensuremath{{\mathsf{Comm}}} \ni c &\Coloneqq
    \ensuremath{{\ensuremath{{\text{\textbf{skip}}}}}}{}
    \mid \ensuremath{{{\lv}~{\coloneq}~{e}}}
    \mid \ensuremath{{{\lv}~{\coloneq}~{\ensuremath{{\ensuremath{{\text{\textbf{rand}}}}}}({e})}}}
    \mid \ensuremath{{{\lv}~{\coloneq}~{{f}\mathchoice{\!}{\!}{}{}\left({e}\right)}}}
  \\
  &\qquad \mid c \mathop{;} c
    \mid \ensuremath{{\ensuremath{{\text{\textbf{if}}}}~({e})~\{{c}\}~\{{c}\}}}
    \mid \ensuremath{{\ensuremath{{\text{\textbf{while}}}}~({e})~\{c\}}}
\end{align*}
Here, \(e\)~is an expression, \(n\)~is a natural number,
\(b\)~is a boolean, \ensuremath{{x \in \ensuremath{{\mathsf{Var}}}}}~is a variable,
\ensuremath{{\left(e, \cdots, e\right)}}~is a tuple of expressions,
\(\oplus\)~is an operator (e.g., \(+\), \(\le\), or \(\land\)),
\lv{}~is a left-value, \ensuremath{{{x}\!\left[{e}\right]}}~is an array access to~\(x\) at
position~\(e\), \ensuremath{{\left[{e}\right]}}~is a memory access dereferencing the
address~\(e\), and \(f\)~is a function name.

\subsection{Interaction-Tree and Probabilistic Semantics}\label{sec:semantics:semantics}

\begin{figure}
  \begin{subfigure}{\linewidth}
    \small
    
  \renewcommand\arraystretch{1.3}
  \[\begin{array}{r @{{}\triangleq{}} l}
    \ensuremath{{
  \llparenthesis\hspace{0.1em}\ensuremath{{\ensuremath{{\text{\textbf{skip}}}}}}\hspace{0.1em}\rrparenthesis ^{\ensuremath{\text{\tiny{\ensuremath{\star}}}}}_{\st}}} & \ensuremath{{\ensuremath{{\mathsf{Ret}}}\mathchoice{\!}{\!}{}{}\left({\st}\right)}}
    \\
    \ensuremath{{
  \llparenthesis\hspace{0.1em}\ensuremath{{{\lv}~{\coloneq}~{e}}}\hspace{0.1em}\rrparenthesis ^{\ensuremath{\text{\tiny{\ensuremath{\star}}}}}_{\st}}} &
      \ensuremath{{{v} \mathop{\ensuremath{{\leftarrow}}} {\ensuremath{{
  \llparenthesis\hspace{0.1em}e\hspace{0.1em}\rrparenthesis ^{\ensuremath{\text{\tiny{\ensuremath{\ensuremath{{\mathsf{Expr}}}}}}}}_{\st}}}} \mathop{;} {\ensuremath{{
  \llparenthesis\hspace{0.1em}{\lv} \leftarrow {v}\hspace{0.1em}\rrparenthesis ^{\ensuremath{\text{\tiny{\ensuremath{\ensuremath{{\mathsf{LVal}}}}}}}}_{\st}}}}}}
    \\
    \ensuremath{{
  \llparenthesis\hspace{0.1em}\ensuremath{{{\lv}~{\coloneq}~{\ensuremath{{\ensuremath{{\text{\textbf{rand}}}}}}({e})}}}\hspace{0.1em}\rrparenthesis ^{\ensuremath{\text{\tiny{\ensuremath{\star}}}}}_{\st}}} &
      \ensuremath{{{n} \mathop{\ensuremath{{\leftarrow}}} {\ensuremath{{
  \llparenthesis\hspace{0.1em}e\hspace{0.1em}\rrparenthesis ^{\ensuremath{\text{\tiny{\ensuremath{\ensuremath{{\mathsf{Expr}}}}}}}}_{\st}}}} \mathop{;} {
        \ensuremath{{{a} \mathop{\ensuremath{{\leftarrow}}} {\ensuremath{{\ensuremath{{\mathsf{trigger}}}\mathchoice{\!}{\!}{}{}\left({\ensuremath{{\mathsf{Rnd}}}(n)}\right)}}} \mathop{;} {
          \ensuremath{{
  \llparenthesis\hspace{0.1em}{\lv} \leftarrow {a}\hspace{0.1em}\rrparenthesis ^{\ensuremath{\text{\tiny{\ensuremath{\ensuremath{{\mathsf{LVal}}}}}}}}_{\st}}}}}}}}}
    \\
    \ensuremath{{
  \llparenthesis\hspace{0.1em}c \mathop{;} c'\hspace{0.1em}\rrparenthesis ^{\ensuremath{\text{\tiny{\ensuremath{\star}}}}}_{\st}}} &
      \ensuremath{{{\st'} \mathop{\ensuremath{{\leftarrow}}} {\ensuremath{{
  \llparenthesis\hspace{0.1em}c\hspace{0.1em}\rrparenthesis ^{\ensuremath{\text{\tiny{\ensuremath{\star}}}}}_{\st}}}} \mathop{;} {\ensuremath{{
  \llparenthesis\hspace{0.1em}c'\hspace{0.1em}\rrparenthesis ^{\ensuremath{\text{\tiny{\ensuremath{\star}}}}}_{\st'}}}}}}
    \\
    \ensuremath{{
  \llparenthesis\hspace{0.1em}\ensuremath{{\ensuremath{{\text{\textbf{if}}}}~({e})~\{{c_{\top}}\}~\{{c_{\bot}}\}}}\hspace{0.1em}\rrparenthesis ^{\ensuremath{\text{\tiny{\ensuremath{\star}}}}}_{\st}}} &
      \ensuremath{{{b} \mathop{\ensuremath{{\leftarrow}}} {\ensuremath{{
  \llparenthesis\hspace{0.1em}e\hspace{0.1em}\rrparenthesis ^{\ensuremath{\text{\tiny{\ensuremath{\ensuremath{{\mathsf{Expr}}}}}}}}_{\st}}}} \mathop{;} {\ensuremath{{
  \llparenthesis\hspace{0.1em}c_b\hspace{0.1em}\rrparenthesis ^{\ensuremath{\text{\tiny{\ensuremath{\star}}}}}_{\st}}}}}}
    \\
    \ensuremath{{
  \llparenthesis\hspace{0.1em}\ensuremath{{\ensuremath{{\text{\textbf{while}}}}~({e})~\{c\}}}\hspace{0.1em}\rrparenthesis ^{\ensuremath{\text{\tiny{\ensuremath{\star}}}}}_{\st}}} &
      \ensuremath{{\ensuremath{{\mathsf{iter}}}\mathchoice{\!}{\!}{}{}\left({
        \ensuremath{{\lambda{\st}.\, { }}}{\ensuremath{{{b} \mathop{\ensuremath{{\leftarrow}}} {\ensuremath{{
  \llparenthesis\hspace{0.1em}e\hspace{0.1em}\rrparenthesis ^{\ensuremath{\text{\tiny{\ensuremath{\ensuremath{{\mathsf{Expr}}}}}}}}_{\st}}}} \mathop{;} {W}}}}}\right)}}(\st)
    \\
    \multicolumn{2}{r}{\ensuremath{
      \text{where } W =
        \begin{cases}
          \ensuremath{{{\st'} \mathop{\ensuremath{{\leftarrow}}} {\ensuremath{{
  \llparenthesis\hspace{0.1em}c\hspace{0.1em}\rrparenthesis ^{\ensuremath{\text{\tiny{\ensuremath{\star}}}}}_{\st}}}} \mathop{;} {\ensuremath{{\ensuremath{{\mathsf{Ret}}}\mathchoice{\!}{\!}{}{}\left({\ensuremath{{\mathsf{inl}\mathchoice{\!}{\!}{}{}\left({\st'}\right)}}}\right)}}}}}
          & \ensuremath{{\text{if } {b}}}\text,\\
          \ensuremath{{\ensuremath{{\mathsf{Ret}}}\mathchoice{\!}{\!}{}{}\left({\ensuremath{{\mathsf{inr}\mathchoice{\!}{\!}{}{}\left({\st}\right)}}}\right)}} & \ensuremath{{\text{otherwise}}}
        \end{cases}
    }}
    \\
    \ensuremath{{
  \llparenthesis\hspace{0.1em}\ensuremath{{{\lv}~{\coloneq}~{{f}\mathchoice{\!}{\!}{}{}\left({e}\right)}}}\hspace{0.1em}\rrparenthesis ^{\ensuremath{\text{\tiny{\ensuremath{\star}}}}}_{\st}}} &
      \renewcommand\arraystretch{1}
      \begin{array}[t]{@{} l}
        v \mathop{\ensuremath{{\leftarrow}}} \ensuremath{{
  \llparenthesis\hspace{0.1em}e\hspace{0.1em}\rrparenthesis ^{\ensuremath{\text{\tiny{\ensuremath{\ensuremath{{\mathsf{Expr}}}}}}}}_{\st}}} \mathop{;}
        \\[\jot]
        \left(\ensuremath{{m}}',v'\right) \mathop{\ensuremath{{\leftarrow}}}
          \ensuremath{{\ensuremath{{\mathsf{trigger}}}\mathchoice{\!}{\!}{}{}\left({\ensuremath{{\mathsf{Call}}}(f,\ensuremath{{\st_{\ensuremath{{m}}}}},v)}\right)}} \mathop{;}
        \\[\jot]
        \ensuremath{{
  \llparenthesis\hspace{0.1em}{\lv} \leftarrow {v'}\hspace{0.1em}\rrparenthesis ^{\ensuremath{\text{\tiny{\ensuremath{\ensuremath{{\mathsf{LVal}}}}}}}}_{\left(\ensuremath{{m}}', \ensuremath{{\st_{\ensuremath{{r}}}}}\right)}}}
      \end{array}
  \end{array}\]
{}
    \caption{Call{} semantics.
      The \ensuremath{{\mathsf{trigger}}}{}~operator introduces a visible event node
      (e.g.,~\ensuremath{{\ensuremath{{\mathsf{Rnd}}}\mathchoice{\!}{\!}{}{}\left({n}\right)}}) and returns the answer (e.g.,~\(a\)).}\label{fig:sem}
  \end{subfigure}

  \begin{subfigure}{\linewidth}
    \small
    
  \begin{align*}
    \ensuremath{{\ensuremath{{\ensuremath{{H}}^{\ensuremath{\text{\tiny{\ensuremath{\hspace{0.1em}\mathsf{c}}}}}}}}\mathchoice{\!}{\!}{}{}\left(\ensuremath{{\mathsf{Call}}}\mathchoice{\!}{\!}{}{}\left({f}, {\ensuremath{{m}}}, {v}\right)\right)}} &\triangleq
      \ensuremath{{{\st} \mathop{\ensuremath{{\leftarrow}}} {
        \ensuremath{{
  \llparenthesis\hspace{0.1em}\fbody{f}\hspace{0.1em}\rrparenthesis ^{\ensuremath{\text{\tiny{\ensuremath{\star}}}}}_{m,\ensuremath{{\{{\farg{f} \mapsto v}\}}}}}}
      } \mathop{;} {
        \ensuremath{{\ensuremath{{\mathsf{Ret}}}\mathchoice{\!}{\!}{}{}\left({\ensuremath{{\st_{\ensuremath{{m}}}}}, \ensuremath{{\st_{\ensuremath{{r}}}}}(\fres{f})}\right)}}
      }}}
      \\
    \ensuremath{{
  \llparenthesis\hspace{0.1em}c\hspace{0.1em}\rrparenthesis ^{}_{\st}}} : \ensuremath{{\mathsf{itree}~{\left(\ensuremath{{\mathsf{Err}}} \mathop{+'} \ensuremath{{\mathsf{Rnd}}}\right)}~{\ensuremath{{\mathsf{S}}}}}}
      &\triangleq \ensuremath{{\ensuremath{{\mathsf{interp\_mrec}}}\mathchoice{\!}{\!}{}{}\left({\ensuremath{{\ensuremath{{H}}^{\ensuremath{\text{\tiny{\ensuremath{\hspace{0.1em}\mathsf{c}}}}}}}}}, {\ensuremath{{
  \llparenthesis\hspace{0.1em}c\hspace{0.1em}\rrparenthesis ^{\ensuremath{\text{\tiny{\ensuremath{\star}}}}}_{\st}}}}\right)}}\text,
  \end{align*}
{}
    \caption{Interprocedural{} semantics.
      The \ensuremath{{\mathsf{interp\_mrec}}}{}~operator answers every \ensuremath{{\mathsf{Call}}}{}~event
      with the \ensuremath{{\ensuremath{{H}}^{\ensuremath{\text{\tiny{\ensuremath{\hspace{0.1em}\mathsf{c}}}}}}}}{}~handler (including those triggered
      by~\ensuremath{{\ensuremath{{H}}^{\ensuremath{\text{\tiny{\ensuremath{\hspace{0.1em}\mathsf{c}}}}}}}}{}).}\label{fig:sem-ER}
  \end{subfigure}
  \caption{Interaction Tree Semantics for \textsc{Jasmin}{}.}\end{figure}

This section presents a probabilistic semantics of \textsc{Jasmin}{} programs
based on ITrees.
The semantics is defined in several layers, as is standard in the ITree
literature.
First, we define the \emph{call{} semantics}, which
treats function calls, errors, and random sampling as uninterpreted
events in the program (e.g., instead of executing the body of a callee,
it triggers an event).
Afterward, we interpret call events and obtain the \emph{interprocedural{}
semantics}, which executes the bodies of callees to answer
call events. Lastly, we interpret the remaining events and achieve a
\emph{probabilistic semantics} that maps commands to distributions over
states.
This layered definition structure allows us to seamlessly move between
levels of abstraction: we use specific layers to reason about relevant
features of the semantics and treat other features abstractly, enabling
high reuse of code and proofs in generic lemmas of the compiler, in our
program logic, and in our definitions of security.
We now describe the layers relevant to this work.
\looseness=-1

The first layer is the call{} semantics, which employs three event
families: \ensuremath{{\mathsf{Call}}}{}~for function calls, \ensuremath{{\mathsf{Err}}}{}~for errors, and
\ensuremath{{\mathsf{Rnd}}}{}~for random sampling, denoted
\ensuremath{{\ensuremath{{
  \llparenthesis\hspace{0.1em}c\hspace{0.1em}\rrparenthesis ^{\ensuremath{\text{\tiny{\ensuremath{\star}}}}}_{\st}}} : \ensuremath{{\mathsf{itree}~{\left(\ensuremath{{\ensuremath{{\mathsf{Call}}} \mathop{+'} \ensuremath{{\mathsf{Err}}} \mathop{+'} \ensuremath{{\mathsf{Rnd}}}}}\right)}~{\ensuremath{{\mathsf{S}}}}}}}}
(where \(c\)~is a command, \ensuremath{{\st \in \ensuremath{{\mathsf{S}}}}}~a state).
A state \ensuremath{{\st \in \ensuremath{{\mathsf{S}}}}} is a pair of a variable map
\ensuremath{{\ensuremath{{r}} : \ensuremath{{\mathsf{Var}}} \to \ensuremath{{\mathsf{Val}}}}} (assigning values to variables) and
a memory \ensuremath{{\ensuremath{{m}} : \ensuremath{{\{{0,1}\}}}^{64} \to \ensuremath{{\{{0,1}\}}}^{8}}} (assigning
bytes to addresses).
The projections \ensuremath{{\st_{\ensuremath{{r}}}}}~and~\ensuremath{{\st_{\ensuremath{{m}}}}} are the variable map and
memory of \st{}, respectively.
The call{} semantics relies on a semantics of expressions
\ensuremath{{\ensuremath{{
  \llparenthesis\hspace{0.1em}e\hspace{0.1em}\rrparenthesis ^{\ensuremath{\text{\tiny{\ensuremath{\ensuremath{{\mathsf{Expr}}}}}}}}_{\st}}} : \ensuremath{{\mathsf{itree}~{\ensuremath{{\mathsf{Err}}}}~{\ensuremath{{\mathsf{Val}}}}}}}} (where \(e\)~is an
expression) and left-values
\ensuremath{{\ensuremath{{
  \llparenthesis\hspace{0.1em}{\lv} \leftarrow {v}\hspace{0.1em}\rrparenthesis ^{\ensuremath{\text{\tiny{\ensuremath{\ensuremath{{\mathsf{LVal}}}}}}}}_{\st}}} : \ensuremath{{\mathsf{itree}~{\ensuremath{{\mathsf{Err}}}}~{\ensuremath{{\mathsf{S}}}}}}}} (where \(v\)~is a
value), which are standard.

\Cref{fig:sem}{} defines the call{} semantics.
The \ensuremath{{\ensuremath{{\text{\textbf{skip}}}}}}{} command returns the state unchanged.
An assignment evaluates the right-hand side and writes the result to the
state---note that either step can trigger an \ensuremath{{\mathsf{Err}}}{}~event.
A random sampling \ensuremath{{{\lv}~{\coloneq}~{\ensuremath{{\ensuremath{{\text{\textbf{rand}}}}}}({e})}}} evaluates~\(e\) to a number~\(n\),
triggers a \ensuremath{{\ensuremath{{\mathsf{Rnd}}}\mathchoice{\!}{\!}{}{}\left({n}\right)}}~event (which expects \(n\)~uniformly sampled
bytes) and writes them to~\lv{}.
(Note that this event is a special case of the one presented in
\cref{sec:crypto:semantics}, as \textsc{Jasmin}{} only supports sampling bytes.)
The cases of sequencing and conditionals use the standard ITree monadic
bind operator.
While loops iteratively (potentially infinitely) execute the body with
the \ensuremath{{\mathsf{iter}}}{} operator. Finally, function calls evaluate their
argument~\(e\), trigger a \ensuremath{{\mathsf{Call}}}{}~event, write the answer to~\lv{}, and
update the memory. Each \ensuremath{{\mathsf{Call}}}{}~event carries function name, memory,
and argument value.

The second layer is the interprocedural{} semantics, which builds atop the
call{} semantics by interpreting \ensuremath{{\mathsf{Call}}}{}~events.
As is standard with ITrees, we define a \emph{handler} for
\ensuremath{{\ensuremath{{\mathsf{Call}}}\mathchoice{\!}{\!}{}{}\left({f, m, v}\right)}}~events, which executes the body of the callee~\(f\) on
an initial state with memory~\(m\) and arguments~\(v\), as shown in
\cref{fig:sem-ER}{} (there, \ensuremath{{\{{\farg{f} \mapsto v}\}}} is the variable map that
contains~\(v\) as the argument of~\(f\)).
\Cref{fig:sem-ER}{} uses this handler and the mutual recursion
operator~\ensuremath{{\mathsf{interp\_mrec}}}{} to define the interprocedural{} semantics,
consuming all \ensuremath{{\mathsf{Call}}}{}~events.
The \ensuremath{{\mathsf{Err}}}{}~and~\ensuremath{{\mathsf{Rnd}}}{} events, on the other hand, remain uninterpreted,
as effects of the program.

\begin{table}
  \caption{Semantics Used in This Paper.}\label{tbl:sems}
  \small
  \renewcommand*\arraystretch{1.1}
  
  \begin{tabular}{l @{\hspace{2em}} l}
    \toprule
    Semantics
    & Symbol
\\
    \midrule
    Call{}
    & \ensuremath{{\ensuremath{{
  \llparenthesis\hspace{0.1em}c\hspace{0.1em}\rrparenthesis ^{\ensuremath{\text{\tiny{\ensuremath{\star}}}}}_{\st}}} : \ensuremath{{\mathsf{itree}~{\left(\ensuremath{{\ensuremath{{\mathsf{Call}}} \mathop{+'} \ensuremath{{\mathsf{Err}}} \mathop{+'} \ensuremath{{\mathsf{Rnd}}}}}\right)}~{\ensuremath{{\mathsf{S}}}}}}}}
\\
    Interprocedural{}
    & \ensuremath{{
        \ensuremath{{
  \llparenthesis\hspace{0.1em}c\hspace{0.1em}\rrparenthesis ^{}_{\st}}} :
          \ensuremath{{\mathsf{itree}~{\left(\ensuremath{{\mathsf{Err}}} \mathop{+'} \ensuremath{{\mathsf{Rnd}}}\right)}~{\ensuremath{{\mathsf{S}}}}}}
       }}
\\
    Probabilistic
    & \ensuremath{{\ensuremath{{\left\llbracket c\right\rrbracket}}_{\st} : \ensuremath{{\mathsf{S}}} \cup \ensuremath{{\{{\st[_{\mathsf{err}}]}\}}} \to \ensuremath{{\mathopen[0, 1\mathclose]}}}}
\\
    \bottomrule
  \end{tabular}
{}
\end{table}

Lastly, we handle \ensuremath{{\mathsf{Err}}}{}~events by returning a distinguished error
state~\st[_{\mathsf{err}}]{}, obtaining an
\ensuremath{{\mathsf{itree}~{\ensuremath{{\mathsf{Rnd}}}}~{\left(\ensuremath{{\mathsf{S}}} \cup \ensuremath{{\{{\st[_{\mathsf{err}}]}\}}}\right)}}}.
Then, we use the semantics from \cref{fig:semp}{} to obtain a semantics that
maps every command~\(c\) and state~\st{} to a distribution over states
or errors
\ensuremath{{\ensuremath{{\left\llbracket c\right\rrbracket}}_{\st} : \ensuremath{{\mathsf{S}}} \cup \ensuremath{{\{{\st[_{\mathsf{err}}]}\}}} \to \ensuremath{{\mathopen[0, 1\mathclose]}}}}.
\Cref{tbl:sems}{} summarizes the semantics notions used in this paper,
each defined in terms of the one above.

\paragraph*{Safety}
A \textsc{Jasmin}{} program is \emph{safe} (relative to some precondition) if
for every function~\(f\) and state~\st{} satisfying the precondition
for~\(f\), its semantics triggers no \ensuremath{{\mathsf{Err}}}{}~events and its
results are fully initialized.
Specifically, the latter condition requires that every cell of returned
arrays is set to a value---\textsc{Jasmin}{} allows programs to take and return
arrays with some cells uninitialized.
Consequently, for safe \textsc{Jasmin}{} programs, the probabilistic semantics
is simply a distribution over states
\ensuremath{{\ensuremath{{\left\llbracket c\right\rrbracket}}_{\st} : \ensuremath{{\mathsf{S}}} \to \ensuremath{{\mathopen[0, 1\mathclose]}}}}.
In contrast to the original \textsc{Jasmin}{} semantics, termination is not
a requirement for safety; in particular, almost-surely-terminating
programs like ML-KEM can be safe and the guarantees of compiler
correctness apply to them.

\subsection{Game-Based Security of \textsc{Jasmin}{} Programs}\label{sec:semantics:security}

Using the semantics presented above, we can instantiate the framework
from \cref{sec:crypto} and give formal definitions of KEM-IND-CCA{}
security for \textsc{Jasmin}{} KEMs.
The challenger induced by the \textsc{Jasmin}{} KEM program~\(P\) is
denoted~\challenger[_P]{} and defined as

  \begin{align*}
    \ensuremath{{\ensuremath{{\operatorname{GenKey}}}\mathchoice{\!}{\!}{}{}\left(\right)}}{} &\triangleq
      \ensuremath{{{\st'} \mathop{\ensuremath{{\leftarrow}}} {
        \ensuremath{{
  \llparenthesis\hspace{0.1em}P(\ensuremath{{\ensuremath{{\text{\texttt{genkey}}}}}})\hspace{0.1em}\rrparenthesis ^{}_{\ensuremath{{m}}, \ensuremath{{\emptyset}}}}}
      } \mathop{;} {
        \ensuremath{{\ensuremath{{\mathsf{Ret}}}\mathchoice{\!}{\!}{}{}\left({\st'(\ensuremath{{\ensuremath{{\text{\texttt{pk}}}}}}), \st'(\ensuremath{{\ensuremath{{\text{\texttt{sk}}}}}})}\right)}}
      }}}
    \\
    \ensuremath{{\ensuremath{{\operatorname{Encap}}}\mathchoice{\!}{\!}{}{}\left({\ensuremath{{\mathit{pk}}}}\right)}} &\triangleq
      \ensuremath{{{\st'} \mathop{\ensuremath{{\leftarrow}}} {
        \ensuremath{{
  \llparenthesis\hspace{0.1em}P(\ensuremath{{\ensuremath{{\text{\texttt{encap}}}}}})\hspace{0.1em}\rrparenthesis ^{}_{\ensuremath{{m}}, \ensuremath{{\{{\ensuremath{{\ensuremath{{\text{\texttt{pk}}}}}} \mapsto \ensuremath{{\mathit{pk}}}}\}}}}}}
      } \mathop{;} {
        \ensuremath{{\ensuremath{{\mathsf{Ret}}}\mathchoice{\!}{\!}{}{}\left({\st'(\ensuremath{{\ensuremath{{\text{\texttt{msg}}}}}}), \st'(\ensuremath{{\ensuremath{{\text{\texttt{ct}}}}}})}\right)}}
      }}}
    \\
    \Decap{\ensuremath{{\mathit{sk}}}}{\ensuremath{{\mathit{c}}}} &\triangleq
      \ensuremath{{{\st'} \mathop{\ensuremath{{\leftarrow}}} {
        \ensuremath{{
  \llparenthesis\hspace{0.1em}P(\ensuremath{{\ensuremath{{\text{\texttt{decap}}}}}})\hspace{0.1em}\rrparenthesis ^{}_{
          \ensuremath{{m}}, \ensuremath{{\{{\ensuremath{{\ensuremath{{\text{\texttt{sk}}}}}} \mapsto \ensuremath{{\mathit{sk}}}, \ensuremath{{\ensuremath{{\text{\texttt{ct}}}}}} \mapsto \ensuremath{{\mathit{c}}}}\}}}
        }}}
      } \mathop{;} {
        \ensuremath{{\ensuremath{{\mathsf{Ret}}}\mathchoice{\!}{\!}{}{}\left({\st'(\ensuremath{{\ensuremath{{\text{\texttt{msg}}}}}})}\right)}}
      }}}
  \end{align*}
{} We write \ensuremath{{\ensuremath{{\text{\texttt{genkey}}}}}}{}, \ensuremath{{\ensuremath{{\text{\texttt{encap}}}}}}{}, and \ensuremath{{\ensuremath{{\text{\texttt{decap}}}}}}{} for the names of the
respective functions in the program, \ensuremath{{\ensuremath{{\text{\texttt{pk}}}}}}{}, \ensuremath{{\ensuremath{{\text{\texttt{sk}}}}}}{}, \ensuremath{{\ensuremath{{\text{\texttt{ct}}}}}}{},
and \ensuremath{{\ensuremath{{\text{\texttt{msg}}}}}}{} for those of the variables, and~\ensuremath{{m}}{} for an initial
memory on which the program is safe.
\looseness=-1

The key generation algorithm first executes the body of the
\ensuremath{{\ensuremath{{\text{\texttt{genkey}}}}}}{}~function in~\(P\) on the initial memory~\ensuremath{{m}}{} and the empty
variable map~\ensuremath{{\emptyset}}{}, and then returns the values of
\ensuremath{{\ensuremath{{\text{\texttt{pk}}}}}}{}~and~\ensuremath{{\ensuremath{{\text{\texttt{sk}}}}}}{} in the final state.
The encapsulation algorithm executes~\ensuremath{{\ensuremath{{\text{\texttt{encap}}}}}}{} on a state that maps
the variable~\ensuremath{{\ensuremath{{\text{\texttt{pk}}}}}}{} to the given value~\ensuremath{{\mathit{pk}}}{} and returns the final
values of \ensuremath{{\ensuremath{{\text{\texttt{msg}}}}}}{}~and~\ensuremath{{\ensuremath{{\text{\texttt{ct}}}}}}{}.
Similarly, the decapsulation algorithm executes~\ensuremath{{\ensuremath{{\text{\texttt{decap}}}}}}{} on a state
that maps the variables \ensuremath{{\ensuremath{{\text{\texttt{sk}}}}}}{}~and~\ensuremath{{\ensuremath{{\text{\texttt{ct}}}}}}{} to the values
\ensuremath{{\mathit{sk}}}{}~and~\ensuremath{{\mathit{c}}}{}, returning the final value of~\ensuremath{{\ensuremath{{\text{\texttt{msg}}}}}}{}.
The ITrees for these functions have only \ensuremath{{\mathsf{Rnd}}}{}~events, coinciding with
the definitions in \cref{sec:crypto:kem}---we prove that no error
events~\ensuremath{{\mathsf{Err}}}{} can happen because we consider safe programs.
Thus, queries are handled by~\ensuremath{{\bowtie}}{} in a similar way as
function calls (i.e., running the code of~\Decapname{}).
Finally, we define the advantage of an adversary~\adversary{} against a
\textsc{Jasmin}{} KEM program~\(P\) as \ensuremath{{\ensuremath{{\operatorname{Adv}}}_{\ensuremath{\text{\tiny{\ensuremath{\ensuremath{{\operatorname{KEM-IND-CCA}}}}}}}}\mathchoice{\!}{\!}{}{}\left({\challenger[_P]}, {\adversary}\right)}} (using
the definition from \cref{sec:crypto:semantics}).

\subsection{Game-Based Security of Assembly Programs}\label{sec:semantics:assembly}

The semantics of assembly programs has a similar signature to the one
for \textsc{Jasmin}{} programs, but it operates on \emph{assembly states}, which
comprise:
a register map,
a memory,
a program counter, and
the name of the function currently executing.
As expected, the semantics is defined as the repeated application of a
small-step transition function from assembly states to assembly states.

The security notion for assembly programs has two key differences with
the one for \textsc{Jasmin}{}.
First, array arguments are not given as values but rather as pointers to
memory; secondly, the initial memory is initialized with the contents of
such arguments.
Reading the results
from final states changes in a similar way.  Concretely, in the case of
a KEM, the challenger for a compilation~\(Q\), denoted~\challenger[_Q]{},
is parameterized by four pointers \ensuremath{{\ensuremath{{\text{\texttt{p}}}}\ensuremath{{\ensuremath{{\text{\texttt{pk}}}}}}}}{}, \ensuremath{{\ensuremath{{\text{\texttt{p}}}}\ensuremath{{\ensuremath{{\text{\texttt{sk}}}}}}}}{}, \ensuremath{{\ensuremath{{\text{\texttt{p}}}}\ensuremath{{\ensuremath{{\text{\texttt{ct}}}}}}}}{}, and \ensuremath{{\ensuremath{{\text{\texttt{p}}}}\ensuremath{{\ensuremath{{\text{\texttt{msg}}}}}}}}{}
(which are 64-bit words in x86-64) corresponding to the variables used
in~\challenger[_P]{}. Thus, in \ensuremath{{\operatorname{Encap}}}{}, instead of initializing a
variable map with the array~\ensuremath{{\mathit{pk}}}{}, we write~\ensuremath{{\mathit{pk}}}{} to memory at
position~\ensuremath{{\ensuremath{{\text{\texttt{p}}}}\ensuremath{{\ensuremath{{\text{\texttt{pk}}}}}}}}{} and initialize the register corresponding to the
variable \ensuremath{{\ensuremath{{\text{\texttt{pk}}}}}}{} to~\ensuremath{{\ensuremath{{\text{\texttt{p}}}}\ensuremath{{\ensuremath{{\text{\texttt{pk}}}}}}}}{}.

 \section{Correctness and KEM-IND-CCA{} Preservation}\label{sec:preservation}

This section overviews the \textsc{Jasmin}{} compiler, states the new compiler
correctness theorem and compares it with the existing one, and
presents its KEM-IND-CCA{} security preservation result.
Afterward, we discuss the application of our results to \textsc{Jasmin}{}'s
ML-KEM implementation.
Finally, we generalize our results to all game-based security
properties, leading to the main result of this work: that the \textsc{Jasmin}{}
compiler preserves game-based security
(\cref{thm:generic-preservation}).

\begin{figure*}
  
  \begin{tikzpicture}[
    xscale=1.8,
    yscale=1.1,
    every node/.style={font=\scriptsize},
    lang/.style={draw,shape=ellipse},
    pass/.style={draw,shape=rectangle,rounded corners},
    check/.style={draw,shape=rectangle,rounded corners,thick,dotted},
    proved/.style={fill=green!30},
    trusted/.style={fill=red!40},
    validated/.style={fill=blue!20},
    ar/.style={-stealth,thick,rounded corners},
    frontend/.style={Black!30,thick,dashed,rounded corners},
    backend/.style={Black!30,thick,dashed,rounded corners},
  ]
    \node[lang]          (src)        at (  2, 5) { \texttt{.jazz} };
    \node[pass,trusted]  (parse)      at (  3, 5) { Parse };
    \node[pass,trusted]  (preprocess) at (  4, 5) { Preprocess };
    \node[check,trusted] (typecheck)  at (  5, 5) { Type Check };
    \node[lang]          (jasmin)     at (  6, 5) { \textsc{Jasmin}{} };

    \node[pass,proved] (i2w) at (6, 4) { Int to Word };
    \node[lang] (subword) at (4.7, 4) { Subword };
    \node[pass,proved] (acopy) at (3.4, 4) { Array Copy };
    \node[pass,proved] (addinit) at (2.15, 4) { \begin{Bcenter}Add Array\\Initialization\end{Bcenter} };
    \node[pass,proved] (spilling) at (1, 4) { Spilling };
    \node[pass,proved] (inlining) at (0, 4) { Inlining };

    \node[pass,proved] (prunning1) at (0, 3) { \begin{Bcenter}Function\\Pruning\end{Bcenter} };
    \node[pass,proved] (constprop) at (1.5, 3) { Constant Propagation };
    \node[pass,proved] (dce1) at (3.3, 3) { \begin{Bcenter}Dead Code\\Elimination\end{Bcenter} };
    \node[pass,proved] (unrolling) at (4.6, 3) { Unrolling };
    \node[pass,proved] (prunning2) at (6, 3) { \begin{Bcenter}Function\\Pruning\end{Bcenter} };

    \node[pass,validated] (split1) at (6, 2) { \begin{Bcenter}Live-Range\\Splitting\end{Bcenter} };
    \node[pass,proved] (rminit) at (4.45, 2) { \begin{Bcenter}Remove Array\\Initialization\end{Bcenter} };
    \node[pass,proved] (refargs) at (2.9, 2) { \begin{Bcenter}Make Reference\\Arguments\end{Bcenter} };
    \node[pass,proved] (regarray) at (1.4, 2) { \begin{Bcenter}Reg. Array\\Expansion\end{Bcenter} };
    \node[lang] (direct) at (0, 2) { Direct };

    \node[pass,validated] (split2) at (0, 1) { \begin{Bcenter}Live-Range\\Splitting\end{Bcenter} };
    \node[pass,proved] (globals) at (1.15, 1) { \begin{Bcenter}Remove\\Globals\end{Bcenter} };
    \node[pass,proved] (loadconst) at (2.3, 1) { \begin{Bcenter}Load\\Immediates\end{Bcenter} };
    \node[pass,proved] (lowering) at (3.5, 1) { \begin{Bcenter}Instruction\\Selection\end{Bcenter} };
    \node[pass,proved] (inlineprop) at (4.75, 1) { \begin{Bcenter}Inline\\Propagation\end{Bcenter} };
    \node[pass,proved] (slh) at (6, 1) { \begin{Bcenter}SLH Instr.\\Selection\end{Bcenter} };

    \node[pass,proved] (stackalloc) at (6, 0) { \begin{Bcenter}Stack\\Allocation\end{Bcenter} };
    \node[lang] (stack) at (4.95, 0) { Stack };
    \node[pass,proved] (loweraddr) at (3.9, 0) { \begin{Bcenter}Lower\\Addresses\end{Bcenter} };
    \node[pass,proved] (unused_results) at (2.55, 0) { \begin{Bcenter}Remove\\Unused Results\end{Bcenter} };
    \node[pass,validated] (regalloc) at (1.2, 0) { \begin{Bcenter}Register\\Renaming\end{Bcenter} };
    \node[pass,proved] (dce2) at (0, 0) { \begin{Bcenter}Dead Code\\Elimination\end{Bcenter} };

    \node[check,proved] (chkonevm)      at (0, -1) { \begin{Bcenter}One\\Varmap\end{Bcenter} };
    \node[lang]         (onevm)         at (1.3, -1) { \begin{Bcenter}One\\Varmap\end{Bcenter} };
    \node[pass,proved]  (linearization) at (2.6, -1) { Linearization };
    \node[lang]         (linear)        at (3.9, -1) { Linear };
    \node[pass,proved]  (tunneling)     at (5, -1) { Tunneling };
    \node[pass,proved]  (asmgen)        at (6, -1) { \begin{Bcenter}Assembly\\Generation\end{Bcenter} };

    \node[lang] (asm) at (6, -2) { ASM };
    \node[pass,trusted] (printing) at (4.9, -2) { \begin{Bcenter}Printing\end{Bcenter} };
    \node[lang] (s) at (4, -2) { \texttt{.s} };

    \draw[ar] (src) -- (parse);
    \draw[ar] (parse) -- (preprocess);
    \draw[ar] (preprocess) -- (typecheck);
    \draw[ar] (typecheck) -- (jasmin);
    \draw[ar] (jasmin) -- (i2w);
    \draw[ar] (i2w) -- (subword);
    \draw[ar] (subword) -- (acopy);
    \draw[ar] (acopy) -- (addinit);
    \draw[ar] (addinit) -- (spilling);
    \draw[ar] (spilling) -- (inlining);
    \draw[ar] (inlining) -- (prunning1);
    \draw[ar] (prunning1) -- (constprop);
    \draw[ar] (constprop) -- (dce1);
    \draw[ar] (dce1) -- (unrolling);
    \draw[ar] (unrolling) -- (prunning2);
    \draw[ar] (unrolling) |- ++(-1, 0.55) -| (constprop);
    \draw[ar] (prunning2) -- (split1);
    \draw[ar] (split1) -- (rminit);
    \draw[ar] (rminit) -- (refargs);
    \draw[ar] (refargs) -- (regarray);
    \draw[ar] (regarray) -- (direct);
    \draw[ar] (direct) -- (split2);
    \draw[ar] (split2) -- (globals);
    \draw[ar] (globals) -- (loadconst);
    \draw[ar] (loadconst) -- (lowering);
    \draw[ar] (lowering) -- (inlineprop);
    \draw[ar] (inlineprop) -- (slh);
    \draw[ar] (slh) -- (stackalloc);
    \draw[ar] (stackalloc) -- (stack);
    \draw[ar] (stack) -- (loweraddr);
    \draw[ar] (loweraddr) -- (unused_results);
    \draw[ar] (unused_results) -- (regalloc);
    \draw[ar] (regalloc) -- (dce2);
    \draw[ar] (dce2) -- (chkonevm);
    \draw[ar] (chkonevm) -- (onevm);
    \draw[ar] (onevm) -- (linearization);
    \draw[ar] (linearization) -- (linear);
    \draw[ar] (linear) -- (tunneling);
    \draw[ar] (tunneling) -- (asmgen);
    \draw[ar] (asmgen) -- (asm);
    \draw[ar] (asm) -- (printing);
    \draw[ar] (printing) -- (s);

    \draw[frontend]
         (-0.6,  4.9)
      -- ( 0.3,  4.9)
      -- ( 0.3,  4.5)
      -- ( 6.6,  4.5)
      -- ( 6.6, -0.4)
      -- (-0.6, -0.4)
      -- cycle
    ;
    \node[anchor=north west] at (-0.6,4.9) {\normalsize Front-End};

    \draw[backend]
         (-0.6, -0.5)
      -- ( 6.6, -0.5)
      -- ( 6.6, -1.55)
      -- ( 0.3, -1.55)
      -- ( 0.3, -1.95)
      -- (-0.6, -1.95)
      -- cycle
    ;
    \node[anchor=south west] at (-0.6, -1.95) {\normalsize Back-End};

    \def\xlegend{7.3}
    \def\ylegend{-1.8}
    \def\ylegendsep{0.5}

    \node[trusted,rounded corners]   (legend_trusted)   at (\xlegend, \ylegend+5*\ylegendsep) {\phantom{M}};
    \node[proved,rounded corners]    (legend_verified)  at (\xlegend, \ylegend+4*\ylegendsep) {\phantom{M}};
    \node[validated,rounded corners] (legend_validated) at (\xlegend, \ylegend+3*\ylegendsep) {\phantom{M}};
    \node[lang]                      (legend_lang)      at (\xlegend, \ylegend+2*\ylegendsep) {\phantom{--}}; \node[pass]                      (legend_trans)     at (\xlegend, \ylegend+1*\ylegendsep) {\phantom{M}};
    \node[check]                     (legend_checker)   at (\xlegend, \ylegend+0*\ylegendsep) {\phantom{M}};

    \node[anchor=west] at (legend_trusted.east)   {Trusted step};
    \node[anchor=west] at (legend_verified.east)  {Verified step};
    \node[anchor=west] at (legend_validated.east) {Validated step};
    \node[anchor=west] at (legend_lang.east)      {Language};
    \node[anchor=west] at (legend_trans.east)     {Transformation};
    \node[anchor=west] at (legend_checker.east)   {Checker};
  \end{tikzpicture}
{}
  \caption{Passes and intermediate languages of the \textsc{Jasmin}{} compiler.}\label{fig:passes}
\end{figure*}

The \textsc{Jasmin}{} compiler is a moderately optimizing compiler that is
designed to be predictable and provide a large degree of control to
the programmer; thus, its primary purpose is to eliminate the high-level
constructs and zero-cost abstractions described in
\cref{sec:semantics:jasmin}.
The compiler performs a reduced number of optimizations, such as dead
code elimination and memory reuse (to minimize stack usage).
\Cref{fig:passes}{} presents the passes of
the \textsc{Jasmin}{} compiler. Ellipses correspond to the different
representations of programs, rectangles to transformations, and dotted
rectangles to checkers---i.e., passes that leave the program unchanged
but may reject it. \tikz[baseline=(X.base)]
  \node[
    shape=rectangle,
    rounded corners,
    fill=red!40,
    inner sep=0.7mm
  ] (X) {Red}; boxes correspond to trusted
steps, \tikz[baseline=(X.base)]
  \node[
    shape=rectangle,
    rounded corners,
    fill=green!30,
    inner sep=0.7mm
  ] (X) {green}; ones to those verified in \textsc{Rocq}{}
directly, and \tikz[baseline=(X.base)]
  \node[
    shape=rectangle,
    rounded corners,
    fill=blue!20,
    inner sep=0.7mm
  ] (X) {blue}; ones to steps verified using
translation validation. The passes inside the top dashed gray box
(second to sixth rows) are the compiler \emph{front-end}, whereas the
ones inside the box below (seventh row) are the \emph{back-end}.
\Cref{appendix:passes} presents a summary of the
functionality of each pass. The compiler supports the x86-64, ARMv7,
and RISC-V architectures.

\subsection{Compiler Correctness Statement}\label{sec:preservation:correctness}

We updated the correctness statement and proofs of the compiler to use
the new ITree semantics.
Crucially, \emph{we did not modify the functionality of the compiler},
i.e., the compiler before and after our changes generates the same
assembly; this means that we proved a stronger statement while retaining
the functionality.
The new correctness theorem for the compiler is stated in terms of a
bisimulation (including \ensuremath{{\mathsf{Rnd}}}{}~events) between the semantics of the
source program and its compilation, as follows.

\begin{theorem}[Compiler Correctness \inartifact{}]\label{thm:compiler-correctness}
  Given a program~\(P\), its compilation~\(Q\), any function~\(f\), and
  source and target states \st{}~and~\ensuremath{{\tau}}{},
  \begin{equation*}
    \csim{\st}{\ensuremath{{\tau}}} \implies
    \ensuremath{{{\ensuremath{{
  \llparenthesis\hspace{0.1em}P(f)\hspace{0.1em}\rrparenthesis ^{}_{\st}}}} \mathrel{\euttname[_{\csimname}]} {\ensuremath{{
  \llparenthesis\hspace{0.1em}Q(f)\hspace{0.1em}\rrparenthesis ^{}_{\ensuremath{{\tau}}}}}}}}
  \end{equation*}
  under the following conditions:
  \begin{enumerate}
  \item\label{hyp:safe-def}
    The source program~\(P\) is safe.
  \item\label{hyp:alloc-ok}
    The initial target memory has enough allocated fresh memory for the
    stack of each function (the compiler statically determines how much
    stack space each function needs).
  \item\label{hyp:wf-arg}
    Each writable input pointer is disjoint from all other writable or
    read-only input pointers and from the \ensuremath{{\text{\texttt{.data}}}} section.
  \end{enumerate}
\end{theorem}

Bisimulation, denoted~\euttname{}, means that the two ITrees are the
same except for a finite number of silent steps.
It is indexed by a relation on results, \csimname{} in this case, that
relates the results (i.e., the leaves).
It is the standard bisimulation notion for LTSs and verified compilers,
and it is provided by the ITree library (we define it in
\cref{appendix:itrees}).

Intuitively, the above theorem states that running~\(f\) on related
source-target states \csim{\st}{\ensuremath{{\tau}}} produces the same behaviors
(\ensuremath{{\mathsf{Rnd}}}{} events) and, furthermore, that the final states (if they exist)
are also related.
We define bisimulation precisely in \cref{sec:rhl:xrutt}, and compare it
with the usual forward and backward compiler simulations in
\cref{sec:preservation:comparison}.

The refinement relation~\csimname{} involves several well-formedness
conditions (e.g., that array values in the source correspond to pointers
in the target) and relates values with the usual \emph{more defined
than} relation.
Specifically, a target value is more defined than a source value if
either they are equal or the latter is undefined.
In the case of arrays, the simulation requires that a pointer in the
target points to a region that is more defined than the corresponding
array in the source.
\looseness=-1

\subsection{Comparison with Original Statement}\label{sec:preservation:comparison}

The original semantics of \textsc{Jasmin}{} programs is a big-step operational
semantics of the form \ensuremath{{\langle{c}, {\st}\rangle \Downarrow {\st'}}}, stating that running a
command~\(c\) on a state~\st{} terminates without errors and reaches a
state~\(\st'\)~\cite{DBLP:conf/ccs/AlmeidaBBBGLOPS17,DBLP:conf/sp/AlmeidaBBGKL0S20}.
Random sampling commands consumed an infinite random tape.
Compiler correctness was stated as a forward simulation between a source
command~\(P\) and its
compilation~\(Q\): for every function~\(f\), source states
\st{}~and~\st['], and target state~\ensuremath{{\tau}}{},
\begin{equation*}
  \ensuremath{{\langle{P(f)}, {\st}\rangle \Downarrow {\st'}}} \land \csim{\st}{\ensuremath{{\tau}}} \implies
  \exists \ensuremath{{\tau}}'.\,
    \ensuremath{{\langle{Q(f)}, {\ensuremath{{\tau}}}\rangle \Downarrow {\ensuremath{{\tau}}'}}} \land \csim{\st'}{\ensuremath{{\tau}}'}
  \text,
\end{equation*}
where~\csimname{} is the relation presented in
\cref{sec:preservation:correctness} with an additional clause ensuring
equal random tapes.

Consequently, the original compiler correctness statement ensures that
safe and terminating \textsc{Jasmin}{} programs are compiled to safe and
terminating programs with equal results.  However, it gives no
immediate guarantees for executions that diverge or for the
probabilistic behavior (i.e., as distributions) of programs.

Our work improves over the original compiler correctness in two key
directions.  First, our semantics covers diverging and almost-surely
terminating executions, and is expressed directly as a distribution
rather than requiring a tape.  Secondly, while our proofs follow the
natural forward reasoning pattern, they directly establish a
bisimulation rather than a forward simulation, with no need for
flipping diagrams.

An alternative strategy to overcome the challenges above would have been
to use a small-step operational semantics, with a randomness tape or
oracle, producing an infinite trace.
Interaction trees dispense us from going through such a semantics, which
comes with its difficulties.
First, since our end goal of preservation of cryptographic security
requires a distributional semantics for both source and target
languages, using a trace semantics would require gluing the compiler
correctness statement with distributional semantics for both \textsc{Jasmin}{}
and assembly.
Additionally, nontermination would require infinite tapes, making the
semantics continuous and requiring further glue to recover the desired
cryptographic security statement.
In contrast, ITrees provide a unified framework to define the semantics,
verify the compiler, and reason about probabilistic behavior and
adversarial interactions.

\subsection{Main Theorem}\label{sec:preservation:theorem}

This section leverages the compiler correctness result and the
framework from \cref{sec:crypto} to show that the \textsc{Jasmin}{} compiler
preserves KEM-IND-CCA{} security. We use \cref{thm:eutt-semp} to prove a
KEM-IND-CCA{} security reduction for the \textsc{Jasmin}{} compiler.
That is, we show that \textsc{Jasmin}{} preserves KEM-IND-CCA{} security when
compiling a KEM, as stated in the following theorem.

\begin{theorem}[KEM-IND-CCA{} Security Preservation \inartifact{}]\label{thm:ind-cca-preservation}
  Given a source program~\(P\) with compilation~\(Q\) and an
  adversary~\adversary{}, we have
  \begin{equation*}
    \ensuremath{{\ensuremath{{\operatorname{Adv}}}_{\ensuremath{\text{\tiny{\ensuremath{\ensuremath{{\operatorname{KEM-IND-CCA}}}}}}}}\mathchoice{\!}{\!}{}{}\left({\challenger[_P]}, {\adversary}\right)}} =
      \ensuremath{{\ensuremath{{\operatorname{Adv}}}_{\ensuremath{\text{\tiny{\ensuremath{\ensuremath{{\operatorname{KEM-IND-CCA}}}}}}}}\mathchoice{\!}{\!}{}{}\left({\challenger[_Q]}, {\adversary}\right)}}\text,
  \end{equation*}
  under the compiler correctness conditions from before.
\end{theorem}

This theorem is an exact reduction: advantages are equal, and the
adversary is the same.

\subsection{Application to ML-KEM}\label{sec:preservation:mlkem}

\citet{DBLP:conf/crypto/AlmeidaOBBDGLLLOPQSS24}
use \textsc{EasyCrypt}{} to prove that a highly optimized \textsc{Jasmin}{}
implementation of ML-KEM (denoted \ensuremath{{\mathsf{ML\mbox{-}KEM}_{\ensuremath{\text{\tiny{\ensuremath{\mathsf{Jasmin}}}}}}}}{}) is provably secure under
the Hashed Module variant of the Learning with Errors
assumption (MLWE)~\cite{DBLP:conf/stoc/Regev05}. By combining the
security proof with an adaptation of the functional correctness proof
in~\citet{DBLP:journals/tches/AlmeidaBBGLL00Q23} (the original uses
CRYSTALS-Kyber~\cite{DBLP:conf/eurosp/BosDKLLSSSS18}), they
establish the following theorem.

\begin{theorem}[Provable Security of \ensuremath{{\mathsf{ML\mbox{-}KEM}_{\ensuremath{\text{\tiny{\ensuremath{\mathsf{Jasmin}}}}}}}}{}]\label{thm:mlkem:ec}
For every adversary~\adversary{} that makes at most
\ensuremath{{d}}{}~queries to the hash function and executes in time at
most~\maxtime{},
\begin{equation*}
  \ensuremath{{\ensuremath{{\operatorname{Adv}}}_{\ensuremath{\text{\tiny{\ensuremath{\ensuremath{{\operatorname{KEM-IND-CCA}}}}}}}}\mathchoice{\!}{\!}{}{}\left({\ensuremath{{\mathsf{ML\mbox{-}KEM}_{\ensuremath{\text{\tiny{\ensuremath{\mathsf{Jasmin}}}}}}}}}, {\adversary}\right)}} \leq \ensuremath{{\ensuremath{{\epsilon}}\mathchoice{\!}{\!}{}{}\left(\maxtime, \ensuremath{{d}}\right)}}
  \text.
\end{equation*}
\end{theorem}

Here \ensuremath{{\epsilon}}{} is a complex expression involving
\begin{enumerate*}
\item the correctness bound of \ensuremath{{\mathsf{ML\mbox{-}KEM}_{\ensuremath{\text{\tiny{\ensuremath{\mathsf{Jasmin}}}}}}}}{}, i.e., the probability that
  decapsulation produces an incorrect output;
\item the advantage of an adversary executing in time at
  most~\maxtime['] (where \maxtime[']~is close to~\maxtime{}) against
  Hashed MLWE\@; and
\item pseudorandomness advantages of variants of SHAKE and of the key
  generation smoothing function used in the implementation.
\end{enumerate*}
The full definition is in~\cite[Theorem
  6]{DBLP:conf/crypto/AlmeidaOBBDGLLLOPQSS24}.  Our work yields a
security proof for \emph{the compilation} of ML-KEM (denoted
\ensuremath{{\mathsf{ML\mbox{-}KEM}_{\ensuremath{\text{\tiny{\ensuremath{\mathsf{ASM}}}}}}}}{}), as stated in the following corollary.

\begin{corollary}[Provable Security of \ensuremath{{\mathsf{ML\mbox{-}KEM}_{\ensuremath{\text{\tiny{\ensuremath{\mathsf{ASM}}}}}}}}{}]\label{cor:mlkem:asm}
For every adversary~\adversary{} that makes at most
\ensuremath{{d}}{}~queries to the hash function and executes in time at
most~\maxtime{},
\begin{equation*}
  \ensuremath{{\ensuremath{{\operatorname{Adv}}}_{\ensuremath{\text{\tiny{\ensuremath{\ensuremath{{\operatorname{KEM-IND-CCA}}}}}}}}\mathchoice{\!}{\!}{}{}\left({\ensuremath{{\mathsf{ML\mbox{-}KEM}_{\ensuremath{\text{\tiny{\ensuremath{\mathsf{ASM}}}}}}}}}, {\adversary}\right)}} \leq \ensuremath{{\ensuremath{{\epsilon}}\mathchoice{\!}{\!}{}{}\left(\maxtime, \ensuremath{{d}}\right)}}
  \text.
\end{equation*}
\end{corollary}

Naturally, since \cref{thm:mlkem:ec} is mechanized in \textsc{EasyCrypt}{}, the
TCB of \cref{cor:mlkem:asm} also includes the assumption that the
former proof implies that the \textsc{Jasmin}{} program is KEM-IND-CCA{} secure.

\subsection{Generalization to Game-Based Security}\label{sec:preservation:cil}

As mentioned in the introduction, our \textsc{Rocq}{} development deals with
a general notion of game-based security; thus, all preceding theorems
are proven by instantiating this general framework.
The framework is based on oracle
systems~\cite{DBLP:conf/ccs/BartheDKL10}, which model interactive
games between adaptive adversaries and a cryptographic scheme through
oracle queries.
They provide a unified representation of all common cryptographic
properties---e.g., KEM-IND-CCA{} and SUF-CMA{}---and can capture
different models of cryptography such as the standard, the random
oracle, and the ideal ciphertext models.
\looseness=-1

An oracle system is a family of stateful probabilistic algorithms
indexed by natural numbers.
A pair of functions \ensuremath{{\ensuremath{{\mathsf{I}}} : \ensuremath{{\mathbb{N}}} \to \ensuremath{{\mathsf{Type}}}}} and
\ensuremath{{\ensuremath{{\mathsf{O}}} : \ensuremath{{\mathbb{N}}} \to \ensuremath{{\mathsf{Type}}}}} give the input and output types of each
oracle, respectively.
Oracle queries are events \ensuremath{{\ensuremath{{\mathsf{Que}}}(n,i)}} such that \ensuremath{{n : \ensuremath{{\mathbb{N}}}}} is an
oracle, \ensuremath{{i : \ensuremath{{\ensuremath{{\mathsf{I}}}_{n}}}}} is an input, and the expected answer is of
type \ensuremath{{\ensuremath{{\mathsf{O}}}_{n}}}.
The challenger in this setting gives an implementation for each oracle,
taking an input of type \ensuremath{{\ensuremath{{\mathsf{I}}}_{n}}} and returning an output of type \ensuremath{{\ensuremath{{\mathsf{O}}}_{n}}},
and updating an oracle memory~\ensuremath{{\mathbb{M}}}{}.
On the other hand, the adversary is a probabilistic algorithm that
performs oracle queries and returns in the unit type.
In symbols,
\begin{align*}
  \challenger &:
    (n : \ensuremath{{\mathbb{N}}}) \to
    \left(\ensuremath{{\ensuremath{{\mathsf{I}}}_{n}}} \times \ensuremath{{\mathbb{M}}}\right) \to
    \ensuremath{{\mathsf{itree}~{\ensuremath{{\mathsf{Rnd}}}}~{\left(\ensuremath{{\ensuremath{{\mathsf{O}}}_{n}}} \times \ensuremath{{\mathbb{M}}}\right)}}}\text,
  \\
  \adversary &:
    \ensuremath{{\mathsf{itree}~{\left(\ensuremath{{\mathsf{Que}}} \mathop{+'} \ensuremath{{\mathsf{Rnd}}}\right)}~{\ensuremath{{\mathsf{Unit}}}}}}\text.
\end{align*}
A game executes the adversary and interprets
\ensuremath{{\mathsf{Que}}}{}~events with the appropriate oracles of the challenger.
The adversary wins the game if the trace of query-answer pairs satisfies
a predicate called the \emph{winning condition}
\ensuremath{{\ensuremath{{\mathsf{win}}} : (n, \ensuremath{{\ensuremath{{\mathsf{I}}}_{n}}}, \ensuremath{{\ensuremath{{\mathsf{O}}}_{n}}})^* \to \ensuremath{{\{{0,1}\}}}}}.

Oracle systems have been used to model a variety of cryptographic
schemes and properties~\cite{DBLP:conf/ccs/BartheDKL10,DBLP:conf/eurocrypt/BellareR06,verifiedzkevm2024vcvio}.
As an illustrative example, the KEM-IND-CCA{} oracle system for a given
KEM consists of four oracles:
\begin{enumerate*}
\item an initialization oracle~\ensuremath{{o_{I}}}{} (with
  \ensuremath{{\ensuremath{{\ensuremath{{\mathsf{I}}}_{\ensuremath{{o_{I}}}}}} = \ensuremath{{\mathsf{Unit}}}}} and \ensuremath{{\ensuremath{{\ensuremath{{\mathsf{O}}}_{\ensuremath{{o_{I}}}}}} = \ensuremath{{\mathcal{PK}}}{}}}),
  which uses~\ensuremath{{\operatorname{GenKey}}}{} to generate a key pair, stores it in the
  memory, and returns the public key;
\item a challenge oracle~\ensuremath{{o_{C}}}{} (with
  \ensuremath{{\ensuremath{{\ensuremath{{\mathsf{I}}}_{\ensuremath{{o_{C}}}}}} = \ensuremath{{\mathsf{Unit}}}}} and
  \ensuremath{{\ensuremath{{\ensuremath{{\mathsf{O}}}_{\ensuremath{{o_{C}}}}}} = \ensuremath{{\mathcal{M}}} \times \ensuremath{{\mathcal{C}}}}}), which
  uses~\ensuremath{{\operatorname{Encap}}}{} with the stored public key, samples a random bit and
  message, stores the bit, and returns the chosen message and
  ciphertext;
\item a decapsulation query oracle~\ensuremath{{o_{D}}}{} (with
  \ensuremath{{\ensuremath{{\ensuremath{{\mathsf{I}}}_{\ensuremath{{o_{D}}}}}} = \ensuremath{{\mathcal{C}}}}} and
  \ensuremath{{\ensuremath{{\ensuremath{{\mathsf{O}}}_{\ensuremath{{o_{D}}}}}} = \ensuremath{{\mathcal{M}}}}}), which uses~\Decapname{} with
  the stored secret key to decapsulate the ciphertext and returns its
  result; and
\item a finalize oracle~\ensuremath{{o_{F}}}{} (with
  \ensuremath{{\ensuremath{{\ensuremath{{\mathsf{I}}}_{\ensuremath{{o_{F}}}}}} = \ensuremath{{\{{0,1}\}}}}} and
  \ensuremath{{\ensuremath{{\ensuremath{{\mathsf{O}}}_{\ensuremath{{o_{F}}}}}} = \ensuremath{{\mathsf{Unit}}}}}), which receives the adversary's
  guess.
\end{enumerate*}
The winning condition checks that the adversary's guess is correct, that
the interactions follow the expected order, and that the adversary did
not query the decapsulation oracle on the challenge ciphertext.

Using our framework for oracle systems, we can generalize
\cref{thm:ind-cca-preservation} to all game-based security properties,
as follows.

\begin{theorem}[Security Preservation \inartifact{}]\label{thm:generic-preservation}
  Given a source program~\(P\) with compilation~\(Q\) and an
  adversary~\adversary{}, we have that
  \begin{equation*}
    \pr{\adversary \text{ wins against } \challenger[_P]} =
    \pr{\adversary \text{ wins against } \challenger[_Q]}\text,
  \end{equation*}
  where \challenger[_P]{}~and~\challenger[_Q]{} are the oracle systems induced
  by \(P\)~and~\(Q\), respectively, and we say that an adversary ``wins
  against'' a challenger if their interaction produces a trace that
  satisfies the winning condition.
\end{theorem}

The above theorem is proven directly from source to target, relying only
on the statement of compiler correctness from
\cref{thm:compiler-correctness} and generic ITree properties (in
particular, it is independent of the compiler and its passes).

 \section{Relational Hoare Logic for Interaction Trees}\label{sec:rhl}
In this section, we present the key proof technique used to verify the
compiler in our new setting: a relational Hoare logic for ITree-based
program semantics. We first present standard rules and then discuss
specialized versions for compiler correctness proofs.

\subsection{Tuples and Validity}\label{sec:rhl:tuples}

Relational Hoare logic tuples relate two commands with respect to pre-
and postconditions as follows.

\begin{definition}[Valid Command Tuple]
  A command tuple is valid, written
  \rhlPQ*{\Ip,\Iq}{c_1}{c_2}{\pre}{\post}, when for every pair of states
  \st[_1]~and~\st[_2],
  \begin{equation*}
    \relpre{\st_1}{\st_2} \implies
    \ensuremath{{{
} \mathrel{\raisebox{-1pt}{\ensuremath{{\overset{\ensuremath{\text{\tiny{\ensuremath{u}}}}}{\approx}}}}_{\Ip, \Iq, \post}} { }}}     {\ensuremath{{
  \llparenthesis\hspace{0.1em}c_1\hspace{0.1em}\rrparenthesis ^{\ensuremath{\text{\tiny{\ensuremath{\star}}}}}_{\st_1}}}}
      {\ensuremath{{
  \llparenthesis\hspace{0.1em}c_2\hspace{0.1em}\rrparenthesis ^{\ensuremath{\text{\tiny{\ensuremath{\star}}}}}_{\st_2}}}}\text,
  \end{equation*}
  where \(c_1\)~and~\(c_2\) are commands,
  \ensuremath{{\pre, \post : \ensuremath{{\mathsf{S}}} \to \ensuremath{{\mathsf{S}}} \to \ensuremath{{\mathsf{Prop}}}}} are state pre- and
  postconditions, and
  \Ip{}~and~\Iq{} form a \emph{handler contract}, where
  \ensuremath{{
    \Ip : \ensuremath{{\forall {\ans[_1] \ans[_2]}.\, { \events[_1]~\ans[_1] \to
      \events[_2]~\ans[_2] \to \ensuremath{{\mathsf{Prop}}}}}}
  }} is an \emph{event precondition} and
  \ensuremath{{
    \Iq : \ensuremath{{\forall {\ans[_1] \ans[_2]}.\, { \events[_1]~\ans[_1] \to
      \ans[_1] \to \events[_2]~\ans[_2] \to \ans[_2] \to \ensuremath{{\mathsf{Prop}}}}}}
  }} an \emph{event postcondition}.
\end{definition}
Intuitively, a command tuple is valid when running the commands
\(c_1\)~and~\(c_2\) on \pre{}-related states gives equivalent behaviors
until either~\(c_1\) performs undefined behavior or both commands
terminate with \post{}-related states.
This equivalence of behaviors, denoted \raisebox{-1pt}{\ensuremath{{\overset{\ensuremath{\text{\tiny{\ensuremath{u}}}}}{\approx}}}}{} and defined in the
next section, lifts the standard notion of bisimulation to ITrees with
events (\ensuremath{{\mathsf{Call}}}{}, \ensuremath{{\mathsf{Err}}}{}, and \ensuremath{{\mathsf{Rnd}}}{} in this case) and undefined
behavior.
It uses \Ip{} to relate events performed by the two commands and \Iq{}
to relate the environment's answers to said events.
The judgment extends to programs as follows.

\begin{definition}[Valid Program Tuple]
  A program tuple is valid, written
  \rhlPQ*{\Ip,\Iq}{P_1}{P_2}{\pre}{\post} when for every pair of
  functions \(f\)~and~\(g\) and states \st[_1]~and~\st[_2],
  \begin{equation*}
    \st_1 \mathrel{\ensuremath{{\pre}}_{f\!,g}} \st_2 \implies
    \ensuremath{{{
} \mathrel{\raisebox{-1pt}{\ensuremath{{\overset{\ensuremath{\text{\tiny{\ensuremath{u}}}}}{\approx}}}}_{\Ip, \Iq, \ensuremath{{\post}}_{f\!,g}}} { }}}     {\ensuremath{{
  \llparenthesis\hspace{0.1em}\fbody{f}\hspace{0.1em}\rrparenthesis ^{}_{\st_1}}}}
      {\ensuremath{{
  \llparenthesis\hspace{0.1em}\fbody{g}\hspace{0.1em}\rrparenthesis ^{}_{\st_2}}}}\text,
  \end{equation*}
  where \(P_1\)~and~\(P_2\) are programs,
  \ensuremath{{\pre_{f,g}, \post_{f,g} : \ensuremath{{\mathsf{S}}} \to \ensuremath{{\mathsf{S}}} \to \ensuremath{{\mathsf{Prop}}}}}
  are the pre- and postconditions for functions (indexed by function
  names \(f\)~and~\(g\)),
  \Ip{}~and~\Iq{} form the handler contract.
\end{definition}
Intuitively, a program tuple establishes that running the bodies of a
pair of functions \ensuremath{{f \in P_1}} and \ensuremath{{g \in P_2}} on
\pre[_{f,g}]-related states produces equivalent behaviors forever or until
either~\(f\) performs undefined behavior or both functions terminate
with \post[_{f,g}]-related states.

In the rest of the paper, we use \rhle{e_1}{e_2}{\pre}{\ensuremath{{\psi}}} (where
\(e_1\)~and~\(e_2\) are expressions, \pre{}~relates states, and
\ensuremath{{\psi}}{}~relates values) to express that evaluating \(e_1\)~and~\(e_2\)
in \pre{}-related states yields \ensuremath{{\psi}}{}-related values.  We use
\rhllv{\lv_1}{\lv_2}{\pre}{\ensuremath{{\phi}}}{\post} (where
\pre{}~and~\post{} relate states and \ensuremath{{\phi}}{} values) when writing
\ensuremath{{\phi}}{}-related values to \pre{}-related states yields \post{}-related
states.

\subsection{Equivalence up-to-Cutoff{}}\label{sec:rhl:xrutt}

This section defines the relation~\raisebox{-1pt}{\ensuremath{{\overset{\ensuremath{\text{\tiny{\ensuremath{u}}}}}{\approx}}}}{} needed by RHL tuples.
We first introduce the standard notion of bisimulation from the ITree
library and then our generalization.

Weak bisimulation, denoted~\euttname[_{\phi}] and sometimes called
equivalence up-to-tau, is a widespread notion of equivalence between
ITrees that relates computations that differ only by a finite number of
silent steps on either side and yield \phi{}-related
results~\cite{DBLP:journals/pacmpl/XiaZHHMPZ20}.

Weak bisimulation is too strong for many settings, as it requires events
in the ITrees to match exactly.
Consequently, the ITree library also provides \emph{heterogeneous
equivalence}, denoted~\ruttname[_{\phi,\Ip,\Iq}], a generalization with
two relations for events~\cite{DBLP:journals/pacmpl/SilverZ21,DBLP:journals/pacmpl/Beck0CZZ24}.
It relates ITrees that depend on distinct return types (denoted
\results[_1]~and~\results[_2], related by~\phi{}) and distinct event
families (denoted \events[_1]~and~\events[_2], related with
\Ip{}~and~\Iq{}).
It is a coinductive-inductive relation in the style of weak
bisimulation.

\begin{figure}
  \small
  
  \begin{mathpar}
    {\inferrule[Ret]{
    r_1 \mathrel{\phi} r_2
  }{
     {\ensuremath{{\ensuremath{{\mathsf{Ret}}}\mathchoice{\!}{\!}{}{}\left({r_1}\right)}}} \mathrel{\raisebox{-1pt}{\ensuremath{{\overset{\ensuremath{\text{\tiny{\ensuremath{e/u}}}}}{\approx}}}}} {\ensuremath{{\ensuremath{{\mathsf{Ret}}}\mathchoice{\!}{\!}{}{}\left({r_2}\right)}}}
}}{}

    {\mprset{fraction={===}}
  \inferrule[Tau]{
     {t_1} \mathrel{\raisebox{-1pt}{\ensuremath{{\overset{\ensuremath{\text{\tiny{\ensuremath{e/u}}}}}{\approx}}}}} {t_2}
  }{
    {\ensuremath{{\ensuremath{{\mathsf{Tau}}}\mathchoice{\!}{\!}{}{}\left({t_1}\right)}}} \mathrel{\raisebox{-1pt}{\ensuremath{{\overset{\ensuremath{\text{\tiny{\ensuremath{e/u}}}}}{\approx}}}}} {\ensuremath{{\ensuremath{{\mathsf{Tau}}}\mathchoice{\!}{\!}{}{}\left({t_2}\right)}}}
}}{}

    \colorbox{RoyalBlue!10}{{\inferrule[Cut\(_L\)]{
    C_1(e_1)
  }{
    \ensuremath{{{ } \mathrel{\raisebox{-1pt}{\ensuremath{{\overset{\ensuremath{\text{\tiny{\ensuremath{u}}}}}{\approx}}}}_{}} {\ensuremath{{\ensuremath{{\mathsf{Vis}}}\mathchoice{\!}{\!}{}{}\left({e_1}, {k_1}\right)}}}}} {t_2}
}}{}}

    \colorbox{RoyalBlue!10}{{\inferrule[Cut\(_R\)]{
    C_2(e_2)
  }{
    \ensuremath{{{ } \mathrel{\raisebox{-1pt}{\ensuremath{{\overset{\ensuremath{\text{\tiny{\ensuremath{u}}}}}{\approx}}}}_{}} {t_1}}} {\ensuremath{{\ensuremath{{\mathsf{Vis}}}\mathchoice{\!}{\!}{}{}\left({e_2}, {k_2}\right)}}}
}}{}}

    \inferrule[\rlap{Tau\(_L\)}]{
    {t_1} \mathrel{\raisebox{-1pt}{\ensuremath{{\overset{\ensuremath{\text{\tiny{\ensuremath{e/u}}}}}{\approx}}}}} {t_2} 
  }{
    {\ensuremath{{\mathsf{Tau}}} (t_1)} \mathrel{\raisebox{-1pt}{\ensuremath{{\overset{\ensuremath{\text{\tiny{\ensuremath{e/u}}}}}{\approx}}}}} {t_2}
}{}

    \inferrule[\rlap{Tau\(_R\)}]{
    {t_1} \mathrel{\raisebox{-1pt}{\ensuremath{{\overset{\ensuremath{\text{\tiny{\ensuremath{e/u}}}}}{\approx}}}}} {t_2} 
  }{
    {t_1} \mathrel{\raisebox{-1pt}{\ensuremath{{\overset{\ensuremath{\text{\tiny{\ensuremath{e/u}}}}}{\approx}}}}} {\ensuremath{{\mathsf{Tau}}} (t_2)}
}{}

    {\mprset{fraction={===}}
  \inferrule[Vis]{
    e_1 \mathrel{\Ip} e_2
    \\
    \ensuremath{{\forall {a_1\,a_2}.\, {
      (e_1, a_1) \mathrel{\Iq} (e_2, a_2) \implies
      {k_1(a_1)} \mathrel{\raisebox{-1pt}{\ensuremath{{\overset{\ensuremath{\text{\tiny{\ensuremath{e/u}}}}}{\approx}}}}} {k_2(a_2)}
    }}}
  }{
    {\ensuremath{{\ensuremath{{\mathsf{Vis}}}\mathchoice{\!}{\!}{}{}\left({e_1}, {k_1}\right)}}} \mathrel{\raisebox{-1pt}{\ensuremath{{\overset{\ensuremath{\text{\tiny{\ensuremath{e/u}}}}}{\approx}}}}} {\ensuremath{{\ensuremath{{\mathsf{Vis}}}\mathchoice{\!}{\!}{}{}\left({e_2}, {k_2}\right)}}}
}}{}
  \end{mathpar}
{}
  \caption{Bisimulation relations \ruttname{} and \raisebox{-1pt}{\ensuremath{{\overset{\ensuremath{\text{\tiny{\ensuremath{u}}}}}{\approx}}}}{}.
    The \colorbox{RoyalBlue!10}{boxed} rules are only for \raisebox{-1pt}{\ensuremath{{\overset{\ensuremath{\text{\tiny{\ensuremath{u}}}}}{\approx}}}}{},
    all others (with \raisebox{-1pt}{\ensuremath{{\overset{\ensuremath{\text{\tiny{\ensuremath{e/u}}}}}{\approx}}}}\,) are valid for both \ruttname{} and
    \raisebox{-1pt}{\ensuremath{{\overset{\ensuremath{\text{\tiny{\ensuremath{u}}}}}{\approx}}}}{}.}\label{fig:xrutt}
\end{figure}

\Cref{fig:xrutt}{} defines~\ruttname{} (we distinguish inductive and
coinductive uses of the premise with single and double bars,
respectively~\cite{DBLP:conf/esop/Leroy06}).
The relation is parameterized by a relation
\ensuremath{{\phi : \results[_1] \to \results[_2] \to \ensuremath{{\mathsf{Prop}}}}},
an event precondition
\ensuremath{{\Ip : \ensuremath{{\forall {\ans[_1] \ans[_2]}.\, { \events[_1]~\ans[_1] \to
    \events[_2]~\ans[_2] \to \ensuremath{{\mathsf{Prop}}}}}}
}}, and its postcondition \ensuremath{{\Iq : \ensuremath{{\forall {\ans[_1] \ans[_2]}.\, { \events[_1]~\ans[_1] \to
    \ans[_1] \to \events[_2]~\ans[_2] \to \ans[_2] \to \ensuremath{{\mathsf{Prop}}}}}}
}}.

The \hyperref[fig:xrutt]{\textsc{Ret}}{} rule relates any pair of \phi{}-related return
values and the \hyperref[fig:xrutt]{\textsc{Tau}}{} rule relates two ITrees that perform a
synchronized silent step.
The \hyperref[fig:xrutt]{\textsc{Tau\(_L\)}}{} and \hyperref[fig:xrutt]{\textsc{Tau\(_R\)}}{} rules rely on their hypothesis
\emph{inductively} rather than coinductively, ensuring that related
ITrees can differ only up to a \emph{finite} number of silent transition
steps.
The \hyperref[fig:xrutt]{\textsc{Vis}}{} rule requires that, given answer types
\ans[_1]~and~\ans[_2] and events
\ensuremath{{e_1 : \events[_1]~\ans[_1]}} and
\ensuremath{{e_2 : \events[_2]~\ans[_2]}}, the events are related by~\Ip{}.
The continuations \(k_1(a_1)\)~and~\(k_2(a_2)\) must also be in
relation, assuming that the answers \(a_1\)~and~\(a_2\) satisfy the
event postcondition~\Iq{}.

\paragraph*{Our Generalization}
Heterogeneous equivalence is still too strong for compiler correctness
proofs, which crucially rely on reasoning about undefined behavior on
one side of the relation.
Compiler correctness typically relates the semantics of a source program
to its compilation's \emph{as long as the source program is safe}.
Assuming safety is often essential in compiler proofs, as it allows us
to disregard invalid behaviors that need not be preserved.
Additionally, safety also encodes important invariants of the compiler
passes (e.g., if the compiler introduces a memory access, that access
must be in-bounds) that subsequent passes rely upon.

As a result, our work further generalizes heterogeneous equivalence to
support error handling in terms of undefined behavior.
It relies on the notion of \emph{cutoff event}, on either side of the
relation, that allows the relation to hold trivially from that point,
regardless of the continuation.
\Cref{fig:xrutt}{} defines the new relation~\raisebox{-1pt}{\ensuremath{{\overset{\ensuremath{\text{\tiny{\ensuremath{u}}}}}{\approx}}}}{}, called equivalence
up-to-cutoff{}, as an extension of~\ruttname{} with two additional
boolean predicate parameters,
\ensuremath{{C_i : \ensuremath{{\forall {\ans}.\, {\events_i~\ans \to \mathtt{bool}}}}}} for
\ensuremath{{i = 1,2}}, which decide which events are cutoffs on each side of
the relation.
The \hyperref[fig:xrutt]{\textsc{Cut\(_L\)}}{} rule
relates~\ensuremath{{\ensuremath{{\ensuremath{{\mathsf{Vis}}}\mathchoice{\!}{\!}{}{}\left({e}, {k}\right)}} : \ensuremath{{\mathsf{itree}~{\events_1}~{\results_1}}}}} with
any~\ensuremath{{t : \ensuremath{{\mathsf{itree}~{\events[_2]}~{\results[_2]}}}}} whenever~\(e\) is a
cutoff event on the left-hand side; \hyperref[fig:xrutt]{\textsc{Cut\(_R\)}}{} is symmetric.
When there are no cutoffs on either side, \raisebox{-1pt}{\ensuremath{{\overset{\ensuremath{\text{\tiny{\ensuremath{u}}}}}{\approx}}}}{} is equivalent to
\ruttname{}.

\subsection{Proof Rules}
Relational Hoare logic rules typically come in three flavors:
\emph{two-sided} rules, \emph{one-sided} rules, and \emph{structural}
rules. This section briefly discusses each group in turn (we omit the
\hyperref[fig:extra-rhl]{\textsc{Skip}}{}, \hyperref[fig:extra-rhl]{\textsc{Seq}}{}, and \hyperref[fig:extra-rhl]{\textsc{Case}}{} rules here, which are
standard and included in \cref{sec:appendix-rhl}).
As usual, \rhlPQ{\Ip,\Iq}{c_1}{c_2}{\pre}{\post}
(resp.\ \rhlPQ{\Ip,\Iq}{P_1}{P_2}{\pre}{\post}) means that an RHL
command (resp.\ program) tuple is derivable.
We leave \Ip{}~and~\Iq{} implicit when they are clear from the
context.

\begin{figure}
  \begin{minipage}{\linewidth}
    \small
    \begin{mathpar}
      \inferrule[Assign]{
  \rhle{e_1}{e_2}{\pre}{\ensuremath{{\phi}}}
  \\
  \rhllv{\lv_1}{\lv_2}{\pre}{\ensuremath{{\phi}}}{\post}
}{
  \rhl
    {\ensuremath{{{\lv_1}~{\coloneq}~{e_1}}}}
    {\ensuremath{{{\lv_2}~{\coloneq}~{e_2}}}}
    {\pre}
    {\post}
}{}

      \inferrule[While]{
  \rhle{e_1}{e_2}{\pre}{=}
  \\
  \rhl{c_1}{c_2}{\pre \land e_1}{\pre}
}{
  \rhl
    {\ensuremath{{\ensuremath{{\text{\textbf{while}}}}~({e_1})~\{c_1\}}}}
    {\ensuremath{{\ensuremath{{\text{\textbf{while}}}}~({e_2})~\{c_2\}}}}
    {\pre}
    {\pre \land \neg e_1}
}{}

      \inferrule[Rand]{
  \rhle{e_1}{e_2}{\pre}{\Ip}
  \\
  \rhllv{\lv_1}{\lv_2}{\pre}{\Iq}{\post}
}{
  \rhl
    {\ensuremath{{{\lv_1}~{\coloneq}~{\ensuremath{{\ensuremath{{\text{\textbf{rand}}}}}}({e_1})}}}}
    {\ensuremath{{{\lv_2}~{\coloneq}~{\ensuremath{{\ensuremath{{\text{\textbf{rand}}}}}}({e_2})}}}}
    {\pre}
    {\post}
}{}

      \inferrule[Call]{
  \rhle{e_1}{e_2}{\left(\pre_{\ensuremath{{r}}} \land \Ip[^{\ensuremath{{m}}}]\right)}{\Ip[^v]}
  \\
  \rhllv{\lv_1}{\lv_2}{\left(\pre_{\ensuremath{{r}}} \land \Iq[^{\ensuremath{{m}}}]\right)}{\Iq[^v]}{\post}
}{
  \rhl
    {\ensuremath{{{\lv_1}~{\coloneq}~{{f}\mathchoice{\!}{\!}{}{}\left({e_1}\right)}}}}
    {\ensuremath{{{\lv_2}~{\coloneq}~{{g}\mathchoice{\!}{\!}{}{}\left({e_2}\right)}}}}
    {\left(\pre_{\ensuremath{{r}}} \land \Ip[^{\ensuremath{{m}}}]\right)}
    {\post}
}{}

      \inferrule[Assign\(_L\)]{\ensuremath{{\forall {\st_1 \st_1' \st_2}.\, {
      \relpre{\st_1}{\st_2} \land
      \ensuremath{{{\ensuremath{{
  \llparenthesis\hspace{0.1em}\ensuremath{{{\lv}~{\coloneq}~{e}}}\hspace{0.1em}\rrparenthesis ^{\ensuremath{\text{\tiny{\ensuremath{\star}}}}}_{\st_1}}}} \mathrel{\euttname[_{}]} {\ensuremath{{\ensuremath{{\mathsf{Ret}}}\mathchoice{\!}{\!}{}{}\left({\st_1'}\right)}}}}} \implies
      \relpost{\st_1'}{\st_2}
    }}}
  }{
    \rhl{\ensuremath{{{\lv}~{\coloneq}~{e}}}}{\ensuremath{{\ensuremath{{\text{\textbf{skip}}}}}}}{\pre}{\post}
}{}

      \inferrule[Trans]{
  \rhlPQ{\Ip_1, \Iq_1}{c_1}{c_2}{\pre_1}{\post_1}
  \\
  \rhlPQ{\Ip_2, \Iq_2}{c_2}{c_3}{\pre_2}{\post_2}
}{
  \rhlPQ
    {\compose{\Ip_1}{\Ip_2}, \compose{\Iq_1}{\Iq_2}}
    {c_1}{c_3}
    {\compose{\pre_1}{\pre_2}}
    {\compose{\post_1}{\post_2}}
}{}

      \inferrule[Rec]{
  \rhlPQ
    {\ensuremath{{\Ip \cup {\pre}}}, \ensuremath{{\Iq \cup {\post}}}}
    {\fbody{f}}
    {\fbody{g}}
    {\ensuremath{{\pre}}_{f\!,g}}
    {\ensuremath{{\post}}_{f\!,g}}
  \quad \text{for every } f \in P_1, g \in P_2
  }{
    \rhlPQ{\Ip,\Iq}{P_1}{P_2}{\pre}{\post}
}{}
    \end{mathpar}
  \end{minipage}
  \caption{Relational Hoare Logic.}\label{fig:rhl}\label{fig:rhlstruct}\label{fig:GOSRHL}\label{fig:rhlrec}
\end{figure}

\subsubsection*{Two-Sided Rules}
The first four rules of \cref{fig:rhl}{} are two-sided rules.
The \hyperref[fig:rhl]{\textsc{Assign}}{} rule states that two assignments are related if
evaluating their expressions \(e_1\)~and~\(e_2\) under the
assumption~\pre{} yields \ensuremath{{\phi}}{}-related values, and writing such values
in \lv[_1]~and~\lv[_2] ensures~\post{}.
The \hyperref[fig:rhl]{\textsc{While}}{} rule is \emph{synchronous}, i.e., it requires that
the guards \(e_1\)~and~\(e_2\) evaluate to the same value, hence the
postcondition for the expressions is equality.
The formula~\pre{} acts as a loop invariant, as in Hoare logic.

The \hyperref[fig:rhl]{\textsc{Rand}}{} rule deals with random sampling. It requires
that the arguments \(e_1\)~and~\(e_2\) satisfy the handler
precondition~\Ip{} (for \ensuremath{{\mathsf{Rnd}}}{} events), and that the handler
postcondition~\Iq{} ensures the postcondition~\post{} on the states
after writing the answers to \lv[_1]~and~\lv[_2].

Finally, the \hyperref[fig:rhl]{\textsc{Call}}{} rule relates two function calls.
The specification for functions is given by the handler contract
\Ip{}~and~\Iq{} for \ensuremath{{\mathsf{Call}}}{} events: it establishes the (relational)
pre- and postconditions for calling each pair of functions.
As such, \Ip{}~and~\Iq{} are relations between pairs of lists of values
and memories.
Without loss of generality, we split them into \Ip[^v]~and~\Ip[^{\ensuremath{{m}}}],
and \Iq[^v]~and~\Iq[^{\ensuremath{{m}}}], for the relations values and memories,
respectively.

The first premise of the \hyperref[fig:rhl]{\textsc{Call}}{} rule states that,
given a precondition on variable maps~\pre[_{\ensuremath{{r}}}], the arguments
\(e_1\)~and~\(e_2\) must satisfy the functions' precondition~\Ip[^v].
The second premise asserts the postcondition~\post{} after writing
the results in states where \emph{the initial} precondition on variable
maps~\pre[_{\ensuremath{{r}}}] holds, as well as the functions' postcondition~\Iq{}.
The precondition~\pre[_{\ensuremath{{r}}}] holds after writing results because
the callee executes on its own local variable map, leaving the caller's
unchanged.

\subsubsection*{One-Sided Rules}
These rules relate structurally dissimilar programs---often, an
event-free command with \ensuremath{{\ensuremath{{\text{\textbf{skip}}}}}}{}.
The \hyperref[fig:GOSRHL]{\textsc{Assign\(_L\)}}{} rule is an example: it relates an assignment on
the left-hand side with \ensuremath{{\ensuremath{{\text{\textbf{skip}}}}}}{} on the right-hand side.
The premise states that the postcondition~\post{} must hold after an
assignment on the state on the left-hand side.

\subsubsection*{Structural Rules}
Our logic satisfies many standard structural rules of program logics,
such as reflexivity, consequence, and transitivity.
For example, the \hyperref[fig:rhlstruct]{\textsc{Trans}}{} rule allows us to compose judgments
transitively.
We use~\composename{} for the composition of two relations, defined as
\ensuremath{{
  x \mathrel{\left(\compose{R}{R'}\right)} z \triangleq
  \ensuremath{{\exists {y}.\, {x \mathrel{R} y \land y \mathrel{R'} z}}}
}}.

\subsubsection*{Recursion and Programs}
\Cref{fig:rhlrec}{} presents the \hyperref[fig:rhlrec]{\textsc{Rec}}{} rule to relate two programs.
The rule lifts and generalizes the classical recursion rule in Hoare
logic~\cite{DBLP:series/txcs/AptBO09} to the relational setting.
As before, we write \fbody{f}~and~\fbody{g} for the bodies of
\(f\)~and~\(g\) in \(P_1\)~and~\(P_2\), respectively.
In the premise, we extend the contract to
\ensuremath{{\Ip \cup {\pre}}}~and~\ensuremath{{\Iq \cup {\post}}}, which relates \ensuremath{{\mathsf{Call}}}{} events using
\pre{}~and~\post{}, and other events using \Ip{}~and~\Iq{}.
The rule requires proving that \emph{the bodies} of every pair of
functions \(f\)~and~\(g\) satisfy their relational specification
\emph{assuming that all function calls satisfy theirs}.
Note that this allows reasoning about mutually recursive functions.

As a result, two programs are related if all pairs of bodies satisfy
their relational specification.
While the premise of this rule involves only the call{} semantics,
the soundness theorem below shows that its conclusion relates the
interprocedural{} semantics of functions, allowing us to reason abstractly
about calls.

\subsubsection*{Soundness}
The following theorem establishes the soundness of our program logic,
relating the call{} and interprocedural{} semantics.

\begin{theorem}[Soundness \inartifact{}]\label{thm:soundness}
Provable judgments are valid, i.e.,
\rhlPQ{\Ip,\Iq}{P_1}{P_2}{\pre}{\post} implies
\rhlPQ*{\Ip,\Iq}{P_1}{P_2}{\pre}{\post}.
\end{theorem}

 \section{Key Proof Techniques for Selected Passes}\label{sec:verification}

This section discusses the key techniques used to verify the compiler.
Recall that the \textsc{Jasmin}{} compiler operates over seven different
languages. The front-end has four internal languages: \textsc{Jasmin}{} (the
source), Subword, Direct, and Stack (see \cref{fig:passes}{}).  These
languages have the same syntax but different semantics, i.e., a
language construct such as a function call has different effects,
becoming lower-level down the compilation chain.  Nevertheless, their
semantics are instances of a general one, parameterized over the
effects of certain constructs, such as function calls. The back-end
begins with the Stack language and comprises three other languages:
One Varmap, Linear and ASM\@. All languages are defined in layers
using ITrees; in particular, all have a layer with error and random
events.

We verify the front-end using RHL\@: this section illustrates how RHL
is used in the proofs of three such passes and distill three key
relevant features of the logic---which are not specific to \textsc{Jasmin}{}
or to its ITree semantics.  We verify the back-end directly: we
discuss one pass that leverages the layered semantics design to
define a mixed-step semantics and enable its proof.

\subsection{Register Renaming and Static Analyses}\label{sec:verification:register-renaming}

The Register Renaming pass renames variables to use machine register
names; for example, \ensuremath{{{x}~{\coloneq}~{y+3}}} gets renamed to
\ensuremath{{{\ensuremath{{\mathsf{rdi}}}}~{\coloneq}~{\ensuremath{{\mathsf{rdi}}}+3}}}.
This pass performs no spilling, and therefore leaves the program
structure unchanged.
We verify this pass in translation validation style: we use an
unverified algorithm implemented in OCaml to generate a renamed program,
and then a verified checker to ensure that the programs are
equivalent modulo renaming.

The checker in the Register Renaming pass checks that two programs are
equivalent modulo renaming.
It analyzes the code to compute a \emph{renaming}, a map from source
variables to target registers, at each program point.
For example, the instruction \ensuremath{{{x}~{\coloneq}~{y + 3}}} is equivalent modulo
renaming to \ensuremath{{{\ensuremath{{\mathsf{rdi}}}}~{\coloneq}~{\ensuremath{{\mathsf{rdi}}} + 3}}} if the initial renaming (before the
assignment) is \ensuremath{{\{{y \mapsto \ensuremath{{\mathsf{rdi}}}}\}}} and the resulting one (after the
assignment) is \ensuremath{{\{{x \mapsto \ensuremath{{\mathsf{rdi}}}, y \mapsto \bot}\}}}.

The correctness of the checker states that the source and target states
are equal up to renaming of register variables.
The simulation relation is parameterized by a renaming
\ensuremath{{\domain : \ensuremath{{\mathsf{Var}}} \rightharpoonup \ensuremath{{\mathsf{Var}}}}}, which is a
partial function from variables to variables.
The simulation relation for this pass states that memories should be
identical, and if the renaming is defined on~\(x\), the variable maps
coincide there:
\begin{equation*}
  \csim[_{\domain}]{\st}{\st'} \triangleq
    \ensuremath{{\st_{\ensuremath{{m}}}}} = \ensuremath{{\st'_{\ensuremath{{m}}}}} \land
    \ensuremath{{\forall {x \in \dom{\domain}}.\, {\st(x) = \st'(\domain(x))}}}\text.
\end{equation*}

\paragraph*{Key Feature: Leveraging Static Analyses}
To verify passes like Register Renaming that depend on static analyses,
we develop a general approach based on simplified RHL rules.
The approach entails an abstract set \ensuremath{{\mathcal{A}}}{}, a logical
interpretation \ensuremath{{
  {\left(\ensuremath{{R}}_{\domain}\right)}_{\domain\in\ensuremath{{\mathcal{A}}}} :
    \ensuremath{{\mathsf{S}}} \times \ensuremath{{\mathsf{S}}} \to \ensuremath{{\mathsf{Prop}}}
}}, a \emph{checker} for expressions \ensuremath{{
  \ensuremath{{\ensuremath{{\mathsf{chk}}}_e}} : \ensuremath{{\mathcal{A}}} \times \ensuremath{{\mathsf{Expr}}} \times \ensuremath{{\mathsf{Expr}}} \times
    \ensuremath{{\mathcal{A}}} \to \ensuremath{{\mathsf{Prop}}}
}} and one for left-values \ensuremath{{
  \ensuremath{{\ensuremath{{\mathsf{chk}}}_{\lv}}} : \ensuremath{{\mathcal{A}}} \times \ensuremath{{\mathsf{LVal}}} \times \ensuremath{{\mathsf{LVal}}} \times \ensuremath{{\mathcal{A}}} \to
  \ensuremath{{\mathsf{Prop}}}
}}.
Informally, the abstract set collects information about the two program
executions, and the interpretation maps elements of the abstract domain
to a logical representation.
The checkers, on the other hand, verify whether an abstract value
validates a desired relation~\ensuremath{{\mathcal{S}}}{} (e.g., that two given
expressions evaluate to the same value) and establish an updated
abstract value.  We identify three properties needed to incorporate
checkers in the logic, namely:
{\small\begin{align}
  \left.\begin{array}{@{} r}
    \ensuremath{{{\st_1} \mathrel{\ensuremath{{R}}_{\domain}} {\st_2}}}
    \\
    \ensuremath{{\ensuremath{{\ensuremath{{\mathsf{chk}}}_e}}\mathchoice{\!}{\!}{}{}\left({\domain}, {e_1}, {e_2},{\domain'}\right)}}
  \end{array}\right\} &\implies
  \ensuremath{{{\st_1} \mathrel{\ensuremath{{R}}_{\domain'}} {\st_2}}}\text,
  \tag{\textsc{Mono}}\label{chkweak}
  \\
  \ensuremath{{\ensuremath{{\ensuremath{{\mathsf{chk}}}_e}}\mathchoice{\!}{\!}{}{}\left({\domain}, {e_1}, {e_2},{\domain'}\right)}} &\implies
  \rhle{e_1}{e_2}{\ensuremath{{{\!} \mathrel{\ensuremath{{R}}_{\domain}} {\!}}}}{\ensuremath{{\mathcal{S}}}}\text,
  \tag{\textsc{Corr}\(_e\)}\label{chke}
  \\
  \ensuremath{{\ensuremath{{\ensuremath{{\mathsf{chk}}}_{\lv}}}\mathchoice{\!}{\!}{}{}\left({\domain}, {\lv_1}, {\lv_2}, {\domain'} \right)}} &\implies
  \rhllv
    {\lv_1}
    {\lv_2}
    {\ensuremath{{{\!} \mathrel{\ensuremath{{R}}_{\domain}} {\!}}}}
    {\ensuremath{{\mathcal{S}}}}
    {\ensuremath{{{\!} \mathrel{\ensuremath{{R}}_{\domain'}} {\!}}}}\text.
  \tag{\textsc{Corr}\(_{\lv}\)}\label{chklv}
\end{align}}
The first property ensures that no information is lost after analyzing
expressions (since evaluating expressions does not change the state) and
allows composing the results of multiple checks.
The correctness of \ensuremath{{\ensuremath{{\mathsf{chk}}}_e}}{} ensures that evaluating
\(e_1\)~and~\(e_2\) in \ensuremath{{{\!} \mathrel{\ensuremath{{R}}_{\domain}} {\!}}}-related states yields
\ensuremath{{\mathcal{S}}}{}-related values.
The correctness of \ensuremath{{\ensuremath{{\mathsf{chk}}}_{\lv}}}{} ensures that if the check with \domain{}
is successful and yields \domain['], updating \domain[]-related states
with \ensuremath{{\mathcal{S}}}{}-related values yields \domain[']-related states.

The properties of checkers can be used to simplify several of the RHL
rules presented in \cref{fig:rhl}{}.
The first two rules of \cref{rhl-chk}{} are simplified versions of the
\hyperref[fig:rhl]{\textsc{Assign}}{} and \hyperref[fig:extra-rhl]{\textsc{Case}}{} rules, whose premises rely on the
checkers to establish the necessary RHL tuples.
Such premises are trivially discharged in compiler transformations like
Register Renaming that replace commands only when the checkers succeed.
In this pass, checking two assignments \ensuremath{{{\lv_1}~{\coloneq}~{e_1}}} and
\ensuremath{{{\lv_2}~{\coloneq}~{e_2}}} succeeds only if there exists a renaming~\domain{}
such that \(e_2\)~is a renaming of \(e_1\), and similarly for the
left-values.

\begin{figure}
  \begin{minipage}{\linewidth}
    \small
    \begin{mathpar}
      \inferrule[Assign\(_{\ensuremath{{\mathsf{chk}}}}\)]{
  \ensuremath{{\ensuremath{{\ensuremath{{\mathsf{chk}}}_e}}\mathchoice{\!}{\!}{}{}\left({\domain}, {e_1}, {e_2},{\domain'}\right)}}
  \\
  \ensuremath{{\ensuremath{{\ensuremath{{\mathsf{chk}}}_{\lv}}}\mathchoice{\!}{\!}{}{}\left({\domain'}, {\lv_1}, {\lv_2}, {\domain''} \right)}}
}{
  \rhl
    {\ensuremath{{{\lv_1}~{\coloneq}~{e_1}}}}
    {\ensuremath{{{\lv_2}~{\coloneq}~{e_2}}}}
    {\ensuremath{{{} \mathrel{\ensuremath{{R}}_{\domain}} {}}}}
    {\ensuremath{{{} \mathrel{\ensuremath{{R}}_{\domain''}} {}}}}
}{}

      \inferrule[Case\(_{\ensuremath{{\mathsf{chk}}}}\)]{
  \ensuremath{{\ensuremath{{\ensuremath{{\mathsf{chk}}}_e}}\mathchoice{\!}{\!}{}{}\left({\domain}, {e_1}, {e_2},{\domain'}\right)}}
  \\
  \rhle{e_1}{e_2}{\ensuremath{{{} \mathrel{\ensuremath{{R}}_{\domain}} {}}}}{=}
  \\\\
  \rhl{c_1}{c_2}{\ensuremath{{{} \mathrel{\ensuremath{{R}}_{\domain'}} {}}}}{\ensuremath{{{} \mathrel{\ensuremath{{R}}_{\domain''}} {}}}}
  \\
  \rhl{c_1'}{c_2'}{\ensuremath{{{} \mathrel{\ensuremath{{R}}_{\domain'}} {}}}}{\ensuremath{{{} \mathrel{\ensuremath{{R}}_{\domain''}} {}}}}
}{
  \rhl
    {\ensuremath{{\ensuremath{{\text{\textbf{if}}}}~({e_1})~\{{c_1}\}~\{{c_1'}\}}}}
    {\ensuremath{{\ensuremath{{\text{\textbf{if}}}}~({e_2})~\{{c_2}\}~\{{c_2'}\}}}}
    {\ensuremath{{{} \mathrel{\ensuremath{{R}}_{\domain}} {}}}}
    {\ensuremath{{{} \mathrel{\ensuremath{{R}}_{\domain''}} {}}}}
}{}

      \inferrule[Case\(_b\)]{
  \ensuremath{{\forall {\st_1 \st_2}.\, {
    \relpre{\st_1}{\st_2} \implies
    \ensuremath{{{\ensuremath{{
  \llparenthesis\hspace{0.1em}e\hspace{0.1em}\rrparenthesis ^{\ensuremath{\text{\tiny{\ensuremath{\ensuremath{{\mathsf{Expr}}}}}}}}_{\st_1}}}} \mathrel{\euttname[_{}]} {\ensuremath{{\ensuremath{{\mathsf{Ret}}}\mathchoice{\!}{\!}{}{}\left({b}\right)}}}}}
  }}}
  \\
  \rhl
    {c_b}
    {c}
    {\pre}
    {\post}
}{
  \rhl
    {\ensuremath{{\ensuremath{{\text{\textbf{if}}}}~({e})~\{{c_{\top}}\}~\{{c_{\bot}}\}}}}
    {c}
    {\pre}
    {\post}
}{}

      \inferrule[Unroll\(_L\)]{
  \rhl
    {\ensuremath{{\ensuremath{{\text{\textbf{if}}}}~({e})~\{{c \mathop{;} \ensuremath{{\ensuremath{{\text{\textbf{while}}}}~({e})~\{c\}}}}\}~\{{\ensuremath{{\ensuremath{{\text{\textbf{skip}}}}}}}\}}}}
    {c'}
    {\pre}
    {\post}
}{
  \rhl
    {\ensuremath{{\ensuremath{{\text{\textbf{while}}}}~({e})~\{c\}}}}
    {c'}
    {\pre}
    {\post}
}{}
    \end{mathpar}
  \end{minipage}
  \caption{Specialized Relational Hoare Logic.}\label{fig:OSRHL}\label{rhl-chk}\end{figure}

\subsection{Constant Branches and One-Sided Rules}\label{sec:verification:constant-branches}

The Constant Propagation pass propagates the values of constants,
simplifies complex expressions, and optimizes control flow constructs
if their guard is constant.
For example, \ensuremath{{\ensuremath{{{i}~{\coloneq}~{3}}}\mathop{;} \ensuremath{{{x}~{\coloneq}~{i-2}}}}}
maps to \ensuremath{{\ensuremath{{{i}~{\coloneq}~{3}}}\mathop{;} \ensuremath{{{x}~{\coloneq}~{1}}}}}.
This transformation optimizes control flow constructs whose guards are
constant---i.e., it removes constant branches.  For example, the
conditional command \ensuremath{{\ensuremath{{\text{\textbf{if}}}}~({x = x})~\{{c}\}~\{{c'}\}}} maps to~\(c\), and
\ensuremath{{\ensuremath{{\text{\textbf{while}}}}~({x < x})~\{c\}}} maps to~\ensuremath{{\ensuremath{{\text{\textbf{skip}}}}}}{}.
We prove it correct by showing that the source and target states are
equal.

\paragraph*{Key Feature: Specialized One-Sided Rules}
We introduce specialized one-sided rules that, for instance, deal
directly with constant branches.
Thus, when a while loop has a false guard, we first use the
\hyperref[fig:OSRHL]{\textsc{Unroll\(_L\)}}{} rule from \cref{fig:OSRHL}{}, which relates a while loop to
its one-step unrolling, and then the \hyperref[fig:OSRHL]{\textsc{Case\(_b\)}}{} rule to relate
it to \ensuremath{{\ensuremath{{\text{\textbf{skip}}}}}}{}.
In general, specialized one-sided rules are tailored for common compiler
transformations.
They pattern-match on the command on the left-hand side (i.e., the
source) and set the command on the right-hand side (i.e., the target) to
the transformation of the former.
Conveniently, the rules have non-relational proof obligations to prove
that the right-hand side correctly implements the source.
Such obligations enjoy good automation in the \textsc{Rocq}{} prover, thanks to
our defining a partial symbolic evaluator for commands that can reduce
loop- and event-free commands to their results; for instance, it
establishes that \ensuremath{{{\ensuremath{{
  \llparenthesis\hspace{0.1em}\ensuremath{{{x}~{\coloneq}~{0}}}\hspace{0.1em}\rrparenthesis ^{\ensuremath{\text{\tiny{\ensuremath{\star}}}}}_{\st}}}} \mathrel{\euttname[_{}]} {\ensuremath{{\ensuremath{{\mathsf{Ret}}}\mathchoice{\!}{\!}{}{}\left({\ensuremath{{{\st}\mathchoice{\!}{\!}{}{}\left[{x} \mapsto {0}\right]}}}\right)}}}}}.
All passes in \textsc{Jasmin}{} that modify the program structure use one-sided
rules; further examples are Instruction Selection to introduce
assignments before complex computations and Dead Code Elimination to
remove useless assignments.
\looseness=-1

\subsection{Inlining and Different Semantics}\label{sec:verification:inlining}

This pass replaces function calls with the body of the callee if the
latter is annotated as \ensuremath{{\text{\texttt{\#inline}}}}{}.
The transformation is performed in the order in which functions are
defined, which means that we perform inlining once per function (as its
callees have already been processed).
The proof for this pass is the most complex in the compiler after our
changes, since a critical step in the argument requires a generic result
on ITrees that is not provable in RHL directly.
Proving the correctness of this pass is challenging because it requires
relating an ITree with \ensuremath{{\ensuremath{{\mathsf{Call}}}\mathchoice{\!}{\!}{}{}\left({f, \ensuremath{{m}}, v}\right)}}~events with an ITree
where some of these events have already been interpreted as the
semantics of the body of~\(f\) on a state containing \ensuremath{{m}}{}~and~\(v\).
This is not expressible as one of our one-sided rules, as their
equivalence holds only \emph{after} interpreting \ensuremath{{\mathsf{Call}}}{}~events.

\paragraph*{Key Feature: Parameterization with Respect to Semantics}
Our solution exploits the parametricity of the logic over program
semantics. Concretely, we introduce a new handler for \ensuremath{{\mathsf{Call}}}{} events,
denoted \ensuremath{{\ensuremath{{H}}^I}}{}, to bring out the effects of inlining in a new
\emph{one-step inlining semantics}.  Using this new semantics, we
proceed in two steps: first, we show that the semantics of the source
program is equivalent to its one-step inlining semantics; and second,
we use RHL to prove that the one-step inlining semantics of the source
program is equivalent to the semantics of the target program.

The new handler and semantics are as follows:
{\small\begin{align*}
  \ensuremath{{\ensuremath{{\ensuremath{{H}}^I}}\mathchoice{\!}{\!}{}{}\left(\ensuremath{{\mathsf{Call}}}\mathchoice{\!}{\!}{}{}\left({f}, {\ensuremath{{m}}}, {v}\right)\right)}} &\triangleq
    \begin{cases}
      \ensuremath{{\ensuremath{{\ensuremath{{H}}^{\ensuremath{\text{\tiny{\ensuremath{\hspace{0.1em}\mathsf{c}}}}}}}}\mathchoice{\!}{\!}{}{}\left(\ensuremath{{\mathsf{Call}}}\mathchoice{\!}{\!}{}{}\left({f}, {\ensuremath{{m}}}, {v}\right)\right)}} & \ensuremath{{\text{if } {{f} \text{ is } {\ensuremath{{\text{\texttt{\#inline}}}}}}}}\text,
      \\
      \ensuremath{{\ensuremath{{\mathsf{trigger}}}\mathchoice{\!}{\!}{}{}\left({\ensuremath{{\mathsf{Call}}}(f,\ensuremath{{m}},v)}\right)}} & \ensuremath{{\text{otherwise}}}\text,
    \end{cases}
  \\
  \ensuremath{{
  \llparenthesis\hspace{0.1em}c\hspace{0.1em}\rrparenthesis ^{\ensuremath{\text{\tiny{\ensuremath{I}}}}}_{\st}}} &\triangleq
    \ensuremath{{\ensuremath{{\mathsf{interp}}}\mathchoice{\!}{\!}{}{}\left({\ensuremath{{\ensuremath{{H}}^I}}}, {\ensuremath{{
  \llparenthesis\hspace{0.1em}c\hspace{0.1em}\rrparenthesis ^{\ensuremath{\text{\tiny{\ensuremath{\star}}}}}_{\st}}}}\right)}}\text.
\end{align*}}
This handler interprets a call to an \ensuremath{{\text{\texttt{\#inline}}}}{} function~\(f\) by
executing its body using the usual call handler~\ensuremath{{\ensuremath{{H}}^{\ensuremath{\text{\tiny{\ensuremath{\hspace{0.1em}\mathsf{c}}}}}}}}{}, and
other calls by re-triggering the original event.
Based on this handler, the one-step inlining semantics interprets calls
to \ensuremath{{\text{\texttt{\#inline}}}}{} functions as ITrees that correspond to their bodies
instead of \ensuremath{{\mathsf{Call}}}{}~events.
The bodies of \ensuremath{{\text{\texttt{\#inline}}}}{} functions may themselves contain function
calls, and these are not replaced, hence the name ``one-step'' inlining
semantics.

Armed with this new semantics, we use existing lemmas from the ITree
library that show that the interprocedural semantics of the source
program (which interprets \emph{every} \ensuremath{{\mathsf{Call}}}{}~event) is equivalent to
interpreting all \ensuremath{{\mathsf{Call}}}{}~events \emph{after} the one-step inlining
semantics.
That is, we have
\ensuremath{{\forall {c, \st}.\, {
  \ensuremath{{{
} \mathrel{\euttname[_{}]} { }}}   {\ensuremath{{
  \llparenthesis\hspace{0.1em}c\hspace{0.1em}\rrparenthesis ^{}_{\st}}}}
    {\ensuremath{{\ensuremath{{\mathsf{interp\_mrec}}}\mathchoice{\!}{\!}{}{}\left({\ensuremath{{\ensuremath{{H}}^{\ensuremath{\text{\tiny{\ensuremath{\hspace{0.1em}\mathsf{c}}}}}}}}}, {\ensuremath{{
  \llparenthesis\hspace{0.1em}c\hspace{0.1em}\rrparenthesis ^{\ensuremath{\text{\tiny{\ensuremath{I}}}}}_{\st}}}}\right)}}}
}}}.
Intuitively, this holds because both sides handle events the same way,
but the right-hand side handles some of them ``earlier''
with~\ensuremath{{
  \llparenthesis\hspace{0.1em}\cdot\hspace{0.1em}\rrparenthesis ^{\ensuremath{\text{\tiny{\ensuremath{I}}}}}_{}}}{}.
\looseness=-1

Finally, we use RHL to prove that the one-step inlining semantics of the
source is equivalent to the semantics of the target.
That is, for every source command~\(c\) whose compilation is
\ensuremath{{\bar{c}}} and state~\st{},
\ensuremath{{
  \ensuremath{{{
} \mathrel{\raisebox{-1pt}{\ensuremath{{\overset{\ensuremath{\text{\tiny{\ensuremath{u}}}}}{\approx}}}}_{}} { }}}   {\ensuremath{{\ensuremath{{\mathsf{interp\_mrec}}}\mathchoice{\!}{\!}{}{}\left({\ensuremath{{\ensuremath{{H}}^{\ensuremath{\text{\tiny{\ensuremath{\hspace{0.1em}\mathsf{c}}}}}}}}}, {\ensuremath{{
  \llparenthesis\hspace{0.1em}c\hspace{0.1em}\rrparenthesis ^{\ensuremath{\text{\tiny{\ensuremath{I}}}}}_{\st}}}}\right)}}}
    {\ensuremath{{
  \llparenthesis\hspace{0.1em}\ensuremath{{\bar{c}}}\hspace{0.1em}\rrparenthesis ^{}_{\st}}}}
}}.
Critically, the left-hand side uses the one-step inlining
semantics~\ensuremath{{
  \llparenthesis\hspace{0.1em}\cdot\hspace{0.1em}\rrparenthesis ^{\ensuremath{\text{\tiny{\ensuremath{I}}}}}_{}}}{} while the right-hand side uses the
call{} semantics~\ensuremath{{
  \llparenthesis\hspace{0.1em}\cdot\hspace{0.1em}\rrparenthesis ^{\ensuremath{\text{\tiny{\ensuremath{\star}}}}}_{}}}{} (recall the definition
of~\ensuremath{{
  \llparenthesis\hspace{0.1em}\cdot\hspace{0.1em}\rrparenthesis ^{}_{}}}{} in \cref{fig:sem-ER}{}).
We can prove this statement only because we generalize the logic and
\cref{thm:soundness} to allow different semantics on the left- and
right-hand sides.
\looseness=-1

To verify passes like Inlining that use different semantics, our \textsc{Rocq}{}
formalization of the semantics and of RHL is substantially more general
than what was presented in \cref{sec:rhl}, \emph{being parametric on the
semantics of either side}.
This generalization allows us to verify the front-end with a single
implementation of RHL, and is essential in the proof of the Inlining
pass.
Besides Inlining, this generalization is needed for passes with
different input and output languages (i.e., Int to Word, Reg.\ Array
Expansion, Stack Allocation, and One Varmap).

\subsection{Linearization and Mixed-Step Semantics}\label{sec:verification:linearization}

The most substantial proof in the compiler back-end is linearization.
This pass transforms programs with structured control-flow into ones
with jumps.
For example, \ensuremath{{\ensuremath{{\text{\textbf{while}}}}~({e})~\{c\}}} transforms into \ensuremath{{
  \ensuremath{{\ensuremath{{\text{\textbf{jmp}}}}\,{\ell}}}\mathop{;} \ell'{:}\ensuremath{{\bar{c}}}\mathop{;} \ell{:}\ensuremath{{\ensuremath{{\text{\textbf{if}}}}\,{e}\,\ensuremath{{\text{\textbf{jmp}}}}\,{\ell'}}}
}}, where \ensuremath{{\bar{c}}}~is the compilation of~\(c\).
The output language of linearization is Linear, an unstructured language
with labels, jumps, conditional jumps, and calls.
Linear states augment source states with a function name and a program
counter, which are used to index the linear program.
The semantics of Linear is the repeated application of a transition
function on states.
\looseness=-1

\paragraph*{Key Feature: Mixed-Step Semantics}
An essential part of the proof of linearization entails synchronizing
function calls in the source program with those in the target.
With the previous big-step semantics, this was part of the (semantics)
induction hypothesis.
However, with the new denotational semantics, we need a means of
reasoning about small-step semantics in a function-modular way: the
proof of Linearization is greatly simplified when reasoning in a
semantics that treats calls as uninterpreted events.

To this end, we define an auxiliary semantics, the \emph{mixed-step
semantics}, that behaves like the small-step one but triggers
\ensuremath{{\mathsf{Call}}}{}~events for function calls and later interprets them analogously
to the source.
This allows us to prove the correctness of the Linearization pass at
the uninterpreted layer of both semantics, treating function calls
abstractly.
Afterward, we show the equivalence between the mixed-step semantics and
the small-step one as an abstract ITree equivalence between the
\ensuremath{{\mathsf{iter}}}{}~and~\ensuremath{{\mathsf{interp\_mrec}}}{} operators.
 \section{Related Work}\label{sec:related}
We structure this section along the different lines of related work.

\paragraph*{Program Semantics}
There is a long line of work that formalizes semantics for
nonterminating and probabilistic computations in proof assistants.
Coinductive program semantics and definitions of recursive functions are
mechanized in~\cite{DBLP:conf/esop/Leroy06,DBLP:conf/tphol/NakataU09,DBLP:conf/mpc/McBride15}.
Semantics of probabilistic programs
are mechanized in~\cite{DBLP:journals/entcs/HurdMM05} (discrete
setting),~\cite{DBLP:conf/cpp/AffeldtCS23,DBLP:conf/itp/HirataM023}
(continuous setting), and~\cite{DBLP:journals/pacmpl/GregersenAHTB24}
(tape-based). We refer
the reader to~\cite{DBLP:books/cu/20/BKS2020} for further information
about (mechanized and non-mechanized) approaches to semantics of
probabilistic programs.
\looseness=-1

\paragraph*{Verified Compilers}
There is a long history of using proof assistants for proving
functional correctness and for security properties of compilers;
see~\cite{handbook:pa}.
We now focus on three prominent examples.

First, CompCert~\cite{2006-Leroy-compcert} is a compiler for C, written
and verified in \textsc{Rocq}{}. Early versions of CompCert used a big-step
semantics and later ones a small-step coinductive semantics. All proofs
are based on simulation diagrams.
Secondly, CakeML~\cite{DBLP:conf/popl/KumarMNO14} is a compiler for
Standard ML, written and verified in HOL\@. The semantics is defined in
a functional big-step style with clocks to deal with divergence, and
proofs are based on simulation diagrams and equational reasoning.
Pancake~\cite{DBLP:journals/corr/abs-2501-08249} is an offspring of
CakeML
that uses an ITree semantics, proven equivalent to the
functional big-step one, to reason about the
interactions between programs and the external world.
Thirdly, Vellvm~\cite{DBLP:conf/popl/ZhaoNMZ12,DBLP:journals/pacmpl/ZakowskiBYZZZ21}
is a compiler for the LLVM framework written and verified in \textsc{Rocq}{}.
The latest version of Vellvm uses ITrees to formalize the semantics of
LLVM-IR\@.
Vellvm uses the Static Single Assignment (SSA) representation
and leverages SSA-specific proof techniques for proving correctness of
program transformations.
\looseness=-1

None of these compilers consider probabilistic semantics.
Our approach could in principle be applied to lift their guarantees to
the probabilistic setting.

\paragraph*{Verified Compilers for Probabilistic Languages}
ProbCompCert~\cite{DBLP:journals/pacmpl/TassarottiT23} is a formally
verified compiler for a fragment of the Stan
language~\cite{DBLP:conf/pldi/BaudartBHMS21}. The
main component of ProbCompCert is a verified density compilation
pipeline performed before compiling with the CompCert backend.
Critically, this compilation pipeline is correct for a probabilistic
semantics---not for the usual deterministic semantics. It would be
interesting to consider whether one could enrich our relational Hoare
logic with probabilistic reasoning rules in the style of probabilistic
relational Hoare logic to validate this pipeline.
\looseness=-1

Zar~\cite{DBLP:journals/pacmpl/Bagnall0023} is a formally verified
compiler from discrete probabilistic programs (with potentially
unbounded loops) to an ITree-based representation. It uses extraction
to generate samplers in OCaml and Python. It would be interesting to
use our framework to generate efficient, formally verified, samplers
written in assembly.

\paragraph*{Secure Compilation}
While the aforementioned works focus on classic compiler correctness
in terms of execution traces, stronger
notions have also been considered. One such notion is
robust compilation~\cite{DBLP:conf/esop/AbateBCD0HPTT20,
  DBLP:journals/toplas/PatrignaniKWC24}, which asks
that program behaviors are preserved under adversarial contexts. A main
difference between robust compilation and our work is that it allows
for source and target contexts to differ, so that the latter are able
to capture low-level attacks. We plan to address low-level attacks in
a separate (but compatible) effort, leveraging recent developments in
hardware-software contracts~\cite{DBLP:conf/sp/GuarnieriKRV21}. A
different line of work studies preservation of information flow
properties: \citet{DBLP:conf/ecoop/SilverHCHZ23}
use the framework of ITrees to prove preservation of
noninterference for an \textsc{Imp}-to-\textsc{Asm} compiler.
\citet{DBLP:journals/pacmpl/BartheBGHLPT20,DBLP:conf/csfw/BartheGL18,DBLP:conf/ccs/BartheGLP21}
prove preservation of constant-time in CompCert, a core imperative
language, and an older version of \textsc{Jasmin}{}, respectively---the proof of
the latter is not maintained.
Constant-time (CT) has also been considered in the context of nondeterminism
(e.g., proving that the Bedrock2 compiler preserves CT under
nondeterminism~\cite{DBLP:journals/pacmpl/ConolyEC25}) and speculation
(e.g.,~\cite{DBLP:journals/pacmpl/WallM25,DBLP:journals/pacmpl/FabianPGB25,DBLP:journals/pacmpl/OlmosBBGL25}, which study or prove speculative
CT for simplified, core compilers).

\paragraph*{(Relational) Programs Logics}
Relational program logics provide a general framework for reasoning
about properties of two programs (i.e., 2-properties). Relational
Hoare logic~\cite{DBLP:conf/popl/Benton04} is a program logic
for reasoning about executions of two programs or two executions of
the same program. Relational separation logic
(RSL)~\cite{DBLP:journals/tcs/Yang07} is a program logic for reasoning
about executions of two heap-manipulating programs; it was developed
concurrently with RHL\@. Iris~\cite{DBLP:conf/popl/JungSSSTBD15} is a
mechanized separation logic formalized in the \textsc{Rocq}{} prover,
with several extensions that support relational reasoning.
One of the primary applications of relational program logics is
compiler correctness; for instance, Iris has been used to prove
correctness of
program transformations~\cite{DBLP:journals/pacmpl/GaherSSJDKKD22}.
Recently, RSL has been used to prove correctness of (an idealization
of) the tail-modulo constructor optimization used by the OCaml
compiler~\cite{DBLP:journals/pacmpl/AllainBCPS25}.

Probabilistic RHL (pRHL)~\cite{DBLP:conf/popl/BartheGB09} is a
relational program logic for probabilistic programs leveraging
\emph{proofs by coupling}.
The logic presented in \cref{sec:rhl} treats random sampling abstractly,
i.e., the proof shows that they are preserved \emph{exactly}.
This drastically simplifies reasoning about transformations and suffices
for all passes in \textsc{Jasmin}{}.
In contrast, pRHL supports sophisticated probabilistic reasoning that
could be used to prove sampling-related optimizations (such as those in
ProbCompCert and Zar).
The logic presented in this paper could admit pRHL rules, when
instantiated with the probabilistic semantics~\ensuremath{{\left\llbracket\cdot\right\rrbracket}}{}.
It would be interesting to explore such extensions to reason about
optimizations~\cite{DBLP:journals/pacmpl/TassarottiT23,DBLP:journals/pacmpl/Bagnall0023,DBLP:journals/pacmpl/LiWZ24} and
expectations~\cite{DBLP:journals/pacmpl/AvanziniBDG25}.
\looseness=-1

Several works develop program logics and related formalisms
on top of ITrees. \citet{Vistrup25} develop a
modular method for defining program logics within Iris,
relying on a generic definition of weakest precondition for ITrees,
alongside effect libraries that can be combined together to deal with
complex languages. \citet{DBLP:journals/pacmpl/SilverZ21} formalize
Dijkstra monads for weakest precondition reasoning with ITrees.
\citet{Silver23} formalize a refinement calculus on
top of ITrees.  Finally, \citet{Michelland24}
develop formally verified abstract interpreters using ITrees.

\paragraph*{Computer-Aided Cryptography}
There is a large body of work that formalizes cryptographic security
in proof assistants~\cite{DBLP:conf/sp/BarbosaBBBCLP21}. In
particular, the \textsc{EasyCrypt}{} proof
assistant~\cite{DBLP:conf/crypto/BartheGHB11,DBLP:conf/fosad/BartheDGKSS13}
has been used to mechanize the security proof of many constructions,
including ML-KEM~\cite{DBLP:conf/crypto/AlmeidaOBBDGLLLOPQSS24}.
\textsc{EasyCrypt}{} is based on pRHL (probabilistic relational Hoare
logic)~\cite{DBLP:conf/popl/BartheGB09}. Other mechanized
frameworks for cryptographic security include
FCF~\cite{DBLP:conf/post/PetcherM15},
CryptHOL~\cite{DBLP:journals/joc/BasinLS20},
SSProve~\cite{DBLP:journals/toplas/HaselwarterRMWASHMS23}, and
IPDL~\cite{DBLP:journals/pacmpl/GancherSFSM23}. Some works are more
specifically focused on formalizing oracle systems. For
instance,~\cite{DBLP:journals/iacr/TumaH24} presents a formalization of
oracle systems in Lean. Their formalization has been further extended
in~\cite{DBLP:journals/iacr/DziembowskiFMS25} to reason about
computational soundness.

\citet{DBLP:conf/fse/AlmeidaBBD16} present a general method for proving
preservation of cryptographic security against adversaries with access
to side-channel
leakage. However, their proof is carried on pen-and-paper, at a high
level of abstraction, and elides many of the important details
addressed by our work.
We discuss side-channels in the context of our work in the conclusion.

Low*~\cite{DBLP:journals/pacmpl/ProtzenkoZRRWBD17} is a low-level
language for writing efficient cryptographic code and verifying it.
It has been used to implement cryptographic schemes, verifying for some
of them a combination of functional correctness, protocol security,
memory safety, and constant-time.
It can be compiled to CompCert's Clight (this compilation is verified,
including constant-time preservation) or to C (which is not verified).
Low* is the language that implements the
HACL*~\cite{DBLP:conf/ccs/PolubelovaBPBFK20} cryptographic library,
which in turn is used by the cryptographic provider
EverCrypt~\cite{DBLP:conf/sp/ProtzenkoPFHPBB20}.
Our work is complementary to these efforts:
\cref{thm:eutt-semp,thm:compiler-correctness} take source-level (like
those of Low* and HACL*) cryptographic guarantees down to assembly
level.
Combining the approaches, which would involve a Low*-to-\textsc{Jasmin}{}
back-end that preserves game-based security, is interesting but further
requires that source-level proofs are done in the standard computational
model with a probabilistic semantics, rather than its idealized,
symbolic, deterministic semantics.

\citet{DBLP:journals/pacmpl/ErbsenPJLGPC24} present the whole-system
verification of a cryptographic server, including RISC-V machine code,
network packets, and functional specification of X25519.
It uses omnisemantics as a unifying semantics
framework~\cite{DBLP:journals/toplas/ChargueraudCEG23},
Bedrock2~\cite{DBLP:conf/pldi/ErbsenGCWC21} to implement, compile, and
verify many software components; Fiat
Cryptography~\cite{DBLP:conf/sp/ErbsenPGSC19} to generate verified, fast
finite-field arithmetic; and
Rupicola~\cite{DBLP:conf/pldi/Pit-ClaudelPJEC22} to compile functional
programs to correct Bedrock2 ones; everything is combined using \textsc{Rocq}{}
as a foundation.
The end result is functional correctness and safety of the entire
cryptographic server: it receives network packets, authenticates them,
and flips a bit in the state.
Both \citet{DBLP:journals/pacmpl/ErbsenPJLGPC24} and our work aim to
bring (orthogonal) guarantees down to assembly level: the focus of the
former is on whole-system functional correctness, while ours is on
preserving game-based probabilistic security.
We discuss extending our work to system-level guarantees in
\cref{sec:introduction}.

 \section{Conclusion and Future Work}\label{sec:conclusion}

This paper lays the foundations for carrying probabilistic correctness
and game-based security guarantees from source \textsc{Jasmin}{} programs
to assembly programs generated by its compiler.
To do so, we mechanize a framework for game-based security, introduce a
new semantics for \textsc{Jasmin}{} and assembly programs, as well as a
relational Hoare logic that allows applying the framework.

In independent future work, we plan to develop an assembly type system
for speculative constant-time (SCT), proving that the type system
enforces SCT, and that SCT entails system-level security under some
standard assumptions in hardware. This effort, which builds on prior
work~\cite{DBLP:conf/ccs/BartheBCLP14,song2025,DBLP:conf/sp/BartheCGKLOPRS21,DBLP:conf/sp/ShivakumarBGLOPST23},
is almost entirely independent of the current work and from the
compiler proof---it only requires proving that the \textsc{Jasmin}{} compiler
preserves declassification, which is very direct. However, it can be
combined with the present paper to show that provable security at source
level and side-channel security at assembly level guarantee system-level
security against powerful attackers that can observe program leakage
and have control over cache and branch predictors.

Another important direction for future work is to apply our
methodology to other post-quantum cryptographic standards, including
ML-DSA and SLH-DSA\@. There are ongoing efforts to prove functional
correctness and game-based security of \textsc{Jasmin}{} implementations of
ML-DSA and SLH-DSA, building on machine-checked security proofs
from~\cite{DBLP:conf/crypto/BarbosaBDDFGHHLW23,DBLP:conf/crypto/BarbosaDGHMS22}.
Our framework will naturally allow these guarantees to be transferred
to assembly programs, once the source level proofs are completed.

 \begin{acks}
We thank the reviewers for their time and insightful feedback.
We are very grateful for the help of Yannick Zakowski and Li-yao Xia,
which was instrumental in the proof of Inlining.
Likewise, we are greatly thankful to Pierre-Yves Strub and Jean-Christophe
Léchenet, who generously helped us with the formalization of the theorems about
probability distributions and Stack Allocation, respectively.
This research was supported
by the
\grantsponsor{DFG}{
  \textit{Deutsche Forschungsgemeinschaft} (DFG, German Research Foundation)}{}
as part of the Excellence Strategy of the German Federal and State Governments
-- \grantnum{DFG}{EXC 2092 CASA - 390781972}; and by the
\grantsponsor{ANR
}{
  \textit{Agence Nationale de la Recherche} (French National Research Agency)
}{}
as part of the France 2030 programme -- \grantnum{ANR}{ANR-22-PECY-0006}.
\end{acks}
 
\bibliographystyle{ACM-Reference-Format}
\bibliography{references.bib}


\begin{thebibliography}{86}


\ifx \showCODEN    \undefined \def \showCODEN     #1{\unskip}     \fi
\ifx \showDOI      \undefined \def \showDOI       #1{#1}\fi
\ifx \showISBNx    \undefined \def \showISBNx     #1{\unskip}     \fi
\ifx \showISBNxiii \undefined \def \showISBNxiii  #1{\unskip}     \fi
\ifx \showISSN     \undefined \def \showISSN      #1{\unskip}     \fi
\ifx \showLCCN     \undefined \def \showLCCN      #1{\unskip}     \fi
\ifx \shownote     \undefined \def \shownote      #1{#1}          \fi
\ifx \showarticletitle \undefined \def \showarticletitle #1{#1}   \fi
\ifx \showURL      \undefined \def \showURL       {\relax}        \fi
\providecommand\bibfield[2]{#2}
\providecommand\bibinfo[2]{#2}
\providecommand\natexlab[1]{#1}
\providecommand\showeprint[2][]{arXiv:#2}

\bibitem[Abate et~al\mbox{.}(2020)]%
        {DBLP:conf/esop/AbateBCD0HPTT20}
\bibfield{author}{\bibinfo{person}{Carmine Abate}, \bibinfo{person}{Roberto
  Blanco}, \bibinfo{person}{{\c{S}}tefan Ciob{\^{a}}c{\u{a}}},
  \bibinfo{person}{Adrien Durier}, \bibinfo{person}{Deepak Garg},
  \bibinfo{person}{Catalin Hritcu}, \bibinfo{person}{Marco Patrignani},
  \bibinfo{person}{{\'{E}}ric Tanter}, {and}
  \bibinfo{person}{J{\'{e}}r{\'{e}}my Thibault}.}
  \bibinfo{year}{2020}\natexlab{}.
\newblock \showarticletitle{Trace-Relating Compiler Correctness and Secure
  Compilation}. In \bibinfo{booktitle}{\emph{Programming Languages and Systems
  - 29th European Symposium on Programming, {ESOP} 2020, Held as Part of the
  European Joint Conferences on Theory and Practice of Software, {ETAPS} 2020,
  Dublin, Ireland, April 25-30, 2020, Proceedings}}
  \emph{(\bibinfo{series}{Lecture Notes in Computer Science},
  Vol.~\bibinfo{volume}{12075})}, \bibfield{editor}{\bibinfo{person}{Peter
  M{\"{u}}ller}} (Ed.). \bibinfo{publisher}{Springer}, \bibinfo{pages}{1--28}.
\newblock
\urldef\tempurl%
\url{https://doi.org/10.1007/978-3-030-44914-8\_1}
\showDOI{\tempurl}


\bibitem[Adkins and Schmieg(2026)]%
        {google-pqc}
\bibfield{author}{\bibinfo{person}{Heather Adkins} {and}
  \bibinfo{person}{Sophie Schmieg}.} \bibinfo{year}{2026}\natexlab{}.
\newblock \bibinfo{booktitle}{\emph{Quantum frontiers may be closer than they
  appear}}.
\newblock
\urldef\tempurl%
\url{https://blog.google/innovation-and-ai/technology/safety-security/cryptography-migration-timeline/}
\showURL{%
\tempurl}
\newblock
\shownote{The Keyword (Google blog)}.


\bibitem[Affeldt et~al\mbox{.}(2023)]%
        {DBLP:conf/cpp/AffeldtCS23}
\bibfield{author}{\bibinfo{person}{Reynald Affeldt}, \bibinfo{person}{Cyril
  Cohen}, {and} \bibinfo{person}{Ayumu Saito}.}
  \bibinfo{year}{2023}\natexlab{}.
\newblock \showarticletitle{Semantics of Probabilistic Programs using s-Finite
  Kernels in Coq}. In \bibinfo{booktitle}{\emph{Proceedings of the 12th {ACM}
  {SIGPLAN} International Conference on Certified Programs and Proofs, {CPP}
  2023, Boston, MA, USA, January 16-17, 2023}},
  \bibfield{editor}{\bibinfo{person}{Robbert Krebbers},
  \bibinfo{person}{Dmitriy Traytel}, \bibinfo{person}{Brigitte Pientka}, {and}
  \bibinfo{person}{Steve Zdancewic}} (Eds.). \bibinfo{publisher}{{ACM}},
  \bibinfo{pages}{3--16}.
\newblock
\urldef\tempurl%
\url{https://doi.org/10.1145/3573105.3575691}
\showDOI{\tempurl}


\bibitem[Allain et~al\mbox{.}(2025)]%
        {DBLP:journals/pacmpl/AllainBCPS25}
\bibfield{author}{\bibinfo{person}{Cl{\'{e}}ment Allain},
  \bibinfo{person}{Fr{\'{e}}d{\'{e}}ric Bour}, \bibinfo{person}{Basile
  Cl{\'{e}}ment}, \bibinfo{person}{Fran{\c{c}}ois Pottier}, {and}
  \bibinfo{person}{Gabriel Scherer}.} \bibinfo{year}{2025}\natexlab{}.
\newblock \showarticletitle{Tail Modulo Cons, OCaml, and Relational Separation
  Logic}.
\newblock \bibinfo{journal}{\emph{Proc. {ACM} Program. Lang.}}
  \bibinfo{volume}{9}, \bibinfo{number}{{POPL}} (\bibinfo{year}{2025}),
  \bibinfo{pages}{2337--2363}.
\newblock
\urldef\tempurl%
\url{https://doi.org/10.1145/3704915}
\showDOI{\tempurl}


\bibitem[Almeida et~al\mbox{.}(2026)]%
        {almeida2026formosa}
\bibfield{author}{\bibinfo{person}{Jos{\'e}~Bacelar Almeida},
  \bibinfo{person}{Gustavo Xavier Delerue~Marinho Alves},
  \bibinfo{person}{Santiago Arranz-Olmos}, \bibinfo{person}{Manuel Barbosa},
  \bibinfo{person}{Francisca Barros}, \bibinfo{person}{Gilles Barthe},
  \bibinfo{person}{Lionel Blatter}, \bibinfo{person}{Chitchanok
  Chuengsatiansup}, \bibinfo{person}{Ignacio Cuevas},
  \bibinfo{person}{Fran{\c{c}}ois Dupressoir}, \bibinfo{person}{Lu{\'\i}s
  Esqu{\'\i}vel}, \bibinfo{person}{Benjamin Gr{\'e}goire},
  \bibinfo{person}{Ruben Gonzalez}, \bibinfo{person}{Jan Jancar},
  \bibinfo{person}{Vincent Hwang}, \bibinfo{person}{Vincent Laporte},
  \bibinfo{person}{Jean-Christophe L{\'e}chenet}, \bibinfo{person}{Ting han
  Lim}, \bibinfo{person}{Cameron Low}, \bibinfo{person}{Tiago Oliveira},
  \bibinfo{person}{Hugo Pacheco}, \bibinfo{person}{Swarn Priya},
  \bibinfo{person}{Miguel Quaresma}, \bibinfo{person}{Rolfe Schmidt},
  \bibinfo{person}{Peter Schwabe}, \bibinfo{person}{Antoine S{\'e}r{\'e}},
  \bibinfo{person}{Basavesh~Ammanaghatta Shivakumar},
  \bibinfo{person}{Pierre-Yves Strub}, \bibinfo{person}{Lucas Tabary-Maujean},
  \bibinfo{person}{Yuval Yarom}, \bibinfo{person}{Zhiyuan Zhang}, {and}
  \bibinfo{person}{Jieyu Zheng}.} \bibinfo{year}{2026}\natexlab{}.
\newblock \bibinfo{booktitle}{\emph{Formosa Crypto: End-to-end formally
  verified crypto software}}.
\newblock Taipei, Taiwan.
\newblock
\urldef\tempurl%
\url{https://realworldcrypto.iacr.org/2026/acceptedtalks.php}
\showURL{%
\tempurl}


\bibitem[Almeida et~al\mbox{.}(2024)]%
        {DBLP:conf/crypto/AlmeidaOBBDGLLLOPQSS24}
\bibfield{author}{\bibinfo{person}{Jos{\'{e}}~Bacelar Almeida},
  \bibinfo{person}{Santiago Arranz-Olmos}, \bibinfo{person}{Manuel Barbosa},
  \bibinfo{person}{Gilles Barthe}, \bibinfo{person}{Fran{\c{c}}ois Dupressoir},
  \bibinfo{person}{Benjamin Gr{\'{e}}goire}, \bibinfo{person}{Vincent Laporte},
  \bibinfo{person}{Jean{-}Christophe L{\'{e}}chenet}, \bibinfo{person}{Cameron
  Low}, \bibinfo{person}{Tiago Oliveira}, \bibinfo{person}{Hugo Pacheco},
  \bibinfo{person}{Miguel Quaresma}, \bibinfo{person}{Peter Schwabe}, {and}
  \bibinfo{person}{Pierre{-}Yves Strub}.} \bibinfo{year}{2024}\natexlab{}.
\newblock \showarticletitle{Formally Verifying Kyber - Episode {V:}
  Machine-Checked {IND-CCA} Security and Correctness of {ML-KEM} in EasyCrypt}.
  In \bibinfo{booktitle}{\emph{Advances in Cryptology - {CRYPTO} 2024 - 44th
  Annual International Cryptology Conference, Santa Barbara, CA, USA, August
  18-22, 2024, Proceedings, Part {II}}} \emph{(\bibinfo{series}{Lecture Notes
  in Computer Science}, Vol.~\bibinfo{volume}{14921})},
  \bibfield{editor}{\bibinfo{person}{Leonid Reyzin} {and}
  \bibinfo{person}{Douglas Stebila}} (Eds.). \bibinfo{publisher}{Springer},
  \bibinfo{pages}{384--421}.
\newblock
\urldef\tempurl%
\url{https://doi.org/10.1007/978-3-031-68379-4\_12}
\showDOI{\tempurl}


\bibitem[Almeida et~al\mbox{.}(2017)]%
        {DBLP:conf/ccs/AlmeidaBBBGLOPS17}
\bibfield{author}{\bibinfo{person}{Jos{\'{e}}~Bacelar Almeida},
  \bibinfo{person}{Manuel Barbosa}, \bibinfo{person}{Gilles Barthe},
  \bibinfo{person}{Arthur Blot}, \bibinfo{person}{Benjamin Gr{\'{e}}goire},
  \bibinfo{person}{Vincent Laporte}, \bibinfo{person}{Tiago Oliveira},
  \bibinfo{person}{Hugo Pacheco}, \bibinfo{person}{Benedikt Schmidt}, {and}
  \bibinfo{person}{Pierre{-}Yves Strub}.} \bibinfo{year}{2017}\natexlab{}.
\newblock \showarticletitle{Jasmin: High-Assurance and High-Speed
  Cryptography}. In \bibinfo{booktitle}{\emph{Proceedings of the 2017 {ACM}
  {SIGSAC} Conference on Computer and Communications Security, {CCS} 2017,
  Dallas, TX, USA, October 30 - November 03, 2017}},
  \bibfield{editor}{\bibinfo{person}{Bhavani Thuraisingham},
  \bibinfo{person}{David Evans}, \bibinfo{person}{Tal Malkin}, {and}
  \bibinfo{person}{Dongyan Xu}} (Eds.). \bibinfo{publisher}{{ACM}},
  \bibinfo{pages}{1807--1823}.
\newblock
\urldef\tempurl%
\url{https://doi.org/10.1145/3133956.3134078}
\showDOI{\tempurl}


\bibitem[Almeida et~al\mbox{.}(2016)]%
        {DBLP:conf/fse/AlmeidaBBD16}
\bibfield{author}{\bibinfo{person}{Jos{\'{e}}~Bacelar Almeida},
  \bibinfo{person}{Manuel Barbosa}, \bibinfo{person}{Gilles Barthe}, {and}
  \bibinfo{person}{Fran{\c{c}}ois Dupressoir}.}
  \bibinfo{year}{2016}\natexlab{}.
\newblock \showarticletitle{Verifiable Side-Channel Security of Cryptographic
  Implementations: Constant-Time {MEE-CBC}}. In \bibinfo{booktitle}{\emph{Fast
  Software Encryption - 23rd International Conference, {FSE} 2016, Bochum,
  Germany, March 20-23, 2016, Revised Selected Papers}}
  \emph{(\bibinfo{series}{Lecture Notes in Computer Science},
  Vol.~\bibinfo{volume}{9783})}, \bibfield{editor}{\bibinfo{person}{Thomas
  Peyrin}} (Ed.). \bibinfo{publisher}{Springer}, \bibinfo{pages}{163--184}.
\newblock
\urldef\tempurl%
\url{https://doi.org/10.1007/978-3-662-52993-5\_9}
\showDOI{\tempurl}


\bibitem[Almeida et~al\mbox{.}(2020)]%
        {DBLP:conf/sp/AlmeidaBBGKL0S20}
\bibfield{author}{\bibinfo{person}{Jos{\'{e}}~Bacelar Almeida},
  \bibinfo{person}{Manuel Barbosa}, \bibinfo{person}{Gilles Barthe},
  \bibinfo{person}{Benjamin Gr{\'{e}}goire}, \bibinfo{person}{Adrien Koutsos},
  \bibinfo{person}{Vincent Laporte}, \bibinfo{person}{Tiago Oliveira}, {and}
  \bibinfo{person}{Pierre{-}Yves Strub}.} \bibinfo{year}{2020}\natexlab{}.
\newblock \showarticletitle{The Last Mile: High-Assurance and High-Speed
  Cryptographic Implementations}. In \bibinfo{booktitle}{\emph{2020 {IEEE}
  Symposium on Security and Privacy, {SP} 2020, San Francisco, CA, USA, May
  18-21, 2020}}. \bibinfo{publisher}{{IEEE}}, \bibinfo{pages}{965--982}.
\newblock
\urldef\tempurl%
\url{https://doi.org/10.1109/SP40000.2020.00028}
\showDOI{\tempurl}


\bibitem[Almeida et~al\mbox{.}(2023)]%
        {DBLP:journals/tches/AlmeidaBBGLL00Q23}
\bibfield{author}{\bibinfo{person}{Jos{\'{e}}~Bacelar Almeida},
  \bibinfo{person}{Manuel Barbosa}, \bibinfo{person}{Gilles Barthe},
  \bibinfo{person}{Benjamin Gr{\'{e}}goire}, \bibinfo{person}{Vincent Laporte},
  \bibinfo{person}{Jean{-}Christophe L{\'{e}}chenet}, \bibinfo{person}{Tiago
  Oliveira}, \bibinfo{person}{Hugo Pacheco}, \bibinfo{person}{Miguel Quaresma},
  \bibinfo{person}{Peter Schwabe}, \bibinfo{person}{Antoine S{\'{e}}r{\'{e}}},
  {and} \bibinfo{person}{Pierre{-}Yves Strub}.}
  \bibinfo{year}{2023}\natexlab{}.
\newblock \showarticletitle{Formally verifying Kyber Episode {IV:}
  Implementation correctness}.
\newblock \bibinfo{journal}{\emph{{IACR} Trans. Cryptogr. Hardw. Embed. Syst.}}
  \bibinfo{volume}{2023}, \bibinfo{number}{3} (\bibinfo{year}{2023}),
  \bibinfo{pages}{164--193}.
\newblock
\urldef\tempurl%
\url{https://doi.org/10.46586/TCHES.V2023.I3.164-193}
\showDOI{\tempurl}


\bibitem[Apt et~al\mbox{.}(2009)]%
        {DBLP:series/txcs/AptBO09}
\bibfield{author}{\bibinfo{person}{{Krzysztof R.} Apt}, \bibinfo{person}{{Frank
  S.} {de Boer}}, {and} \bibinfo{person}{Ernst{-}R{\"{u}}diger Olderog}.}
  \bibinfo{year}{2009}\natexlab{}.
\newblock \bibinfo{booktitle}{\emph{Verification of Sequential and Concurrent
  Programs}}.
\newblock \bibinfo{publisher}{Springer}.
\newblock
\urldef\tempurl%
\url{https://doi.org/10.1007/978-1-84882-745-5}
\showDOI{\tempurl}


\bibitem[Arranz{-}Olmos et~al\mbox{.}(2025)]%
        {DBLP:journals/pacmpl/OlmosBBGL25}
\bibfield{author}{\bibinfo{person}{Santiago Arranz{-}Olmos},
  \bibinfo{person}{Gilles Barthe}, \bibinfo{person}{Lionel Blatter},
  \bibinfo{person}{Benjamin Gr{\'{e}}goire}, {and} \bibinfo{person}{Vincent
  Laporte}.} \bibinfo{year}{2025}\natexlab{}.
\newblock \showarticletitle{Preservation of Speculative Constant-Time by
  Compilation}.
\newblock \bibinfo{journal}{\emph{Proc. {ACM} Program. Lang.}}
  \bibinfo{volume}{9}, \bibinfo{number}{{POPL}} (\bibinfo{year}{2025}),
  \bibinfo{pages}{1293--1325}.
\newblock
\urldef\tempurl%
\url{https://doi.org/10.1145/3704880}
\showDOI{\tempurl}


\bibitem[{Arranz-Olmos} et~al\mbox{.}(2026)]%
        {public-artifact}
\bibfield{author}{\bibinfo{person}{Santiago {Arranz-Olmos}},
  \bibinfo{person}{Gilles Barthe}, \bibinfo{person}{Lionel Blatter},
  \bibinfo{person}{Benjamin Gr{\'{e}}goire}, \bibinfo{person}{Vincent Laporte},
  {and} \bibinfo{person}{Paolo Torrini}.} \bibinfo{year}{2026}\natexlab{}.
\newblock \bibinfo{booktitle}{\emph{{Artifact for "KEM-IND-CCA-Preserving
  Compilation of Jasmin's ML-KEM"}}}.
\newblock
\urldef\tempurl%
\url{https://doi.org/10.5281/zenodo.21968144}
\showDOI{\tempurl}


\bibitem[Avanzini et~al\mbox{.}(2025)]%
        {DBLP:journals/pacmpl/AvanziniBDG25}
\bibfield{author}{\bibinfo{person}{Martin Avanzini}, \bibinfo{person}{Gilles
  Barthe}, \bibinfo{person}{Davide Davoli}, {and} \bibinfo{person}{Benjamin
  Gr{\'{e}}goire}.} \bibinfo{year}{2025}\natexlab{}.
\newblock \showarticletitle{A Quantitative Probabilistic Relational Hoare
  Logic}.
\newblock \bibinfo{journal}{\emph{Proc. {ACM} Program. Lang.}}
  \bibinfo{volume}{9}, \bibinfo{number}{{POPL}} (\bibinfo{year}{2025}),
  \bibinfo{pages}{1167--1195}.
\newblock
\urldef\tempurl%
\url{https://doi.org/10.1145/3704876}
\showDOI{\tempurl}


\bibitem[Bagnall et~al\mbox{.}(2023)]%
        {DBLP:journals/pacmpl/Bagnall0023}
\bibfield{author}{\bibinfo{person}{Alexander Bagnall}, \bibinfo{person}{Gordon
  Stewart}, {and} \bibinfo{person}{Anindya Banerjee}.}
  \bibinfo{year}{2023}\natexlab{}.
\newblock \showarticletitle{Formally Verified Samplers from Probabilistic
  Programs with Loops and Conditioning}.
\newblock \bibinfo{journal}{\emph{Proc. {ACM} Program. Lang.}}
  \bibinfo{volume}{7}, \bibinfo{number}{{PLDI}} (\bibinfo{year}{2023}),
  \bibinfo{pages}{1--24}.
\newblock
\urldef\tempurl%
\url{https://doi.org/10.1145/3591220}
\showDOI{\tempurl}


\bibitem[Barbosa et~al\mbox{.}(2021)]%
        {DBLP:conf/sp/BarbosaBBBCLP21}
\bibfield{author}{\bibinfo{person}{Manuel Barbosa}, \bibinfo{person}{Gilles
  Barthe}, \bibinfo{person}{Karthik Bhargavan}, \bibinfo{person}{Bruno
  Blanchet}, \bibinfo{person}{Cas Cremers}, \bibinfo{person}{Kevin Liao}, {and}
  \bibinfo{person}{Bryan Parno}.} \bibinfo{year}{2021}\natexlab{}.
\newblock \showarticletitle{SoK: Computer-Aided Cryptography}. In
  \bibinfo{booktitle}{\emph{42nd {IEEE} Symposium on Security and Privacy, {SP}
  2021, San Francisco, CA, USA, 24-27 May 2021}}. \bibinfo{publisher}{{IEEE}},
  \bibinfo{pages}{777--795}.
\newblock
\urldef\tempurl%
\url{https://doi.org/10.1109/SP40001.2021.00008}
\showDOI{\tempurl}


\bibitem[Barbosa et~al\mbox{.}(2023a)]%
        {DBLP:conf/crypto/BarbosaBDDFGHHLW23}
\bibfield{author}{\bibinfo{person}{Manuel Barbosa}, \bibinfo{person}{Gilles
  Barthe}, \bibinfo{person}{Christian Doczkal}, \bibinfo{person}{Jelle Don},
  \bibinfo{person}{Serge Fehr}, \bibinfo{person}{Benjamin Gr{\'{e}}goire},
  \bibinfo{person}{Yu{-}Hsuan Huang}, \bibinfo{person}{Andreas H{\"{u}}lsing},
  \bibinfo{person}{Yi Lee}, {and} \bibinfo{person}{Xiaodi Wu}.}
  \bibinfo{year}{2023}\natexlab{a}.
\newblock \showarticletitle{Fixing and Mechanizing the Security Proof of
  Fiat-Shamir with Aborts and Dilithium}. In \bibinfo{booktitle}{\emph{Advances
  in Cryptology - {CRYPTO} 2023 - 43rd Annual International Cryptology
  Conference, {CRYPTO} 2023, Santa Barbara, CA, USA, August 20-24, 2023,
  Proceedings, Part {V}}} \emph{(\bibinfo{series}{Lecture Notes in Computer
  Science})}, \bibfield{editor}{\bibinfo{person}{Helena Handschuh} {and}
  \bibinfo{person}{Anna Lysyanskaya}} (Eds.). \bibinfo{publisher}{Springer},
  \bibinfo{pages}{358--389}.
\newblock
\urldef\tempurl%
\url{https://doi.org/10.1007/978-3-031-38554-4\_12}
\showDOI{\tempurl}


\bibitem[Barbosa et~al\mbox{.}(2023b)]%
        {DBLP:conf/crypto/BarbosaDGHMS22}
\bibfield{author}{\bibinfo{person}{Manuel Barbosa},
  \bibinfo{person}{Fran{\c{c}}ois Dupressoir}, \bibinfo{person}{Benjamin
  Gr{\'{e}}goire}, \bibinfo{person}{Andreas H{\"{u}}lsing},
  \bibinfo{person}{Matthias Meijers}, {and} \bibinfo{person}{Pierre{-}Yves
  Strub}.} \bibinfo{year}{2023}\natexlab{b}.
\newblock \showarticletitle{Machine-Checked Security for $\mathrm{XMSS}$ as in
  {RFC} 8391 and $\mathrm{SPHINCS}^{+}$}. In \bibinfo{booktitle}{\emph{Advances
  in Cryptology - {CRYPTO} 2023 - 43rd Annual International Cryptology
  Conference, {CRYPTO} 2023, Santa Barbara, CA, USA, August 20-24, 2023,
  Proceedings, Part {V}}} \emph{(\bibinfo{series}{Lecture Notes in Computer
  Science}, Vol.~\bibinfo{volume}{14085})},
  \bibfield{editor}{\bibinfo{person}{Helena Handschuh} {and}
  \bibinfo{person}{Anna Lysyanskaya}} (Eds.). \bibinfo{publisher}{Springer},
  \bibinfo{pages}{421--454}.
\newblock
\urldef\tempurl%
\url{https://doi.org/10.1007/978-3-031-38554-4\_14}
\showDOI{\tempurl}


\bibitem[Barthe(2026)]%
        {handbook:pa}
\bibfield{author}{\bibinfo{person}{Gilles Barthe}.}
  \bibinfo{year}{2026}\natexlab{}.
\newblock \showarticletitle{Formalization of security}. In
  \bibinfo{booktitle}{\emph{Proof Assistants and Their Applications in
  Mathematics and Computer Science}}, \bibfield{editor}{\bibinfo{person}{Jasmin
  Blanchette} {and} \bibinfo{person}{Assia Mahboubi}} (Eds.).
  \bibinfo{publisher}{Springer}.
\newblock


\bibitem[Barthe et~al\mbox{.}(2014)]%
        {DBLP:conf/ccs/BartheBCLP14}
\bibfield{author}{\bibinfo{person}{Gilles Barthe}, \bibinfo{person}{Gustavo
  Betarte}, \bibinfo{person}{Juan~Diego Campo}, \bibinfo{person}{Carlos~Daniel
  Luna}, {and} \bibinfo{person}{David Pichardie}.}
  \bibinfo{year}{2014}\natexlab{}.
\newblock \showarticletitle{System-level Non-interference for Constant-time
  Cryptography}. In \bibinfo{booktitle}{\emph{Proceedings of the 2014 {ACM}
  {SIGSAC} Conference on Computer and Communications Security, Scottsdale, AZ,
  USA, November 3-7, 2014}}, \bibfield{editor}{\bibinfo{person}{Gail{-}Joon
  Ahn}, \bibinfo{person}{Moti Yung}, {and} \bibinfo{person}{Ninghui Li}}
  (Eds.). \bibinfo{publisher}{{ACM}}, \bibinfo{pages}{1267--1279}.
\newblock
\urldef\tempurl%
\url{https://doi.org/10.1145/2660267.2660283}
\showDOI{\tempurl}


\bibitem[Barthe et~al\mbox{.}(2020a)]%
        {DBLP:journals/pacmpl/BartheBGHLPT20}
\bibfield{author}{\bibinfo{person}{Gilles Barthe}, \bibinfo{person}{Sandrine
  Blazy}, \bibinfo{person}{Benjamin Gr{\'{e}}goire},
  \bibinfo{person}{R{\'{e}}mi Hutin}, \bibinfo{person}{Vincent Laporte},
  \bibinfo{person}{David Pichardie}, {and} \bibinfo{person}{Alix Trieu}.}
  \bibinfo{year}{2020}\natexlab{a}.
\newblock \showarticletitle{Formal verification of a constant-time preserving
  {C} compiler}.
\newblock \bibinfo{journal}{\emph{Proc. {ACM} Program. Lang.}}
  \bibinfo{volume}{4}, \bibinfo{number}{{POPL}} (\bibinfo{year}{2020}),
  \bibinfo{pages}{7:1--7:30}.
\newblock
\urldef\tempurl%
\url{https://doi.org/10.1145/3371075}
\showDOI{\tempurl}


\bibitem[Barthe et~al\mbox{.}(2021a)]%
        {DBLP:conf/sp/BartheCGKLOPRS21}
\bibfield{author}{\bibinfo{person}{Gilles Barthe}, \bibinfo{person}{Sunjay
  Cauligi}, \bibinfo{person}{Benjamin Gr{\'{e}}goire}, \bibinfo{person}{Adrien
  Koutsos}, \bibinfo{person}{Kevin Liao}, \bibinfo{person}{Tiago Oliveira},
  \bibinfo{person}{Swarn Priya}, \bibinfo{person}{Tamara Rezk}, {and}
  \bibinfo{person}{Peter Schwabe}.} \bibinfo{year}{2021}\natexlab{a}.
\newblock \showarticletitle{High-Assurance Cryptography in the Spectre Era}. In
  \bibinfo{booktitle}{\emph{42nd {IEEE} Symposium on Security and Privacy, {SP}
  2021, San Francisco, CA, USA, 24-27 May 2021}}. \bibinfo{publisher}{{IEEE}},
  \bibinfo{pages}{1884--1901}.
\newblock
\urldef\tempurl%
\url{https://doi.org/10.1109/SP40001.2021.00046}
\showDOI{\tempurl}


\bibitem[Barthe et~al\mbox{.}(2010)]%
        {DBLP:conf/ccs/BartheDKL10}
\bibfield{author}{\bibinfo{person}{Gilles Barthe}, \bibinfo{person}{Marion
  Daubignard}, \bibinfo{person}{Bruce~M. Kapron}, {and}
  \bibinfo{person}{Yassine Lakhnech}.} \bibinfo{year}{2010}\natexlab{}.
\newblock \showarticletitle{Computational indistinguishability logic}. In
  \bibinfo{booktitle}{\emph{Proceedings of the 17th {ACM} Conference on
  Computer and Communications Security, {CCS} 2010, Chicago, Illinois, USA,
  October 4-8, 2010}}, \bibfield{editor}{\bibinfo{person}{Ehab Al{-}Shaer},
  \bibinfo{person}{Angelos~D. Keromytis}, {and} \bibinfo{person}{Vitaly
  Shmatikov}} (Eds.). \bibinfo{publisher}{{ACM}}, \bibinfo{pages}{375--386}.
\newblock
\urldef\tempurl%
\url{https://doi.org/10.1145/1866307.1866350}
\showDOI{\tempurl}


\bibitem[Barthe et~al\mbox{.}(2013)]%
        {DBLP:conf/fosad/BartheDGKSS13}
\bibfield{author}{\bibinfo{person}{Gilles Barthe},
  \bibinfo{person}{Fran{\c{c}}ois Dupressoir}, \bibinfo{person}{Benjamin
  Gr{\'{e}}goire}, \bibinfo{person}{C{\'{e}}sar Kunz},
  \bibinfo{person}{Benedikt Schmidt}, {and} \bibinfo{person}{Pierre{-}Yves
  Strub}.} \bibinfo{year}{2013}\natexlab{}.
\newblock \showarticletitle{EasyCrypt: {A} Tutorial}. In
  \bibinfo{booktitle}{\emph{Foundations of Security Analysis and Design {VII} -
  {FOSAD} 2012/2013 Tutorial Lectures}} \emph{(\bibinfo{series}{Lecture Notes
  in Computer Science}, Vol.~\bibinfo{volume}{8604})},
  \bibfield{editor}{\bibinfo{person}{Alessandro Aldini},
  \bibinfo{person}{Javier L{\'{o}}pez}, {and} \bibinfo{person}{Fabio
  Martinelli}} (Eds.). \bibinfo{publisher}{Springer},
  \bibinfo{pages}{146--166}.
\newblock
\urldef\tempurl%
\url{https://doi.org/10.1007/978-3-319-10082-1\_6}
\showDOI{\tempurl}


\bibitem[Barthe et~al\mbox{.}(2011)]%
        {DBLP:conf/crypto/BartheGHB11}
\bibfield{author}{\bibinfo{person}{Gilles Barthe}, \bibinfo{person}{Benjamin
  Gr{\'{e}}goire}, \bibinfo{person}{Sylvain Heraud}, {and}
  \bibinfo{person}{Santiago Zanella{-}B{\'{e}}guelin}.}
  \bibinfo{year}{2011}\natexlab{}.
\newblock \showarticletitle{Computer-Aided Security Proofs for the Working
  Cryptographer}. In \bibinfo{booktitle}{\emph{Advances in Cryptology -
  {CRYPTO} 2011 - 31st Annual Cryptology Conference, Santa Barbara, CA, USA,
  August 14-18, 2011. Proceedings}} \emph{(\bibinfo{series}{Lecture Notes in
  Computer Science}, Vol.~\bibinfo{volume}{6841})},
  \bibfield{editor}{\bibinfo{person}{Phillip Rogaway}} (Ed.).
  \bibinfo{publisher}{Springer}, \bibinfo{pages}{71--90}.
\newblock
\urldef\tempurl%
\url{https://doi.org/10.1007/978-3-642-22792-9\_5}
\showDOI{\tempurl}


\bibitem[Barthe et~al\mbox{.}(2018)]%
        {DBLP:conf/csfw/BartheGL18}
\bibfield{author}{\bibinfo{person}{Gilles Barthe}, \bibinfo{person}{Benjamin
  Gr{\'{e}}goire}, {and} \bibinfo{person}{Vincent Laporte}.}
  \bibinfo{year}{2018}\natexlab{}.
\newblock \showarticletitle{Secure Compilation of Side-Channel Countermeasures:
  The Case of Cryptographic "Constant-Time"}. In \bibinfo{booktitle}{\emph{31st
  {IEEE} Computer Security Foundations Symposium, {CSF} 2018, Oxford, United
  Kingdom, July 9-12, 2018}}. \bibinfo{publisher}{{IEEE} Computer Society},
  \bibinfo{pages}{328--343}.
\newblock
\urldef\tempurl%
\url{https://doi.org/10.1109/CSF.2018.00031}
\showDOI{\tempurl}


\bibitem[Barthe et~al\mbox{.}(2021b)]%
        {DBLP:conf/ccs/BartheGLP21}
\bibfield{author}{\bibinfo{person}{Gilles Barthe}, \bibinfo{person}{Benjamin
  Gr{\'{e}}goire}, \bibinfo{person}{Vincent Laporte}, {and}
  \bibinfo{person}{Swarn Priya}.} \bibinfo{year}{2021}\natexlab{b}.
\newblock \showarticletitle{Structured Leakage and Applications to
  Cryptographic Constant-Time and Cost}. In \bibinfo{booktitle}{\emph{{CCS}
  '21: 2021 {ACM} {SIGSAC} Conference on Computer and Communications Security,
  Virtual Event, Republic of Korea, November 15 - 19, 2021}},
  \bibfield{editor}{\bibinfo{person}{Yongdae Kim}, \bibinfo{person}{Jong Kim},
  \bibinfo{person}{Giovanni Vigna}, {and} \bibinfo{person}{Elaine Shi}} (Eds.).
  \bibinfo{publisher}{{ACM}}, \bibinfo{pages}{462--476}.
\newblock
\urldef\tempurl%
\url{https://doi.org/10.1145/3460120.3484761}
\showDOI{\tempurl}


\bibitem[Barthe et~al\mbox{.}(2009)]%
        {DBLP:conf/popl/BartheGB09}
\bibfield{author}{\bibinfo{person}{Gilles Barthe}, \bibinfo{person}{Benjamin
  Gr{\'{e}}goire}, {and} \bibinfo{person}{Santiago Zanella{-}B{\'{e}}guelin}.}
  \bibinfo{year}{2009}\natexlab{}.
\newblock \showarticletitle{Formal certification of code-based cryptographic
  proofs}. In \bibinfo{booktitle}{\emph{Proceedings of the 36th {ACM}
  {SIGPLAN-SIGACT} Symposium on Principles of Programming Languages, {POPL}
  2009, Savannah, GA, USA, January 21-23, 2009}},
  \bibfield{editor}{\bibinfo{person}{Zhong Shao} {and}
  \bibinfo{person}{Benjamin~C. Pierce}} (Eds.). \bibinfo{publisher}{{ACM}},
  \bibinfo{pages}{90--101}.
\newblock
\urldef\tempurl%
\url{https://doi.org/10.1145/1480881.1480894}
\showDOI{\tempurl}


\bibitem[Barthe et~al\mbox{.}(2020b)]%
        {DBLP:books/cu/20/BKS2020}
\bibfield{editor}{\bibinfo{person}{Gilles Barthe},
  \bibinfo{person}{Joost{-}Pieter Katoen}, {and} \bibinfo{person}{Alexandra
  Silva}} (Eds.). \bibinfo{year}{2020}\natexlab{b}.
\newblock \bibinfo{booktitle}{\emph{Foundations of Probabilistic Programming}}.
\newblock \bibinfo{publisher}{Cambridge University Press}.
\newblock
\showISBNx{9781108770750}
\urldef\tempurl%
\url{https://doi.org/10.1017/9781108770750}
\showDOI{\tempurl}


\bibitem[Basin et~al\mbox{.}(2020)]%
        {DBLP:journals/joc/BasinLS20}
\bibfield{author}{\bibinfo{person}{David~A. Basin}, \bibinfo{person}{Andreas
  Lochbihler}, {and} \bibinfo{person}{S.~Reza Sefidgar}.}
  \bibinfo{year}{2020}\natexlab{}.
\newblock \showarticletitle{CryptHOL: Game-Based Proofs in Higher-Order Logic}.
\newblock \bibinfo{journal}{\emph{J. Cryptol.}} \bibinfo{volume}{33},
  \bibinfo{number}{2} (\bibinfo{year}{2020}), \bibinfo{pages}{494--566}.
\newblock
\urldef\tempurl%
\url{https://doi.org/10.1007/S00145-019-09341-Z}
\showDOI{\tempurl}


\bibitem[Baudart et~al\mbox{.}(2021)]%
        {DBLP:conf/pldi/BaudartBHMS21}
\bibfield{author}{\bibinfo{person}{Guillaume Baudart}, \bibinfo{person}{Javier
  Burroni}, \bibinfo{person}{Martin Hirzel}, \bibinfo{person}{Louis Mandel},
  {and} \bibinfo{person}{Avraham Shinnar}.} \bibinfo{year}{2021}\natexlab{}.
\newblock \showarticletitle{Compiling Stan to generative probabilistic
  languages and extension to deep probabilistic programming}. In
  \bibinfo{booktitle}{\emph{{PLDI} '21: 42nd {ACM} {SIGPLAN} International
  Conference on Programming Language Design and Implementation, Virtual Event,
  Canada, June 20-25, 2021}}, \bibfield{editor}{\bibinfo{person}{Stephen~N.
  Freund} {and} \bibinfo{person}{Eran Yahav}} (Eds.).
  \bibinfo{publisher}{{ACM}}, \bibinfo{pages}{497--510}.
\newblock
\urldef\tempurl%
\url{https://doi.org/10.1145/3453483.3454058}
\showDOI{\tempurl}


\bibitem[Beck et~al\mbox{.}(2024)]%
        {DBLP:journals/pacmpl/Beck0CZZ24}
\bibfield{author}{\bibinfo{person}{Calvin Beck}, \bibinfo{person}{Irene Yoon},
  \bibinfo{person}{Hanxi Chen}, \bibinfo{person}{Yannick Zakowski}, {and}
  \bibinfo{person}{Steve Zdancewic}.} \bibinfo{year}{2024}\natexlab{}.
\newblock \showarticletitle{A Two-Phase Infinite/Finite Low-Level Memory Model:
  Reconciling Integer-Pointer Casts, Finite Space, and undef at the {LLVM} {IR}
  Level of Abstraction}.
\newblock \bibinfo{journal}{\emph{Proc. {ACM} Program. Lang.}}
  \bibinfo{volume}{8}, \bibinfo{number}{{ICFP}} (\bibinfo{year}{2024}),
  \bibinfo{pages}{789--817}.
\newblock
\urldef\tempurl%
\url{https://doi.org/10.1145/3674652}
\showDOI{\tempurl}


\bibitem[Bellare and Rogaway(2006)]%
        {DBLP:conf/eurocrypt/BellareR06}
\bibfield{author}{\bibinfo{person}{Mihir Bellare} {and}
  \bibinfo{person}{Phillip Rogaway}.} \bibinfo{year}{2006}\natexlab{}.
\newblock \showarticletitle{The Security of Triple Encryption and a Framework
  for Code-Based Game-Playing Proofs}. In \bibinfo{booktitle}{\emph{Advances in
  Cryptology - {EUROCRYPT} 2006, 25th Annual International Conference on the
  Theory and Applications of Cryptographic Techniques, St. Petersburg, Russia,
  May 28 - June 1, 2006, Proceedings}} \emph{(\bibinfo{series}{Lecture Notes in
  Computer Science}, Vol.~\bibinfo{volume}{4004})},
  \bibfield{editor}{\bibinfo{person}{Serge Vaudenay}} (Ed.).
  \bibinfo{publisher}{Springer}, \bibinfo{pages}{409--426}.
\newblock
\urldef\tempurl%
\url{https://doi.org/10.1007/11761679\_25}
\showDOI{\tempurl}


\bibitem[Benton(2004)]%
        {DBLP:conf/popl/Benton04}
\bibfield{author}{\bibinfo{person}{Nick Benton}.}
  \bibinfo{year}{2004}\natexlab{}.
\newblock \showarticletitle{Simple relational correctness proofs for static
  analyses and program transformations}. In
  \bibinfo{booktitle}{\emph{Proceedings of the 31st {ACM} {SIGPLAN-SIGACT}
  Symposium on Principles of Programming Languages, {POPL} 2004, Venice, Italy,
  January 14-16, 2004}}, \bibfield{editor}{\bibinfo{person}{Neil~D. Jones}
  {and} \bibinfo{person}{Xavier Leroy}} (Eds.). \bibinfo{publisher}{{ACM}},
  \bibinfo{pages}{14--25}.
\newblock
\urldef\tempurl%
\url{https://doi.org/10.1145/964001.964003}
\showDOI{\tempurl}


\bibitem[Bos et~al\mbox{.}(2018)]%
        {DBLP:conf/eurosp/BosDKLLSSSS18}
\bibfield{author}{\bibinfo{person}{Joppe~W. Bos}, \bibinfo{person}{L{\'{e}}o
  Ducas}, \bibinfo{person}{Eike Kiltz}, \bibinfo{person}{Tancr{\`{e}}de
  Lepoint}, \bibinfo{person}{Vadim Lyubashevsky}, \bibinfo{person}{John~M.
  Schanck}, \bibinfo{person}{Peter Schwabe}, \bibinfo{person}{Gregor Seiler},
  {and} \bibinfo{person}{Damien Stehl{\'{e}}}.}
  \bibinfo{year}{2018}\natexlab{}.
\newblock \showarticletitle{{CRYSTALS} - Kyber: {A} CCA-Secure
  Module-Lattice-Based {KEM}}. In \bibinfo{booktitle}{\emph{2018 {IEEE}
  European Symposium on Security and Privacy, EuroS{\&}P 2018, London, United
  Kingdom, April 24-26, 2018}}. \bibinfo{publisher}{{IEEE}},
  \bibinfo{pages}{353--367}.
\newblock
\urldef\tempurl%
\url{https://doi.org/10.1109/EUROSP.2018.00032}
\showDOI{\tempurl}


\bibitem[Cauligi et~al\mbox{.}(2020)]%
        {DBLP:conf/pldi/CauligiDGTSRB20}
\bibfield{author}{\bibinfo{person}{Sunjay Cauligi}, \bibinfo{person}{Craig
  Disselkoen}, \bibinfo{person}{Klaus von Gleissenthall},
  \bibinfo{person}{Dean~M. Tullsen}, \bibinfo{person}{Deian Stefan},
  \bibinfo{person}{Tamara Rezk}, {and} \bibinfo{person}{Gilles Barthe}.}
  \bibinfo{year}{2020}\natexlab{}.
\newblock \showarticletitle{Constant-time foundations for the new spectre era}.
  In \bibinfo{booktitle}{\emph{Proceedings of the 41st {ACM} {SIGPLAN}
  International Conference on Programming Language Design and Implementation,
  {PLDI} 2020, London, UK, June 15-20, 2020}},
  \bibfield{editor}{\bibinfo{person}{Alastair~F. Donaldson} {and}
  \bibinfo{person}{Emina Torlak}} (Eds.). \bibinfo{publisher}{{ACM}},
  \bibinfo{pages}{913--926}.
\newblock
\urldef\tempurl%
\url{https://doi.org/10.1145/3385412.3385970}
\showDOI{\tempurl}


\bibitem[Chargu{\'{e}}raud et~al\mbox{.}(2023)]%
        {DBLP:journals/toplas/ChargueraudCEG23}
\bibfield{author}{\bibinfo{person}{Arthur Chargu{\'{e}}raud},
  \bibinfo{person}{Adam Chlipala}, \bibinfo{person}{Andres Erbsen}, {and}
  \bibinfo{person}{Samuel Gruetter}.} \bibinfo{year}{2023}\natexlab{}.
\newblock \showarticletitle{Omnisemantics: Smooth Handling of Nondeterminism}.
\newblock \bibinfo{journal}{\emph{{ACM} Trans. Program. Lang. Syst.}}
  \bibinfo{volume}{45}, \bibinfo{number}{1} (\bibinfo{year}{2023}),
  \bibinfo{pages}{5:1--5:43}.
\newblock
\urldef\tempurl%
\url{https://doi.org/10.1145/3579834}
\showDOI{\tempurl}


\bibitem[Conoly et~al\mbox{.}(2025)]%
        {DBLP:journals/pacmpl/ConolyEC25}
\bibfield{author}{\bibinfo{person}{Owen Conoly}, \bibinfo{person}{Andres
  Erbsen}, {and} \bibinfo{person}{Adam Chlipala}.}
  \bibinfo{year}{2025}\natexlab{}.
\newblock \showarticletitle{Smooth, Integrated Proofs of Cryptographic Constant
  Time for Nondeterministic Programs and Compilers}.
\newblock \bibinfo{journal}{\emph{Proc. {ACM} Program. Lang.}}
  \bibinfo{volume}{9}, \bibinfo{number}{{PLDI}} (\bibinfo{year}{2025}),
  \bibinfo{pages}{1692--1715}.
\newblock
\urldef\tempurl%
\url{https://doi.org/10.1145/3729318}
\showDOI{\tempurl}


\bibitem[D'Silva et~al\mbox{.}(2015)]%
        {DBLP:conf/sp/DSilvaPS15}
\bibfield{author}{\bibinfo{person}{Vijay D'Silva}, \bibinfo{person}{Mathias
  Payer}, {and} \bibinfo{person}{Dawn~Xiaodong Song}.}
  \bibinfo{year}{2015}\natexlab{}.
\newblock \showarticletitle{The Correctness-Security Gap in Compiler
  Optimization}. In \bibinfo{booktitle}{\emph{2015 {IEEE} Symposium on Security
  and Privacy Workshops, {SPW} 2015, San Jose, CA, USA, May 21-22, 2015}}.
  \bibinfo{publisher}{{IEEE} Computer Society}, \bibinfo{pages}{73--87}.
\newblock
\urldef\tempurl%
\url{https://doi.org/10.1109/SPW.2015.33}
\showDOI{\tempurl}


\bibitem[Dziembowski et~al\mbox{.}(2025)]%
        {DBLP:journals/iacr/DziembowskiFMS25}
\bibfield{author}{\bibinfo{person}{Stefan Dziembowski},
  \bibinfo{person}{Grzegorz Fabianski}, \bibinfo{person}{Daniele Micciancio},
  {and} \bibinfo{person}{Rafal Stefanski}.} \bibinfo{year}{2025}\natexlab{}.
\newblock \showarticletitle{Computationally-Sound Symbolic Cryptography in
  Lean}.
\newblock \bibinfo{journal}{\emph{{IACR} Cryptol. ePrint Arch.}}
  (\bibinfo{year}{2025}), \bibinfo{pages}{1700}.
\newblock
\urldef\tempurl%
\url{https://eprint.iacr.org/2025/1700}
\showURL{%
\tempurl}


\bibitem[Erbsen et~al\mbox{.}(2021)]%
        {DBLP:conf/pldi/ErbsenGCWC21}
\bibfield{author}{\bibinfo{person}{Andres Erbsen}, \bibinfo{person}{Samuel
  Gruetter}, \bibinfo{person}{Joonwon Choi}, \bibinfo{person}{Clark Wood},
  {and} \bibinfo{person}{Adam Chlipala}.} \bibinfo{year}{2021}\natexlab{}.
\newblock \showarticletitle{Integration verification across software and
  hardware for a simple embedded system}. In \bibinfo{booktitle}{\emph{{PLDI}
  '21: 42nd {ACM} {SIGPLAN} International Conference on Programming Language
  Design and Implementation, Virtual Event, Canada, June 20-25, 2021}},
  \bibfield{editor}{\bibinfo{person}{Stephen~N. Freund} {and}
  \bibinfo{person}{Eran Yahav}} (Eds.). \bibinfo{publisher}{{ACM}},
  \bibinfo{pages}{604--619}.
\newblock
\urldef\tempurl%
\url{https://doi.org/10.1145/3453483.3454065}
\showDOI{\tempurl}


\bibitem[Erbsen et~al\mbox{.}(2019)]%
        {DBLP:conf/sp/ErbsenPGSC19}
\bibfield{author}{\bibinfo{person}{Andres Erbsen}, \bibinfo{person}{Jade
  Philipoom}, \bibinfo{person}{Jason Gross}, \bibinfo{person}{Robert Sloan},
  {and} \bibinfo{person}{Adam Chlipala}.} \bibinfo{year}{2019}\natexlab{}.
\newblock \showarticletitle{Simple High-Level Code for Cryptographic Arithmetic
  - With Proofs, Without Compromises}. In \bibinfo{booktitle}{\emph{2019 {IEEE}
  Symposium on Security and Privacy, {SP} 2019, San Francisco, CA, USA, May
  19-23, 2019}}. \bibinfo{publisher}{{IEEE}}, \bibinfo{pages}{1202--1219}.
\newblock
\urldef\tempurl%
\url{https://doi.org/10.1109/SP.2019.00005}
\showDOI{\tempurl}


\bibitem[Erbsen et~al\mbox{.}(2024)]%
        {DBLP:journals/pacmpl/ErbsenPJLGPC24}
\bibfield{author}{\bibinfo{person}{Andres Erbsen}, \bibinfo{person}{Jade
  Philipoom}, \bibinfo{person}{Dustin Jamner}, \bibinfo{person}{Ashley Lin},
  \bibinfo{person}{Samuel Gruetter}, \bibinfo{person}{Cl{\'{e}}ment
  Pit{-}Claudel}, {and} \bibinfo{person}{Adam Chlipala}.}
  \bibinfo{year}{2024}\natexlab{}.
\newblock \showarticletitle{Foundational Integration Verification of a
  Cryptographic Server}.
\newblock \bibinfo{journal}{\emph{Proc. {ACM} Program. Lang.}}
  \bibinfo{volume}{8}, \bibinfo{number}{{PLDI}} (\bibinfo{year}{2024}),
  \bibinfo{pages}{1704--1729}.
\newblock
\urldef\tempurl%
\url{https://doi.org/10.1145/3656446}
\showDOI{\tempurl}


\bibitem[Fabian et~al\mbox{.}(2025)]%
        {DBLP:journals/pacmpl/FabianPGB25}
\bibfield{author}{\bibinfo{person}{Xaver Fabian}, \bibinfo{person}{Marco
  Patrignani}, \bibinfo{person}{Marco Guarnieri}, {and}
  \bibinfo{person}{Michael Backes}.} \bibinfo{year}{2025}\natexlab{}.
\newblock \showarticletitle{Do You Even Lift? Strengthening Compiler Security
  Guarantees against Spectre Attacks}.
\newblock \bibinfo{journal}{\emph{Proc. {ACM} Program. Lang.}}
  \bibinfo{volume}{9}, \bibinfo{number}{{POPL}} (\bibinfo{year}{2025}),
  \bibinfo{pages}{893--922}.
\newblock
\urldef\tempurl%
\url{https://doi.org/10.1145/3704867}
\showDOI{\tempurl}


\bibitem[{Formosa-Crypto Project}(2026)]%
        {formosa-mlkem}
\bibfield{author}{\bibinfo{person}{{Formosa-Crypto Project}}.}
  \bibinfo{year}{2026}\natexlab{}.
\newblock \bibinfo{booktitle}{\emph{Formosa {ML-KEM}: A Formally Verified
  Implementation of {ML-KEM}}}.
\newblock
\urldef\tempurl%
\url{https://github.com/formosa-crypto/formosa-mlkem}
\showURL{%
\tempurl}


\bibitem[G{\"{a}}her et~al\mbox{.}(2022)]%
        {DBLP:journals/pacmpl/GaherSSJDKKD22}
\bibfield{author}{\bibinfo{person}{Lennard G{\"{a}}her},
  \bibinfo{person}{Michael Sammler}, \bibinfo{person}{Simon Spies},
  \bibinfo{person}{Ralf Jung}, \bibinfo{person}{Hoang{-}Hai Dang},
  \bibinfo{person}{Robbert Krebbers}, \bibinfo{person}{Jeehoon Kang}, {and}
  \bibinfo{person}{Derek Dreyer}.} \bibinfo{year}{2022}\natexlab{}.
\newblock \showarticletitle{Simuliris: a separation logic framework for
  verifying concurrent program optimizations}.
\newblock \bibinfo{journal}{\emph{Proc. {ACM} Program. Lang.}}
  \bibinfo{volume}{6}, \bibinfo{number}{{POPL}} (\bibinfo{year}{2022}),
  \bibinfo{pages}{1--31}.
\newblock
\urldef\tempurl%
\url{https://doi.org/10.1145/3498689}
\showDOI{\tempurl}


\bibitem[Gancher et~al\mbox{.}(2023)]%
        {DBLP:journals/pacmpl/GancherSFSM23}
\bibfield{author}{\bibinfo{person}{Joshua Gancher}, \bibinfo{person}{Kristina
  Sojakova}, \bibinfo{person}{Xiong Fan}, \bibinfo{person}{Elaine Shi}, {and}
  \bibinfo{person}{Greg Morrisett}.} \bibinfo{year}{2023}\natexlab{}.
\newblock \showarticletitle{A Core Calculus for Equational Proofs of
  Cryptographic Protocols}.
\newblock \bibinfo{journal}{\emph{Proc. {ACM} Program. Lang.}}
  \bibinfo{volume}{7}, \bibinfo{number}{{POPL}} (\bibinfo{year}{2023}),
  \bibinfo{pages}{866--892}.
\newblock
\urldef\tempurl%
\url{https://doi.org/10.1145/3571223}
\showDOI{\tempurl}


\bibitem[Gregersen et~al\mbox{.}(2024)]%
        {DBLP:journals/pacmpl/GregersenAHTB24}
\bibfield{author}{\bibinfo{person}{Simon~Oddershede Gregersen},
  \bibinfo{person}{Alejandro Aguirre}, \bibinfo{person}{Philipp~G.
  Haselwarter}, \bibinfo{person}{Joseph Tassarotti}, {and}
  \bibinfo{person}{Lars Birkedal}.} \bibinfo{year}{2024}\natexlab{}.
\newblock \showarticletitle{Asynchronous Probabilistic Couplings in
  Higher-Order Separation Logic}.
\newblock \bibinfo{journal}{\emph{Proc. {ACM} Program. Lang.}}
  \bibinfo{volume}{8}, \bibinfo{number}{{POPL}} (\bibinfo{year}{2024}),
  \bibinfo{pages}{753--784}.
\newblock
\urldef\tempurl%
\url{https://doi.org/10.1145/3632868}
\showDOI{\tempurl}


\bibitem[Guarnieri et~al\mbox{.}(2021)]%
        {DBLP:conf/sp/GuarnieriKRV21}
\bibfield{author}{\bibinfo{person}{Marco Guarnieri}, \bibinfo{person}{Boris
  K{\"{o}}pf}, \bibinfo{person}{Jan Reineke}, {and} \bibinfo{person}{Pepe
  Vila}.} \bibinfo{year}{2021}\natexlab{}.
\newblock \showarticletitle{Hardware-Software Contracts for Secure
  Speculation}. In \bibinfo{booktitle}{\emph{42nd {IEEE} Symposium on Security
  and Privacy, {SP} 2021, San Francisco, CA, USA, 24-27 May 2021}}.
  \bibinfo{publisher}{{IEEE}}, \bibinfo{pages}{1868--1883}.
\newblock
\urldef\tempurl%
\url{https://doi.org/10.1109/SP40001.2021.00036}
\showDOI{\tempurl}


\bibitem[Haselwarter et~al\mbox{.}(2023)]%
        {DBLP:journals/toplas/HaselwarterRMWASHMS23}
\bibfield{author}{\bibinfo{person}{Philipp~G. Haselwarter},
  \bibinfo{person}{Exequiel Rivas}, \bibinfo{person}{Antoine {Van Muylder}},
  \bibinfo{person}{Th{\'{e}}o Winterhalter}, \bibinfo{person}{Carmine Abate},
  \bibinfo{person}{Nikolaj Sidorenco}, \bibinfo{person}{Catalin Hritcu},
  \bibinfo{person}{Kenji Maillard}, {and} \bibinfo{person}{Bas Spitters}.}
  \bibinfo{year}{2023}\natexlab{}.
\newblock \showarticletitle{SSProve: {A} Foundational Framework for Modular
  Cryptographic Proofs in Coq}.
\newblock \bibinfo{journal}{\emph{{ACM} Trans. Program. Lang. Syst.}}
  \bibinfo{volume}{45}, \bibinfo{number}{3} (\bibinfo{year}{2023}),
  \bibinfo{pages}{15:1--15:61}.
\newblock
\urldef\tempurl%
\url{https://doi.org/10.1145/3594735}
\showDOI{\tempurl}


\bibitem[Hirata et~al\mbox{.}(2023)]%
        {DBLP:conf/itp/HirataM023}
\bibfield{author}{\bibinfo{person}{Michikazu Hirata}, \bibinfo{person}{Yasuhiko
  Minamide}, {and} \bibinfo{person}{Tetsuya Sato}.}
  \bibinfo{year}{2023}\natexlab{}.
\newblock \showarticletitle{Semantic Foundations of Higher-Order Probabilistic
  Programs in Isabelle/HOL}. In \bibinfo{booktitle}{\emph{14th International
  Conference on Interactive Theorem Proving, {ITP} 2023, July 31 to August 4,
  2023, Bia{\l}ystok, Poland}} \emph{(\bibinfo{series}{LIPIcs},
  Vol.~\bibinfo{volume}{268})}, \bibfield{editor}{\bibinfo{person}{Adam
  Naumowicz} {and} \bibinfo{person}{Ren{\'{e}} Thiemann}} (Eds.).
  \bibinfo{publisher}{Schloss Dagstuhl - Leibniz-Zentrum f{\"{u}}r Informatik},
  \bibinfo{pages}{18:1--18:18}.
\newblock
\urldef\tempurl%
\url{https://doi.org/10.4230/LIPICS.ITP.2023.18}
\showDOI{\tempurl}


\bibitem[Hurd et~al\mbox{.}(2004)]%
        {DBLP:journals/entcs/HurdMM05}
\bibfield{author}{\bibinfo{person}{Joe Hurd}, \bibinfo{person}{Annabelle
  McIver}, {and} \bibinfo{person}{Carroll Morgan}.}
  \bibinfo{year}{2004}\natexlab{}.
\newblock \showarticletitle{Probabilistic Guarded Commands Mechanized in
  \emph{HOL}}. In \bibinfo{booktitle}{\emph{Proceedings of the Second Workshop
  on Quantitative Aspects of Programming Languages, {QAPL} 2004, Barcelona,
  Spain, March 27-28, 2004}} \emph{(\bibinfo{series}{Electronic Notes in
  Theoretical Computer Science}, Vol.~\bibinfo{volume}{112})},
  \bibfield{editor}{\bibinfo{person}{Antonio Cerone} {and}
  \bibinfo{person}{Alessandra~Di Pierro}} (Eds.).
  \bibinfo{publisher}{Elsevier}, \bibinfo{pages}{95--111}.
\newblock
\urldef\tempurl%
\url{https://doi.org/10.1016/J.ENTCS.2004.01.021}
\showDOI{\tempurl}


\bibitem[Jung et~al\mbox{.}(2015)]%
        {DBLP:conf/popl/JungSSSTBD15}
\bibfield{author}{\bibinfo{person}{Ralf Jung}, \bibinfo{person}{David Swasey},
  \bibinfo{person}{Filip Sieczkowski}, \bibinfo{person}{Kasper Svendsen},
  \bibinfo{person}{Aaron Turon}, \bibinfo{person}{Lars Birkedal}, {and}
  \bibinfo{person}{Derek Dreyer}.} \bibinfo{year}{2015}\natexlab{}.
\newblock \showarticletitle{Iris: Monoids and Invariants as an Orthogonal Basis
  for Concurrent Reasoning}. In \bibinfo{booktitle}{\emph{Proceedings of the
  42nd Annual {ACM} {SIGPLAN-SIGACT} Symposium on Principles of Programming
  Languages, {POPL} 2015, Mumbai, India, January 15-17, 2015}},
  \bibfield{editor}{\bibinfo{person}{Sriram~K. Rajamani} {and}
  \bibinfo{person}{David Walker}} (Eds.). \bibinfo{publisher}{{ACM}},
  \bibinfo{pages}{637--650}.
\newblock
\urldef\tempurl%
\url{https://doi.org/10.1145/2676726.2676980}
\showDOI{\tempurl}


\bibitem[Kumar et~al\mbox{.}(2014)]%
        {DBLP:conf/popl/KumarMNO14}
\bibfield{author}{\bibinfo{person}{Ramana Kumar}, \bibinfo{person}{Magnus~O.
  Myreen}, \bibinfo{person}{Michael Norrish}, {and} \bibinfo{person}{Scott
  Owens}.} \bibinfo{year}{2014}\natexlab{}.
\newblock \showarticletitle{CakeML: a verified implementation of {ML}}. In
  \bibinfo{booktitle}{\emph{The 41st Annual {ACM} {SIGPLAN-SIGACT} Symposium on
  Principles of Programming Languages, {POPL} '14, San Diego, CA, USA, January
  20-21, 2014}}, \bibfield{editor}{\bibinfo{person}{Suresh Jagannathan} {and}
  \bibinfo{person}{Peter Sewell}} (Eds.). \bibinfo{publisher}{{ACM}},
  \bibinfo{pages}{179--192}.
\newblock
\urldef\tempurl%
\url{https://doi.org/10.1145/2535838.2535841}
\showDOI{\tempurl}


\bibitem[Leroy(2006a)]%
        {DBLP:conf/esop/Leroy06}
\bibfield{author}{\bibinfo{person}{Xavier Leroy}.}
  \bibinfo{year}{2006}\natexlab{a}.
\newblock \showarticletitle{Coinductive Big-Step Operational Semantics}. In
  \bibinfo{booktitle}{\emph{Programming Languages and Systems, 15th European
  Symposium on Programming, {ESOP} 2006, Held as Part of the Joint European
  Conferences on Theory and Practice of Software, {ETAPS} 2006, Vienna,
  Austria, March 27-28, 2006, Proceedings}} \emph{(\bibinfo{series}{Lecture
  Notes in Computer Science}, Vol.~\bibinfo{volume}{3924})},
  \bibfield{editor}{\bibinfo{person}{Peter Sestoft}} (Ed.).
  \bibinfo{publisher}{Springer}, \bibinfo{pages}{54--68}.
\newblock
\urldef\tempurl%
\url{https://doi.org/10.1007/11693024\_5}
\showDOI{\tempurl}


\bibitem[Leroy(2006b)]%
        {2006-Leroy-compcert}
\bibfield{author}{\bibinfo{person}{Xavier Leroy}.}
  \bibinfo{year}{2006}\natexlab{b}.
\newblock \showarticletitle{Formal certification of a compiler back-end, or:
  programming a compiler with a proof assistant}. In
  \bibinfo{booktitle}{\emph{33rd ACM symposium on Principles of Programming
  Languages}}. \bibinfo{publisher}{ACM Press}, \bibinfo{pages}{42--54}.
\newblock


\bibitem[Li et~al\mbox{.}(2024)]%
        {DBLP:journals/pacmpl/LiWZ24}
\bibfield{author}{\bibinfo{person}{Jianlin Li}, \bibinfo{person}{Eric Wang},
  {and} \bibinfo{person}{Yizhou Zhang}.} \bibinfo{year}{2024}\natexlab{}.
\newblock \showarticletitle{Compiling Probabilistic Programs for Variable
  Elimination with Information Flow}.
\newblock \bibinfo{journal}{\emph{Proc. {ACM} Program. Lang.}}
  \bibinfo{volume}{8}, \bibinfo{number}{{PLDI}} (\bibinfo{year}{2024}),
  \bibinfo{pages}{1755--1780}.
\newblock
\urldef\tempurl%
\url{https://doi.org/10.1145/3656448}
\showDOI{\tempurl}


\bibitem[McBride(2015)]%
        {DBLP:conf/mpc/McBride15}
\bibfield{author}{\bibinfo{person}{Conor McBride}.}
  \bibinfo{year}{2015}\natexlab{}.
\newblock \showarticletitle{Turing-Completeness Totally Free}. In
  \bibinfo{booktitle}{\emph{Mathematics of Program Construction - 12th
  International Conference, {MPC} 2015, K{\"{o}}nigswinter, Germany, June 29 -
  July 1, 2015. Proceedings}} \emph{(\bibinfo{series}{Lecture Notes in Computer
  Science}, Vol.~\bibinfo{volume}{9129})},
  \bibfield{editor}{\bibinfo{person}{Ralf Hinze} {and} \bibinfo{person}{Janis
  Voigtl{\"{a}}nder}} (Eds.). \bibinfo{publisher}{Springer},
  \bibinfo{pages}{257--275}.
\newblock
\urldef\tempurl%
\url{https://doi.org/10.1007/978-3-319-19797-5\_13}
\showDOI{\tempurl}


\bibitem[Michelland et~al\mbox{.}(2024)]%
        {Michelland24}
\bibfield{author}{\bibinfo{person}{S\'{e}bastien Michelland},
  \bibinfo{person}{Yannick Zakowski}, {and} \bibinfo{person}{Laure Gonnord}.}
  \bibinfo{year}{2024}\natexlab{}.
\newblock \showarticletitle{Abstract Interpreters: A Monadic Approach to
  Modular Verification}.
\newblock \bibinfo{journal}{\emph{Proc. ACM Program. Lang.}}
  \bibinfo{volume}{8}, \bibinfo{number}{ICFP} (\bibinfo{year}{2024}),
  \bibinfo{numpages}{28}~pages.
\newblock
\urldef\tempurl%
\url{https://doi.org/10.1145/3674646}
\showDOI{\tempurl}


\bibitem[Moody et~al\mbox{.}(2024)]%
        {NISTpqctransition}
\bibfield{author}{\bibinfo{person}{Dustin Moody}, \bibinfo{person}{Ray
  Perlner}, \bibinfo{person}{Andrew Regenscheid}, \bibinfo{person}{Angela
  Robinson}, {and} \bibinfo{person}{David Cooper}.}
  \bibinfo{year}{2024}\natexlab{}.
\newblock \bibinfo{booktitle}{\emph{Transition to Post-Quantum Cryptography
  Standards}}.
\newblock \bibinfo{type}{NIST Interagency Report} NIST IR 8547 (Initial Public
  Draft). \bibinfo{institution}{National Institute of Standards and
  Technology}.
\newblock
\urldef\tempurl%
\url{https://doi.org/10.6028/NIST.IR.8547.ipd}
\showDOI{\tempurl}


\bibitem[Nakata and Uustalu(2009)]%
        {DBLP:conf/tphol/NakataU09}
\bibfield{author}{\bibinfo{person}{Keiko Nakata} {and} \bibinfo{person}{Tarmo
  Uustalu}.} \bibinfo{year}{2009}\natexlab{}.
\newblock \showarticletitle{Trace-Based Coinductive Operational Semantics for
  While}. In \bibinfo{booktitle}{\emph{Theorem Proving in Higher Order Logics,
  22nd International Conference, TPHOLs 2009, Munich, Germany, August 17-20,
  2009. Proceedings}} \emph{(\bibinfo{series}{Lecture Notes in Computer
  Science}, Vol.~\bibinfo{volume}{5674})},
  \bibfield{editor}{\bibinfo{person}{Stefan Berghofer}, \bibinfo{person}{Tobias
  Nipkow}, \bibinfo{person}{Christian Urban}, {and} \bibinfo{person}{Makarius
  Wenzel}} (Eds.). \bibinfo{publisher}{Springer}, \bibinfo{pages}{375--390}.
\newblock
\urldef\tempurl%
\url{https://doi.org/10.1007/978-3-642-03359-9\_26}
\showDOI{\tempurl}


\bibitem[{National Institute of Standards and Technology}(2024)]%
        {NIST-FIPS-203-2024}
\bibfield{author}{\bibinfo{person}{{National Institute of Standards and
  Technology}}.} \bibinfo{year}{2024}\natexlab{}.
\newblock \bibinfo{booktitle}{\emph{Module-Lattice-Based Key-Encapsulation
  Mechanism Standard}}.
\newblock \bibinfo{type}{Federal Information Processing Standards Publication}
  203. \bibinfo{institution}{National Institute of Standards and Technology},
  \bibinfo{address}{Gaithersburg, MD}.
\newblock
\urldef\tempurl%
\url{https://doi.org/10.6028/NIST.FIPS.203}
\showDOI{\tempurl}
\newblock
\shownote{Published August 13, 2024. Also available as PDF:
  https://nvlpubs.nist.gov/nistpubs/FIPS/NIST.FIPS.203.pdf}.


\bibitem[Patrignani et~al\mbox{.}(2024)]%
        {DBLP:journals/toplas/PatrignaniKWC24}
\bibfield{author}{\bibinfo{person}{Marco Patrignani}, \bibinfo{person}{Robert
  K{\"{u}}nnemann}, \bibinfo{person}{Riad~S. Wahby}, {and}
  \bibinfo{person}{Ethan Cecchetti}.} \bibinfo{year}{2024}\natexlab{}.
\newblock \showarticletitle{Universal Composability Is Robust Compilation}.
\newblock \bibinfo{journal}{\emph{{ACM} Trans. Program. Lang. Syst.}}
  \bibinfo{volume}{46}, \bibinfo{number}{4} (\bibinfo{year}{2024}),
  \bibinfo{pages}{13:1--13:64}.
\newblock
\urldef\tempurl%
\url{https://doi.org/10.1145/3698234}
\showDOI{\tempurl}


\bibitem[Petcher and Morrisett(2015)]%
        {DBLP:conf/post/PetcherM15}
\bibfield{author}{\bibinfo{person}{Adam Petcher} {and} \bibinfo{person}{Greg
  Morrisett}.} \bibinfo{year}{2015}\natexlab{}.
\newblock \showarticletitle{The Foundational Cryptography Framework}. In
  \bibinfo{booktitle}{\emph{Principles of Security and Trust - 4th
  International Conference, {POST} 2015, Held as Part of the European Joint
  Conferences on Theory and Practice of Software, {ETAPS} 2015, London, UK,
  April 11-18, 2015, Proceedings}} \emph{(\bibinfo{series}{Lecture Notes in
  Computer Science}, Vol.~\bibinfo{volume}{9036})},
  \bibfield{editor}{\bibinfo{person}{Riccardo Focardi} {and}
  \bibinfo{person}{Andrew~C. Myers}} (Eds.). \bibinfo{publisher}{Springer},
  \bibinfo{pages}{53--72}.
\newblock
\urldef\tempurl%
\url{https://doi.org/10.1007/978-3-662-46666-7\_4}
\showDOI{\tempurl}


\bibitem[Pit{-}Claudel et~al\mbox{.}(2022)]%
        {DBLP:conf/pldi/Pit-ClaudelPJEC22}
\bibfield{author}{\bibinfo{person}{Cl{\'{e}}ment Pit{-}Claudel},
  \bibinfo{person}{Jade Philipoom}, \bibinfo{person}{Dustin Jamner},
  \bibinfo{person}{Andres Erbsen}, {and} \bibinfo{person}{Adam Chlipala}.}
  \bibinfo{year}{2022}\natexlab{}.
\newblock \showarticletitle{Relational compilation for performance-critical
  applications: extensible proof-producing translation of functional models
  into low-level code}. In \bibinfo{booktitle}{\emph{{PLDI} '22: 43rd {ACM}
  {SIGPLAN} International Conference on Programming Language Design and
  Implementation, San Diego, CA, USA, June 13 - 17, 2022}},
  \bibfield{editor}{\bibinfo{person}{Ranjit Jhala} {and} \bibinfo{person}{Isil
  Dillig}} (Eds.). \bibinfo{publisher}{{ACM}}, \bibinfo{pages}{918--933}.
\newblock
\urldef\tempurl%
\url{https://doi.org/10.1145/3519939.3523706}
\showDOI{\tempurl}


\bibitem[Polubelova et~al\mbox{.}(2020)]%
        {DBLP:conf/ccs/PolubelovaBPBFK20}
\bibfield{author}{\bibinfo{person}{Marina Polubelova},
  \bibinfo{person}{Karthikeyan Bhargavan}, \bibinfo{person}{Jonathan
  Protzenko}, \bibinfo{person}{Benjamin Beurdouche}, \bibinfo{person}{Aymeric
  Fromherz}, \bibinfo{person}{Natalia Kulatova}, {and}
  \bibinfo{person}{Santiago Zanella{-}B{\'{e}}guelin}.}
  \bibinfo{year}{2020}\natexlab{}.
\newblock \showarticletitle{HACLxN: Verified Generic {SIMD} Crypto (for all
  your favourite platforms)}. In \bibinfo{booktitle}{\emph{{CCS} '20: 2020
  {ACM} {SIGSAC} Conference on Computer and Communications Security, Virtual
  Event, USA, November 9-13, 2020}}, \bibfield{editor}{\bibinfo{person}{Jay
  Ligatti}, \bibinfo{person}{Xinming Ou}, \bibinfo{person}{Jonathan Katz},
  {and} \bibinfo{person}{Giovanni Vigna}} (Eds.). \bibinfo{publisher}{{ACM}},
  \bibinfo{pages}{899--918}.
\newblock
\urldef\tempurl%
\url{https://doi.org/10.1145/3372297.3423352}
\showDOI{\tempurl}


\bibitem[Protzenko et~al\mbox{.}(2020)]%
        {DBLP:conf/sp/ProtzenkoPFHPBB20}
\bibfield{author}{\bibinfo{person}{Jonathan Protzenko}, \bibinfo{person}{Bryan
  Parno}, \bibinfo{person}{Aymeric Fromherz}, \bibinfo{person}{Chris
  Hawblitzel}, \bibinfo{person}{Marina Polubelova},
  \bibinfo{person}{Karthikeyan Bhargavan}, \bibinfo{person}{Benjamin
  Beurdouche}, \bibinfo{person}{Joonwon Choi}, \bibinfo{person}{Antoine
  Delignat{-}Lavaud}, \bibinfo{person}{C{\'{e}}dric Fournet},
  \bibinfo{person}{Natalia Kulatova}, \bibinfo{person}{Tahina Ramananandro},
  \bibinfo{person}{Aseem Rastogi}, \bibinfo{person}{Nikhil Swamy},
  \bibinfo{person}{Christoph~M. Wintersteiger}, {and} \bibinfo{person}{Santiago
  Zanella{-}B{\'{e}}guelin}.} \bibinfo{year}{2020}\natexlab{}.
\newblock \showarticletitle{EverCrypt: {A} Fast, Verified, Cross-Platform
  Cryptographic Provider}. In \bibinfo{booktitle}{\emph{2020 {IEEE} Symposium
  on Security and Privacy, {SP} 2020, San Francisco, CA, USA, May 18-21,
  2020}}. \bibinfo{publisher}{{IEEE}}, \bibinfo{pages}{983--1002}.
\newblock
\urldef\tempurl%
\url{https://doi.org/10.1109/SP40000.2020.00114}
\showDOI{\tempurl}


\bibitem[Protzenko et~al\mbox{.}(2017)]%
        {DBLP:journals/pacmpl/ProtzenkoZRRWBD17}
\bibfield{author}{\bibinfo{person}{Jonathan Protzenko},
  \bibinfo{person}{Jean~Karim Zinzindohou{\'{e}}}, \bibinfo{person}{Aseem
  Rastogi}, \bibinfo{person}{Tahina Ramananandro}, \bibinfo{person}{Peng Wang},
  \bibinfo{person}{Santiago Zanella{-}B{\'{e}}guelin}, \bibinfo{person}{Antoine
  Delignat{-}Lavaud}, \bibinfo{person}{Catalin Hritcu},
  \bibinfo{person}{Karthikeyan Bhargavan}, \bibinfo{person}{C{\'{e}}dric
  Fournet}, {and} \bibinfo{person}{Nikhil Swamy}.}
  \bibinfo{year}{2017}\natexlab{}.
\newblock \showarticletitle{Verified low-level programming embedded in {F}}.
\newblock \bibinfo{journal}{\emph{Proc. {ACM} Program. Lang.}}
  \bibinfo{volume}{1}, \bibinfo{number}{{ICFP}} (\bibinfo{year}{2017}),
  \bibinfo{pages}{17:1--17:29}.
\newblock
\urldef\tempurl%
\url{https://doi.org/10.1145/3110261}
\showDOI{\tempurl}


\bibitem[Regev(2005)]%
        {DBLP:conf/stoc/Regev05}
\bibfield{author}{\bibinfo{person}{Oded Regev}.}
  \bibinfo{year}{2005}\natexlab{}.
\newblock \showarticletitle{On lattices, learning with errors, random linear
  codes, and cryptography}. In \bibinfo{booktitle}{\emph{Proceedings of the
  37th Annual {ACM} Symposium on Theory of Computing, Baltimore, MD, USA, May
  22-24, 2005}}, \bibfield{editor}{\bibinfo{person}{Harold~N. Gabow} {and}
  \bibinfo{person}{Ronald Fagin}} (Eds.). \bibinfo{publisher}{{ACM}},
  \bibinfo{pages}{84--93}.
\newblock
\urldef\tempurl%
\url{https://doi.org/10.1145/1060590.1060603}
\showDOI{\tempurl}


\bibitem[Shivakumar et~al\mbox{.}(2023)]%
        {DBLP:conf/sp/ShivakumarBGLOPST23}
\bibfield{author}{\bibinfo{person}{Basavesh~Ammanaghatta Shivakumar},
  \bibinfo{person}{Gilles Barthe}, \bibinfo{person}{Benjamin Gr{\'{e}}goire},
  \bibinfo{person}{Vincent Laporte}, \bibinfo{person}{Tiago Oliveira},
  \bibinfo{person}{Swarn Priya}, \bibinfo{person}{Peter Schwabe}, {and}
  \bibinfo{person}{Lucas Tabary{-}Maujean}.} \bibinfo{year}{2023}\natexlab{}.
\newblock \showarticletitle{Typing High-Speed Cryptography against Spectre v1}.
  In \bibinfo{booktitle}{\emph{44th {IEEE} Symposium on Security and Privacy,
  {SP} 2023, San Francisco, CA, USA, May 21-25, 2023}}.
  \bibinfo{publisher}{{IEEE}}, \bibinfo{pages}{1094--1111}.
\newblock
\urldef\tempurl%
\url{https://doi.org/10.1109/SP46215.2023.10179418}
\showDOI{\tempurl}


\bibitem[Silver et~al\mbox{.}(2023a)]%
        {DBLP:conf/ecoop/SilverHCHZ23}
\bibfield{author}{\bibinfo{person}{Lucas Silver}, \bibinfo{person}{Paul He},
  \bibinfo{person}{Ethan Cecchetti}, \bibinfo{person}{Andrew~K. Hirsch}, {and}
  \bibinfo{person}{Steve Zdancewic}.} \bibinfo{year}{2023}\natexlab{a}.
\newblock \showarticletitle{Semantics for Noninterference with Interaction
  Trees}. In \bibinfo{booktitle}{\emph{37th European Conference on
  Object-Oriented Programming, {ECOOP} 2023, July 17-21, 2023, Seattle,
  Washington, United States}} \emph{(\bibinfo{series}{LIPIcs},
  Vol.~\bibinfo{volume}{263})}, \bibfield{editor}{\bibinfo{person}{Karim Ali}
  {and} \bibinfo{person}{Guido Salvaneschi}} (Eds.).
  \bibinfo{publisher}{Schloss Dagstuhl - Leibniz-Zentrum f{\"{u}}r Informatik},
  \bibinfo{pages}{29:1--29:29}.
\newblock
\urldef\tempurl%
\url{https://doi.org/10.4230/LIPICS.ECOOP.2023.29}
\showDOI{\tempurl}


\bibitem[Silver et~al\mbox{.}(2023b)]%
        {Silver23}
\bibfield{author}{\bibinfo{person}{Lucas Silver}, \bibinfo{person}{Eddy
  Westbrook}, \bibinfo{person}{Matthew Yacavone}, {and} \bibinfo{person}{Ryan
  Scott}.} \bibinfo{year}{2023}\natexlab{b}.
\newblock \showarticletitle{{Interaction Tree Specifications: A Framework for
  Specifying Recursive, Effectful Computations That Supports Auto-Active
  Verification}}. In \bibinfo{booktitle}{\emph{37th European Conference on
  Object-Oriented Programming (ECOOP 2023)}} \emph{(\bibinfo{series}{Leibniz
  International Proceedings in Informatics (LIPIcs)},
  Vol.~\bibinfo{volume}{263})}, \bibfield{editor}{\bibinfo{person}{Karim Ali}
  {and} \bibinfo{person}{Guido Salvaneschi}} (Eds.).
  \bibinfo{publisher}{Schloss Dagstuhl -- Leibniz-Zentrum f{\"u}r Informatik},
  \bibinfo{pages}{30:1--30:26}.
\newblock
\urldef\tempurl%
\url{https://doi.org/10.4230/LIPIcs.ECOOP.2023.30}
\showDOI{\tempurl}


\bibitem[Silver and Zdancewic(2021)]%
        {DBLP:journals/pacmpl/SilverZ21}
\bibfield{author}{\bibinfo{person}{Lucas Silver} {and} \bibinfo{person}{Steve
  Zdancewic}.} \bibinfo{year}{2021}\natexlab{}.
\newblock \showarticletitle{Dijkstra monads forever: termination-sensitive
  specifications for interaction trees}.
\newblock \bibinfo{journal}{\emph{Proc. {ACM} Program. Lang.}}
  \bibinfo{volume}{5}, \bibinfo{number}{{POPL}} (\bibinfo{year}{2021}),
  \bibinfo{pages}{1--28}.
\newblock
\urldef\tempurl%
\url{https://doi.org/10.1145/3434307}
\showDOI{\tempurl}


\bibitem[Song et~al\mbox{.}(2025)]%
        {song2025}
\bibfield{author}{\bibinfo{person}{Shixin Song}, \bibinfo{person}{Tingzhen
  Dong}, \bibinfo{person}{Kosi Nwabueze}, \bibinfo{person}{Julian Zanders},
  \bibinfo{person}{Andres Erbsen}, \bibinfo{person}{Adam Chlipala}, {and}
  \bibinfo{person}{Mengjia Yan}.} \bibinfo{year}{2025}\natexlab{}.
\newblock \showarticletitle{Securing Cryptographic Software via Typed Assembly
  Language (Extended Version)}.
\newblock \bibinfo{journal}{\emph{CoRR}}  \bibinfo{volume}{abs/2509.08727}
  (\bibinfo{year}{2025}).
\newblock
\urldef\tempurl%
\url{https://doi.org/10.48550/ARXIV.2509.08727}
\showDOI{\tempurl}
\showeprint[arXiv]{2509.08727}


\bibitem[Tassarotti and Tristan(2023)]%
        {DBLP:journals/pacmpl/TassarottiT23}
\bibfield{author}{\bibinfo{person}{Joseph Tassarotti} {and}
  \bibinfo{person}{Jean{-}Baptiste Tristan}.} \bibinfo{year}{2023}\natexlab{}.
\newblock \showarticletitle{Verified Density Compilation for a Probabilistic
  Programming Language}.
\newblock \bibinfo{journal}{\emph{Proc. {ACM} Program. Lang.}}
  \bibinfo{volume}{7}, \bibinfo{number}{{PLDI}} (\bibinfo{year}{2023}),
  \bibinfo{pages}{615--637}.
\newblock
\urldef\tempurl%
\url{https://doi.org/10.1145/3591245}
\showDOI{\tempurl}


\bibitem[Tuma and Hopper(2024)]%
        {DBLP:journals/iacr/TumaH24}
\bibfield{author}{\bibinfo{person}{Devon Tuma} {and} \bibinfo{person}{Nicholas
  Hopper}.} \bibinfo{year}{2024}\natexlab{}.
\newblock \showarticletitle{VCVio: {A} Formally Verified Forking Lemma and
  Fiat-Shamir Transform, via a Flexible and Expressive Oracle Representation}.
\newblock \bibinfo{journal}{\emph{{IACR} Cryptol. ePrint Arch.}}
  (\bibinfo{year}{2024}), \bibinfo{pages}{1819}.
\newblock
\urldef\tempurl%
\url{https://eprint.iacr.org/2024/1819}
\showURL{%
\tempurl}


\bibitem[van~der Wall and Meyer(2025)]%
        {DBLP:journals/pacmpl/WallM25}
\bibfield{author}{\bibinfo{person}{S{\"{o}}ren van~der Wall} {and}
  \bibinfo{person}{Roland Meyer}.} \bibinfo{year}{2025}\natexlab{}.
\newblock \showarticletitle{{SNIP:} Speculative Execution and Non-Interference
  Preservation for Compiler Transformations}.
\newblock \bibinfo{journal}{\emph{Proc. {ACM} Program. Lang.}}
  \bibinfo{volume}{9}, \bibinfo{number}{{POPL}} (\bibinfo{year}{2025}),
  \bibinfo{pages}{1506--1535}.
\newblock
\urldef\tempurl%
\url{https://doi.org/10.1145/3704887}
\showDOI{\tempurl}


\bibitem[{Verified-zkEVM}(2024)]%
        {verifiedzkevm2024vcvio}
\bibfield{author}{\bibinfo{person}{{Verified-zkEVM}}.}
  \bibinfo{year}{2024}\natexlab{}.
\newblock \bibinfo{booktitle}{\emph{{VCV-io}: Formalized Cryptography Proofs in
  {Lean 4}}}.
\newblock Verified-zkEVM.
\newblock
\urldef\tempurl%
\url{https://github.com/Verified-zkEVM/VCV-io}
\showURL{%
\tempurl}
\newblock
\shownote{Accessed: 2026-04-22}.


\bibitem[Vistrup et~al\mbox{.}(2025)]%
        {Vistrup25}
\bibfield{author}{\bibinfo{person}{Max Vistrup}, \bibinfo{person}{Michael
  Sammler}, {and} \bibinfo{person}{Ralf Jung}.}
  \bibinfo{year}{2025}\natexlab{}.
\newblock \showarticletitle{Program Logics \`{a} la Carte}.
\newblock \bibinfo{journal}{\emph{Proc. ACM Program. Lang.}}
  \bibinfo{volume}{9}, \bibinfo{number}{POPL} (\bibinfo{year}{2025}),
  \bibinfo{numpages}{32}~pages.
\newblock
\urldef\tempurl%
\url{https://doi.org/10.1145/3704847}
\showDOI{\tempurl}


\bibitem[Xia et~al\mbox{.}(2020)]%
        {DBLP:journals/pacmpl/XiaZHHMPZ20}
\bibfield{author}{\bibinfo{person}{Li{-}yao Xia}, \bibinfo{person}{Yannick
  Zakowski}, \bibinfo{person}{Paul He}, \bibinfo{person}{Chung{-}Kil Hur},
  \bibinfo{person}{Gregory Malecha}, \bibinfo{person}{Benjamin~C. Pierce},
  {and} \bibinfo{person}{Steve Zdancewic}.} \bibinfo{year}{2020}\natexlab{}.
\newblock \showarticletitle{Interaction trees: representing recursive and
  impure programs in Coq}.
\newblock \bibinfo{journal}{\emph{Proc. {ACM} Program. Lang.}}
  \bibinfo{volume}{4}, \bibinfo{number}{{POPL}} (\bibinfo{year}{2020}),
  \bibinfo{pages}{51:1--51:32}.
\newblock
\urldef\tempurl%
\url{https://doi.org/10.1145/3371119}
\showDOI{\tempurl}


\bibitem[Xu et~al\mbox{.}(2023)]%
        {DBLP:conf/uss/XuLDDLWPM23}
\bibfield{author}{\bibinfo{person}{Jianhao Xu}, \bibinfo{person}{Kangjie Lu},
  \bibinfo{person}{Zhengjie Du}, \bibinfo{person}{Zhu Ding},
  \bibinfo{person}{Linke Li}, \bibinfo{person}{Qiushi Wu},
  \bibinfo{person}{Mathias Payer}, {and} \bibinfo{person}{Bing Mao}.}
  \bibinfo{year}{2023}\natexlab{}.
\newblock \showarticletitle{Silent Bugs Matter: {A} Study of
  Compiler-Introduced Security Bugs}. In \bibinfo{booktitle}{\emph{32nd
  {USENIX} Security Symposium, {USENIX} Security 2023, Anaheim, CA, USA, August
  9-11, 2023}}, \bibfield{editor}{\bibinfo{person}{Joseph~A. Calandrino} {and}
  \bibinfo{person}{Carmela Troncoso}} (Eds.). \bibinfo{publisher}{{USENIX}
  Association}, \bibinfo{pages}{3655--3672}.
\newblock
\urldef\tempurl%
\url{https://www.usenix.org/conference/usenixsecurity23/presentation/xu-jianhao}
\showURL{%
\tempurl}


\bibitem[Yang(2007)]%
        {DBLP:journals/tcs/Yang07}
\bibfield{author}{\bibinfo{person}{Hongseok Yang}.}
  \bibinfo{year}{2007}\natexlab{}.
\newblock \showarticletitle{Relational separation logic}.
\newblock \bibinfo{journal}{\emph{Theor. Comput. Sci.}} \bibinfo{volume}{375},
  \bibinfo{number}{1-3} (\bibinfo{year}{2007}), \bibinfo{pages}{308--334}.
\newblock
\urldef\tempurl%
\url{https://doi.org/10.1016/J.TCS.2006.12.036}
\showDOI{\tempurl}


\bibitem[Zakowski et~al\mbox{.}(2021)]%
        {DBLP:journals/pacmpl/ZakowskiBYZZZ21}
\bibfield{author}{\bibinfo{person}{Yannick Zakowski}, \bibinfo{person}{Calvin
  Beck}, \bibinfo{person}{Irene Yoon}, \bibinfo{person}{Ilia Zaichuk},
  \bibinfo{person}{Vadim Zaliva}, {and} \bibinfo{person}{Steve Zdancewic}.}
  \bibinfo{year}{2021}\natexlab{}.
\newblock \showarticletitle{Modular, compositional, and executable formal
  semantics for {LLVM} {IR}}.
\newblock \bibinfo{journal}{\emph{Proc. {ACM} Program. Lang.}}
  \bibinfo{volume}{5}, \bibinfo{number}{{ICFP}} (\bibinfo{year}{2021}),
  \bibinfo{pages}{1--30}.
\newblock
\urldef\tempurl%
\url{https://doi.org/10.1145/3473572}
\showDOI{\tempurl}


\bibitem[Zakowski et~al\mbox{.}(2020)]%
        {Zakowski20}
\bibfield{author}{\bibinfo{person}{Yannick Zakowski}, \bibinfo{person}{Paul
  He}, \bibinfo{person}{Chung-Kil Hur}, {and} \bibinfo{person}{Steve
  Zdancewic}.} \bibinfo{year}{2020}\natexlab{}.
\newblock \showarticletitle{An equational theory for weak bisimulation via
  generalized parameterized coinduction}. In
  \bibinfo{booktitle}{\emph{Proceedings of the 9th ACM SIGPLAN International
  Conference on Certified Programs and Proofs}} \emph{(\bibinfo{series}{CPP
  2020})}. \bibinfo{publisher}{Association for Computing Machinery},
  \bibinfo{pages}{71--84}.
\newblock
\urldef\tempurl%
\url{https://doi.org/10.1145/3372885.3373813}
\showDOI{\tempurl}


\bibitem[Zhao et~al\mbox{.}(2025)]%
        {DBLP:journals/corr/abs-2501-08249}
\bibfield{author}{\bibinfo{person}{Junming Zhao}, \bibinfo{person}{Alessandro
  Legnani}, \bibinfo{person}{Tiana J.~Tsang Ung}, \bibinfo{person}{H. Truong},
  \bibinfo{person}{Tsun~Wang Sau}, \bibinfo{person}{Miki Tanaka},
  \bibinfo{person}{Johannes~{\AA}man Pohjola}, \bibinfo{person}{Thomas Sewell},
  \bibinfo{person}{Rob Sison}, \bibinfo{person}{Syeda Hira},
  \bibinfo{person}{Magnus Myreen}, \bibinfo{person}{Michael Norrish}, {and}
  \bibinfo{person}{Gernot Heiser}.} \bibinfo{year}{2025}\natexlab{}.
\newblock \showarticletitle{Verifying Device Drivers with Pancake}.
\newblock \bibinfo{journal}{\emph{CoRR}}  \bibinfo{volume}{abs/2501.08249}
  (\bibinfo{year}{2025}).
\newblock
\urldef\tempurl%
\url{https://doi.org/10.48550/ARXIV.2501.08249}
\showDOI{\tempurl}
\showeprint[arXiv]{2501.08249}


\bibitem[Zhao et~al\mbox{.}(2012)]%
        {DBLP:conf/popl/ZhaoNMZ12}
\bibfield{author}{\bibinfo{person}{Jianzhou Zhao}, \bibinfo{person}{Santosh
  Nagarakatte}, \bibinfo{person}{Milo M.~K. Martin}, {and}
  \bibinfo{person}{Steve Zdancewic}.} \bibinfo{year}{2012}\natexlab{}.
\newblock \showarticletitle{Formalizing the {LLVM} intermediate representation
  for verified program transformations}. In
  \bibinfo{booktitle}{\emph{Proceedings of the 39th {ACM} {SIGPLAN-SIGACT}
  Symposium on Principles of Programming Languages, {POPL} 2012, Philadelphia,
  Pennsylvania, USA, January 22-28, 2012}},
  \bibfield{editor}{\bibinfo{person}{John Field} {and} \bibinfo{person}{Michael
  Hicks}} (Eds.). \bibinfo{publisher}{{ACM}}, \bibinfo{pages}{427--440}.
\newblock
\urldef\tempurl%
\url{https://doi.org/10.1145/2103656.2103709}
\showDOI{\tempurl}


\end{thebibliography}

\appendix

\section*{Open Science}

We provide the \textsc{Rocq}{} mechanization of our framework and the modified
version of the \textsc{Jasmin}{} compiler with its new proofs, which will
become publicly available and open-source \cite{\ifanon{anonymous-artifact}{public-artifact}}{}.

This is the only artifact that is needed to evaluate the paper's core
contributions.
It can be accessed via the URL in the references \cite{\ifanon{anonymous-artifact}{public-artifact}}{}.
The mapping from the paper's contributions to specific definitions and
theorems in the artifact is listed in the \texttt{README.md} file.

\section{Background on Interaction Trees}\label{appendix:itrees}

Interaction Trees~\cite{DBLP:journals/pacmpl/XiaZHHMPZ20}
provide a convenient framework to reason about interactive and
nonterminating computations. Informally, ITrees represent
computations as potentially infinite trees, where each non-leaf node
is either a silent step or a visible event. Interaction trees
are an attractive option for us because, on the one hand, unlike most
operational semantics, they offer a semantics that is executable (as a
functional definition) and compositional (we can reason inductively on
the language syntax rather than on the more complex semantic datatype).
On the other hand, unlike old-style denotational semantics, ITrees
support modularity (we can deal with different types of events
separately), and they allow for a comparatively simple style of
coinductive reasoning, supported by a rich equational theory and relying
on parameterized coinduction~\cite{Zakowski20}.

\paragraph{Interpretation}
Interaction trees can be interpreted modularly with respect to the
disjoint union of event families (written
\ensuremath{{\events_1 \mathop{+'} \events_2}}) and the
subevent relation (written \ensuremath{{{\events_1} \sqsubseteq {\events_2}}}). The
polymorphic operator \ensuremath{{
  \ensuremath{{\mathsf{trigger}}} : \ensuremath{{\forall {\ans}.\, {\events'~\ans \to \ensuremath{{\mathsf{itree}~{\events}~{\ans}}}}}}
}} allows lifting any subevent in \events['] to an ITree of \events{},
given \ensuremath{{{\events'} \sqsubseteq {\events}}}.
It is defined as
\begin{equation*}
  \ensuremath{{\ensuremath{{\mathsf{trigger}}}\mathchoice{\!}{\!}{}{}\left({e}\right)}} \triangleq \ensuremath{{\ensuremath{{\mathsf{Vis}}}\mathchoice{\!}{\!}{}{}\left({e}, {\ensuremath{{\lambda{x}.\, {\ensuremath{{\ensuremath{{\mathsf{Ret}}}\mathchoice{\!}{\!}{}{}\left({x}\right)}}}}}}\right)}}\text.
\end{equation*}
The \ensuremath{{
  \ensuremath{{\mathsf{iter}}} : \ensuremath{{\forall {\ans~I}.\, {
    \left(I \to \ensuremath{{M}}~\left(I+\ans\right)\right) \to I \to \ensuremath{{M}}~\ans
}}}}} operator can be seen as a while loop in a monad~\ensuremath{{M}}{}, repeatedly
running the body (the first argument) until it produces a value of the
result type~\ans{}.
Mappings from ITrees into a monad~\ensuremath{{M}}{} can be defined via
\emph{event handlers}, which are functions
\ensuremath{{\ensuremath{{H}} : \ensuremath{{\forall {\ans}.\, {\events~\ans \to \ensuremath{{M}}~\ans}}}}} for some
event subfamily \events{}.
A generic interpreter operator,
\begin{equation*}
  \ensuremath{{\mathsf{interp}}} :
    \left(\ensuremath{{\forall {\ans}.\, {\events~\ans \to \ensuremath{{M}}~\ans}}}\right) \to
    \ensuremath{{\forall {\ans}.\, {\ensuremath{{\mathsf{itree}~{\events}~{\ans}}} \to \ensuremath{{M}}~\ans}}}\text,
\end{equation*}
based on \ensuremath{{\mathsf{iter}}}{}, can be used to fold a handler corecursively on
an ITree.
When the event handler itself is recursive, we can use the operator
\begin{multline*}
  \ensuremath{{\mathsf{interp\_mrec}}} :
  \left(
    \ensuremath{{\forall {\ans}.\, {
    \events'~\ans \to
    \ensuremath{{\mathsf{itree}~{\left(\events' +' \events\right)}~{\ans}}}
  }}}\right) \to
    \\
  \ensuremath{{\forall {\ans}.\, {
    \ensuremath{{\mathsf{itree}~{\left(\events' +' \events\right)}~{\ans}}} \to
    \ensuremath{{\mathsf{itree}~{\events}~{\ans}}}
  }}}\text.
\end{multline*}
Since ITrees can be themselves interpretation targets, we can write
modular interpreters, each for a subfamily of events, and compose them
sequentially.

\paragraph{Bisimulation} A basic generalization of
weak bisimilarity for ITrees, denoted \euttname{}, was introduced
in~\cite{DBLP:journals/pacmpl/XiaZHHMPZ20} as a coinductive-inductive
relation, parametrized by a relation over results. It matches the
standard notion of bisimulation for LTSs, and is equivalent to
\ruttname{} where the relation over events is equality.

 \section{Summary of \textsc{Jasmin}{} Compiler Passes}\label{appendix:passes}
\Cref{tbl:passes}{} presents a one-sentence summary of each pass in the
\textsc{Jasmin}{} compiler.

  \begin{table*}
    \caption{A summary of \textsc{Jasmin}{} compiler passes.}\label{tbl:passes}
    \small
    \renewcommand{\arraystretch}{1.15}
    \begin{tabular}{l l}
      \toprule
      Compiler pass & Summary
      \\\midrule
      Preprocess
      & Resolve syntactic sugar and imports.
      \\
      Int to Word
      & Replace word-sized integers with machine words.
      \\
      Array Copy
      & Expand array copies into \ensuremath{{\text{\texttt{for}}}}{} loops.
      \\
      Add Array Initialization
      & Insert array initializations at the beginning of functions.
      \\
      Spilling
      & Replace spill and unspill operators with copies to and from
        fresh variables.
      \\
      Inlining
      & Replace (some) function calls with their bodies.
      \\
      Function Pruning
      & Remove unused functions.
      \\
      Constant Propagation
      & Propagate the values of variables and optimize expressions with
        constants.
      \\
      Dead Code Elimination
      & Remove unnecessary assignments.
      \\
      Unrolling
      & Unroll \ensuremath{{\text{\texttt{for}}}}{} loops.
      \\
      Live-Range Splitting
      & Transform the program into SSA form.
      \\
      Remove Array Initialization
      & Remove array initializations.
      \\
      Make Reference Arguments
      & Use intermediate pointer variables instead of array arguments.
      \\
      Register Array Expansion
      & Expand register arrays into variables.
      \\
      Remove Globals
      & Move local global values to the top-level.
      \\
      Load Immediates
      & Store immediates in temporary registers.
      \\
      Instruction Selection
      & Replace assignments with architecture-specific operations.
      \\
      Inline Propagation
      & Propagate the values of inline variables.
      \\
      SLH Instruction Selection
      & Replace SLH operators with architecture-specific operations.
      \\
      Stack Allocation
      & Replace stack variables with concrete stack positions.
      \\
      Lower Addresses
      & Use only addresses allowed by the architecture.
      \\
      Remove Unused Results
      & Remove unused results of functions.
      \\
      Register Renaming
      & Rename all variables to use architecture-specific registers.
      \\
      One Varmap
      & Verify that local variables can be lifted to a shared state.
      \\
      Linearization
      & Replace structured control flow with jumps.
      \\
      Tunneling
      & Simplify chains of jumps.
      \\
      Assembly Generation
      & Use concrete registers, flags, and conditions instead of
        variables.
      \\
      \bottomrule
    \end{tabular}
  \end{table*}
{}
 \section{Remaining Rules of Relational Hoare Logic}\label{sec:appendix-rhl}

\begin{figure*}
  \begin{minipage}{\linewidth}
    \small
    \begin{mathpar}
      \inferrule[Skip]{
}{
  \rhl{\ensuremath{{\ensuremath{{\text{\textbf{skip}}}}}}}{\ensuremath{{\ensuremath{{\text{\textbf{skip}}}}}}}{\pre}{\pre}
}{}

      \inferrule[Seq]{
  \rhl{c_1}{c_2}{\pre}{\pre'}
  \\\\
  \rhl{c_1'}{c_2'}{\pre'}{\pre''}
}{
  \rhl
    {c_1 \mathop{;} c_1'}
    {c_2 \mathop{;} c_2'}
    {\pre}
    {\pre''}
}{}

      \inferrule[Case]{
  \rhle{e_1}{e_2}{\pre}{=}
  \\
  \rhl{c_1}{c_2}{\pre \land e_1}{\post}
  \\\\
  \rhl{c_1'}{c_2'}{\pre \land \neg e_1}{\post}
}{
  \rhl
    {\ensuremath{{\ensuremath{{\text{\textbf{if}}}}~({e_1})~\{{c_1}\}~\{{c_1'}\}}}}
    {\ensuremath{{\ensuremath{{\text{\textbf{if}}}}~({e_2})~\{{c_2}\}~\{{c_2'}\}}}}
    {\pre}
    {\post}
}{}

      \inferrule[Conseq]{
  \rhlPQ{\Ip['], \Iq[']}{c_1}{c_2}{\pre'}{\post'}
  \\
  \pre \implies \pre'
  \\\\
  \Ip['] \implies \Ip
  \\
  \post' \implies \post
  \\
  \Iq \implies \Iq[']
}{
  \rhlPQ{\Ip,\Iq}{c_1}{c_2}{\pre}{\post}
}{}
    \end{mathpar}
  \end{minipage}
  \caption{Extra rules of relational Hoare logic.}\label{fig:extra-rhl}
\end{figure*}

\Cref{fig:extra-rhl} presents the remaining rules.
The rule of consequence (or weakening), \hyperref[fig:extra-rhl]{\textsc{Conseq}}{}, allows us to
strengthen the precondition (resp.\ the event postcondition) and weaken
the postcondition (resp.\ the event precondition) of an RHL judgment.

 \section{Proof Effort}\label{appendix:proof-effort}

\begin{table*}
  \caption{Reduction in proof effort of the \textsc{Jasmin}{} compiler.
    The ``Shared'' column is the size---in lines of code (LOC)---of
    the auxiliary lemmas and definitions shared by both the previous
    and new proofs.
    The ``Previous'' and ``New'' columns present the size of
    preexisting and new proofs, respectively.
    The ``Reduction'' column is the reduction percentage including the
    ``Shared'' parts not modified by our work (i.e., one minus ``New''
    plus ``Shared'' over ``Previous'' plus ``Shared'').}\label{tbl:loc}
  \small
  \renewcommand{\arraystretch}{1.12}
  
  \begin{tabular}{
      l
      S[table-format=4.0,table-number-alignment=right]
      *{2}{S[table-format=2.0,table-number-alignment=right]}
      S[table-format=2.1,table-number-alignment=right]
    }
    \toprule
    & \multicolumn{3}{c}{LOC} & \%
    \\
    \cmidrule(r){2-4} \cmidrule(r){5-5}
    Compiler pass
    & \multicolumn{1}{l}{Shared}
    & \multicolumn{1}{l}{Previous}
    & \multicolumn{1}{l}{New}
    & \multicolumn{1}{l}{Reduction}
    \\
    \midrule
    Int to Word & 214 & 195 & 84 & 27.1
    \\
    Array Copy & 337 & 214 & 107 & 19.4
    \\
    Add Array Initialization & 352 & 263 & 137 & 20.5
    \\
    Spilling & 416 & 266 & 142 & 18.2
    \\
    Inlining & 255 & 339 & 283 & 9.4
    \\
    Function Pruning & 190 & 224 & 86 & 33.3
    \\
    Constant Propagation & 969 & 341 & 277 & 4.9
    \\
    Dead Code Elimination & 288 & 361 & 177 & 28.4
    \\
    Unrolling & 46 & 168 & 71 & 45.3
    \\
    Remove Array Initialization & 431 & 232 & 89 & 21.6
    \\
    Reference Arguments & 439 & 251 & 132 & 17.2
    \\
    Register Array Expansion & 674 & 243 & 219 & 2.6
    \\
    Remove Globals & 490 & 330 & 188 & 17.3
    \\
    Load Immediates & 142 & 277 & 100 & 42.2
    \\
    x86-64 Instruction Selection & 1681 & 254 & 90 & 8.5
    \\
    ARMv7 Instruction Selection & 1558 & 387 & 110 & 14.2
    \\
    RISC-V Instruction Selection & 583 & 173 & 72 & 13.4
    \\
    SLH Instruction Selection & 795 & 400 & 296 & 8.7
    \\
    Inline Propagation & 538 & 218 & 142 & 10.1
    \\
    Stack Allocation & 10688 & 845 & 695 & 1.3
    \\
    Lower Addresses & 202 & 225 & 84 & 33.0
    \\
    Register Renaming & 433 & 445 & 198 & 28.1
    \\[1ex]
    Total & 21721 & 6651 & 3779 & 10.1
    \\
    \bottomrule
  \end{tabular}
{}
\end{table*}

The \textsc{Jasmin}{} extensions presented in this work involved adapting the
proofs of all passes of the compiler.
This section briefly discusses the proof effort involved, highlighting
the application of RHL for compiler proofs.
All in all, the \textsc{Jasmin}{} framework comprises 110K lines of \textsc{Rocq}{} and
31.7K lines of OCaml code.
The \textsc{Rocq}{} part includes the compiler passes, their proofs, and the
proof of generic assumptions on the architectures.

The proof effort in the compiler has been reduced as a result of our
work; as a rough estimate, the size of the compiler front-end proofs has
decreased by 10\%.
Pleasingly, many preexisting proofs about expressions, straight-line
code, and different well-formedness conditions extend to the new setting
with negligible effort.

\Cref{tbl:loc}{} shows some statistics on the proof effort before and after
our work.
We show the instance of the Instruction Selection pass for each
supported architecture separately.
The Live-Range Splitting pass is absent because it is the composition of
Register Renaming and Dead Code Elimination, and therefore already
accounted for.
Similarly, we omit Remove Unused Results since it is implemented inside
Dead Code Elimination.
 
\end{document}